# A Survey on Silicon Photonics for Deep Learning

FEBIN P SUNNY, COLORADO STATE UNIVERSITY
EBADOLLAH TAHERI, COLORADO STATE UNIVERSITY
MAHDI NIKDAST, COLORADO STATE UNIVERSITY
SUDEEP PASRICHA, COLORADO STATE UNIVERSITY

Deep learning has led to unprecedented successes in solving some very difficult problems in domains such as computer vision, natural language processing, and general pattern recognition. These achievements are the culmination of decades-long research into better training techniques and deeper neural network models, as well as improvements in hardware platforms that are used to train and execute the deep neural network models. Many application-specific integrated circuit (ASIC) hardware accelerators for deep learning have garnered interest in recent years due to their improved performance and energy-efficiency over conventional CPU and GPU architectures. However, these accelerators are constrained by fundamental bottlenecks due to 1) the slowdown in CMOS scaling, which has limited computational and performance-per-watt capabilities of emerging electronic processors, and 2) the use of metallic interconnects for data movement, which do not scale well and are a major cause of bandwidth, latency, and energy inefficiencies in almost every contemporary processor. Silicon photonics has emerged as a promising CMOS-compatible alternative to realize a new generation of deep learning accelerators that can use light for both communication and computation. This article surveys the landscape of silicon photonics to accelerate deep learning, with a coverage of developments across design abstractions in a bottom-up manner, to convey both the capabilities and limitations of the silicon photonics paradigm in the context of deep learning acceleration.

## 1 INTRODUCTION

Deep Learning, which is a sub-field of Artificial Intelligence (AI), has been at the heart of many unprecedented successes in recent years for solving very difficult problems in the domains of computer vision, natural language processing, time series predictions, and understanding big data. This development is remarkable considering how most researchers had abandoned the idea of using deep learning in the 1990s, due to the difficulties in training such models. But seminal work by Hinton et al. in 2006 showed how it was possible to train a deep neural network to recognize handwritten digits with state-of-the-art precision (>98%) [1]. They called their technique "Deep Learning." It did not take long for the scientific community to take notice, and in the following years many researchers showed that deep learning was not only possible, but capable of achieving remarkable performance for solving many problems that no other machine learning techniques could match. Indeed, today deep learning models are at the heart of smart technological solutions that we all use regularly, such as web search engines, music and video recommendation engines, speech recognition in virtual assistants, and object detection in Internet-of-Things (IoT) cameras. Many emerging applications such as self-driving cars [2], autonomous robotics [3], fake news detection [4], pandemic growth and trend prediction [5], network anomaly detection [6], and real-time language translation [7] are being powered by increasingly sophisticated deep learning models.

This research is supported by grants from NSF (CCF-1813370, CCF-2006788).

Authors' addresses: F. Sunny, E. Taheri, M. Nikdast, and S. Pasricha (corresponding author), Department of Electrical and Computer Engineering, Colorado State University, Fort Collins, CO 80523-1373; email: { febin.sunny, ebad.taheri, mahdi.nikdast, sudeep}@colostate.edu. Both F. Sunny and E. Taheri contributed equally to this manuscript.

The magic behind deep learning owes much to our brain's architecture. As far back as 1943, the neurophysiologist Warren McCulloch and mathematician Walter Pitts presented a simplified model of how biological neurons work together in animal brains to perform complex computations [8]. This was the first artificial neural network (ANN) architecture and it inspired a race to build intelligent machines that could rival and eventually surpass the capabilities of the human brain. The introduction of the perceptron in 1957 by Frank Rosenblatt was another landmark, showing how the simple ANN could be trained to solve classification problems [9]. However, the limited capabilities of hardware to run even moderately complex ANNs led researchers to abandon the study of ANNs in the late 1960s. Even though new architectures and better training techniques emerged in the 1980s and early 1990s, progress was limited due to several factors, a crucial one of which was the lack of powerful machines to train and run these models. Fortunately, over the past decade, ever improving capabilities of Complementary Metal Oxide Semiconductor (CMOS) fabrication technology have enabled extremely powerful TFLOPs-class Graphics Processing Unit (GPU) and CPU processing chips with billions of transistors in small form factors that have made it possible to train and use deep ANN (i.e., multi-layer perceptron (MLP)) architectures in a timely and cost-effective manner. Coupled with the availability of large datasets in the IoT and Big Data era, theoretical advances in training algorithms, and the emergence of new deep ANN architectures such as convolutional neural networks (CNNs), deep learning has now established its dominance over other machine learning models for many problems of interest in the domains of computer vision, natural language processing, and general pattern recognition.

With researchers creating deeper and more complex MLP and CNN architectures to push deep learning performance levels to new heights, the underlying hardware platform must consistently deliver better performance levels while also satisfying strict power dissipation limits. This endeavor to achieve higher performance-per-watt has driven hardware architects to design application-specific integrated circuit (ASIC) accelerators for deep learning that have much higher performance-per-watt than conventional general-purpose CPUs and GPUs. IBM's 4096 core TrueNorth chip that was released in 2014 was one of the earliest high-profile ASIC deep learning accelerators [10]. Since then, many other accelerators have become available, including Intel's Loihi [11] and Google's Tensor Processing Units (TPU) [12]. Several academic efforts have also led to the design of new types of ASIC and FPGA-based deep learning accelerators [13]–[17]. Even conventional GPUs and CPUs have evolved to speed up deep learning model execution, e.g., Nvidia GPUs now include tensor cores [18], and CPUs support increasingly advanced vector instructions [19], both of which are designed to accelerate common matrix and vector operations in deep learning processing. Beyond digital domain solutions, accelerators have also been proposed that work in the analog domain [20]–[22] or the analog-digital mixed signal domain [23]–[25].

Unfortunately, these electronic accelerator architectures are beginning to face fundamental limits in the post Moore's law era where processing capabilities are no longer improving as they did over the past several decades [26]. In particular, moving data electronically on metallic wires in these accelerators is a major bandwidth and energy bottleneck [27]. Photonic interconnects offer one of the most promising solutions to overcome these data movement challenges. Photonic links have already replaced metallic ones for light-speed information transmission at almost every hierarchy level of computing, and are now being considered for integration at the chip-scale [28]. The advent of silicon photonics, which allowed for cost-effective integration of optical components based on CMOS electronics manufacturing, has been one of the major catalysts for chip-scale photonic interconnects [29]. Even more remarkable is the fact that various computations required in deep learning, such as matrix-vector multiplications, can be performed entirely in the optical domain [30]. Thus, we are close to a point where it will become possible to realize deep learning accelerators that utilize silicon photonics for both communication and computation. Such silicon photonics based deep learning accelerators can provide unprecedented levels of energy efficiency and parallelism. For instance, with multiply and accumulate (MAC) operations that dominate deep learning computations, photonics-based accelerators can achieve energy footprint efficiency (defined as (MAC/s/mm$^2$) / (joules/MAC)) that is almost 1000× better compared to the most energy efficient electronic accelerators today [31]. Moreover, the operational bandwidth of photonic MACs can approach the photodetection rate, typically in the range of hundreds of GHz. This is far superior to electronic systems today that operate at a clock rate of a few GHz [32].

In this article, we survey the landscape of silicon photonics for accelerating deep learning model training and inference. Prior surveys on a related theme have either focused on surveying performance and energy aspects of a specific type of photonic neural network architecture (e.g., reservoir computing architectures [33]–[35] and Broadcast-and-Weight (B&W) architectures [31], [36]–[38]), or created a simplified classification based on implemented neural-network models (e.g., MLPs, CNNs) [39]. In contrast, in this article we provide a different and more comprehensive tutorial of developments in silicon photonics based deep learning acceleration, with a bottom-up classification across design-layer abstractions: from lower-level fabrication alternatives and devices, to the spectrum of neuron microarchitectures, and covering a variety of integrated neural network architectures at the system level. Our aim is to provide an overview of the plethora of design choices available with silicon photonics towards the realization of photonic deep learning accelerators, along with a discussion of their advantages and limitations. The ability to utilize CMOS-compatible materials, such as germanium (Ge) and silicon nitride (SiN), has enabled new variants of photodiodes, modulators, couplers, and lasers with very interesting performance-energy-reliability tradeoffs. These tradeoffs also exist for different fundamental device types, such as Mach–Zehnder Interferometers (MZIs) and Microring Resonators (MRs), which can be used as the building blocks of photonic artificial neurons. Many different types of photonics-based artificial neuron microarchitectures have been proposed, such as the noncoherent B&W architecture [40] and the coherent artificial linear neuron (COLN) [36]. Such neurons can be cascaded together while respecting photonic signal loss profiles and Signal-to-Noise Ratio (SNR) goals, to construct larger photonics-based neural network fabrics. We believe that such a classification across the design abstractions in a bottom-up manner provides an intuitive and useful way to understand the capabilities and limitations of the silicon photonics paradigm in the context of deep learning acceleration.

The rest of this article is organized as follows. Section 2 starts out with a brief discussion of deep learning models. Section 3 presents an overview of fundamental silicon photonic devices that are widely used in photonic neural networks and relevant for accelerating deep learning models. Section 4 describes various types of artificial neuron architectures designed with silicon photonic components. These neuron architectures form the building blocks of photonic neural network architectures which are discussed in Section 5. Lastly. Section 6 wraps up with a discussion of outstanding challenges and opportunities with silicon photonics for deep learning acceleration.

## 2 AN OVERVIEW OF DEEP LEARNING

Deep learning is a subset of Machine Learning (ML), which itself is a subset of the broader field of AI. Deep learning aims to emulate the deep architecture of a human brain, which has billions of interconnected neurons acting as computational units. Human brains also work hierarchically, starting from simpler concepts and then combining them to learn more abstract ideas. This mode of learning is reflected in deep learning models which break down input data into features and then recombine them to perform the task at hand (e.g., detection, classification). Once relevant features have been learned by a deep learning model in the training phase, the model can be applied to tasks of a similar nature, with no human intervention.

As mentioned earlier, deep learning has gained a lot of attention in recent years. But the concept is not new. The idea to make machines as intelligent as humans is the very basis of the analytical engine conceived by Charles Babbage in 1837. The field of AI and the research into making machines capable of thinking like humans started as far back as the mid-20$^{th}$ century, with the computational model for neural networks and neuron operation developed by Warren McCulloch and Walter Pitts in 1943 [8]. The perceptron algorithm was invented by psychologist Frank Rosenblatt in his seminal 1957 paper [9]. Leveraging this algorithm, Rosenblatt created the first single-layer perceptron (see figure 1 (a)) that is an electronic computational device adhering to the biological principles behind how the human brain functions.

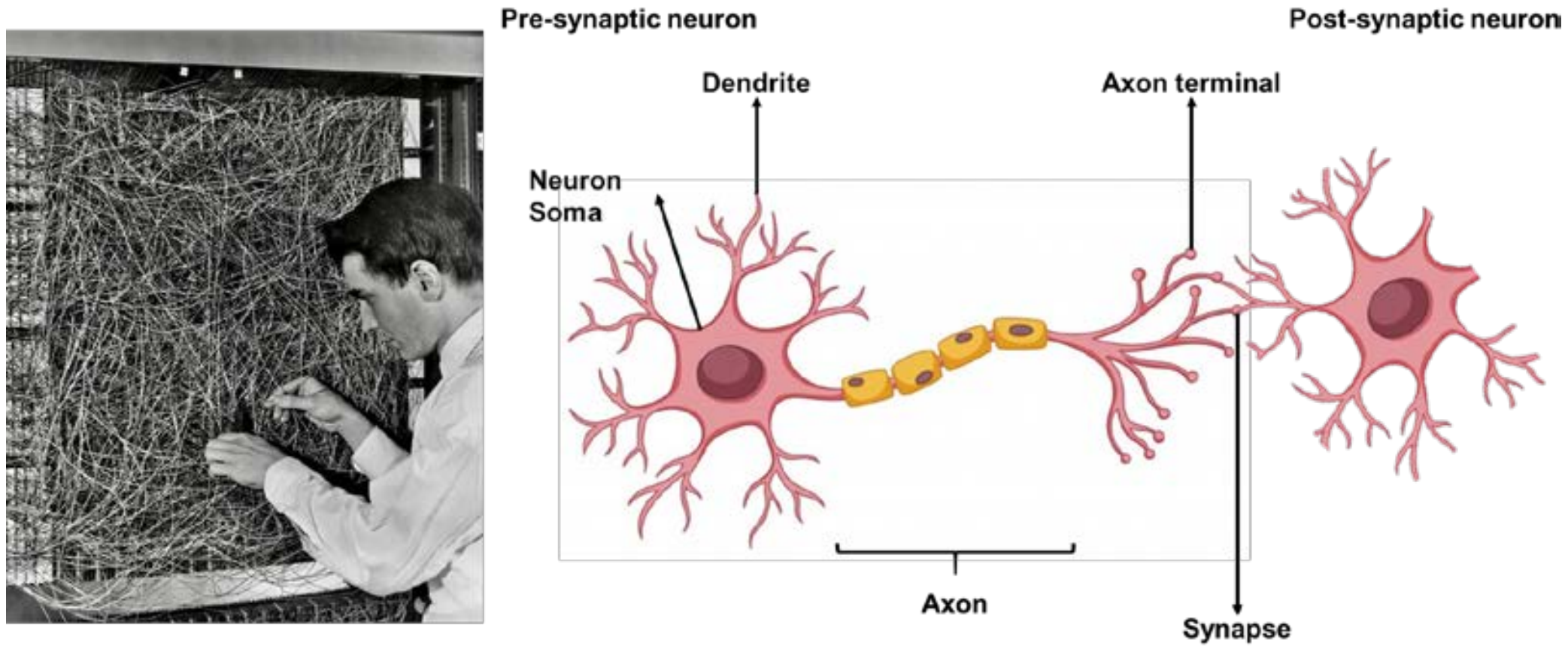

(a) (b)

Figure 1: (a) Frank Rosenblatt with his Mark-1 single-layer perceptron; (b) A depiction of the neuron and the synaptic connection to another neuron. This is a simplistic depiction of a neuron, showing only the most basic components.

## 2.1 Neuron Models

The human brain is comprised of around 86 billion neurons [41], each interconnected using dendrites and axon connections (see figure 1(b)). The biological neuron, which is the main processing component in the brain, consists of a soma, dendrite, axon, and synapse. The soma or cell body of a neuron contains the nucleus and other structures common to living cells. These structures support the chemical processing within the neuron. The dendrites are extensions from the neuron soma and act as receivers or inputs into the neuron. The axons form the "tails" of the neuron and carry signals away from the soma. The axon can further split into branches to achieve incredible interconnectivity. Depending on the type of neuron, this interconnectivity can reach up to 100,000 fan-out connections, a number that is inconceivable to achieve today with CMOS logic gates. The connection between neurons via their extremities occurs at contact points called synapses (figure 1 (b)). Neural signals are transmitted in the form of electrical impulses along these interconnections made of dendrites, axons, and synapses. These connections between the neurons along synapses can strengthen or weaken over time depending on the activity in the synapses. This is referred to as synaptic plasticity. Synaptic plasticity is also hypothesised to be a key component in encoding memories in the brain [42].

The McCulloch-Pitts model represents a very simplified model of this biological neuron [8]. It is comprised of a summation unit and then a threshold gate, as shown in figure 2. The summation unit can have N inputs, with each input assigned a weight value. The products of the inputs and their corresponding weights are summed at the summation unit (Σ), and this sum is passed onto the threshold gate. If the summed signal exceeds the threshold, the gate generates a signal and the neuron generates (or fires) an output signal. The McCulloch-Pitts model adapted a linear threshold for their threshold gate, so the neuron either fires or not depending on the output from the summation unit, making it a binary output neuron. In more modern terms, this linear threshold in the model is called the model's activation function.

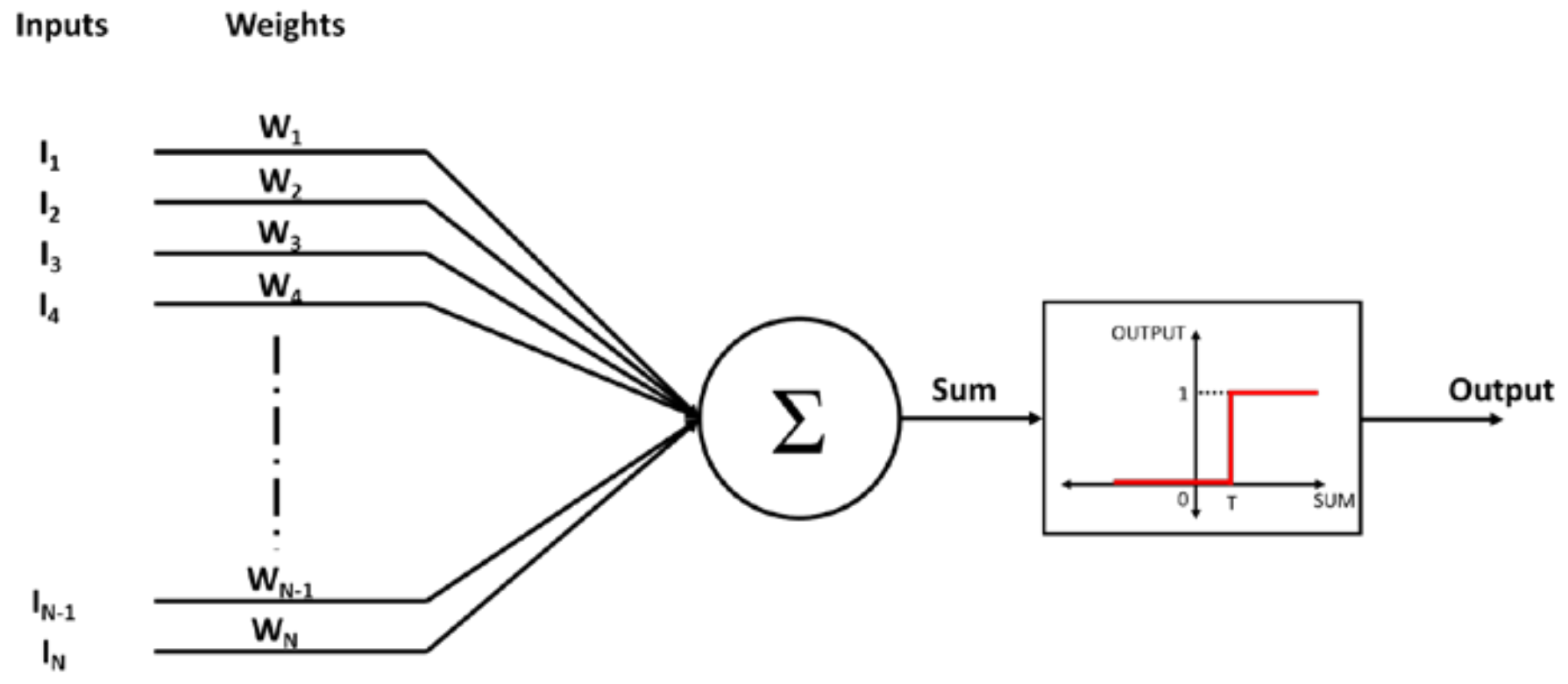

Figure 2: The McCulloch-Pitts computational model of the biological neuron [8]. A linear activation function is used in the model. The neuron fires only when the sum value crosses a threshold, T, making the McCulloch-Pitts model a binary output neuron.

This binary model is a powerful tool and can be used to reach solutions for simple binary classification problems. But for more complex tasks, more complex activation functions and neuron models are necessary. There are other neuron models that mimic the biophysical characteristics of the neuron such as the Hodgkin-Huxley Model [43] and many others [44]–[46]. Using such models requires complex calculations of their biological interactions, which can be computationally taxing. To circumvent this, computationally efficient Integrate-and-Fire (IF) neurons are preferred. IF neurons are, incidentally, one of the oldest neuron models to appear in literature [47]. The Leaky Integrate-and-Fire (LIF) neuron [48] is another extremely popular neuron model, due to its simplicity while being able to achieve complex functionalities in deep neural networks. Another neuron model, which emulates biophysical characteristics of the neuron, much like the Hodgkin-Huxley model, but with lower computational complexity, is the Izhikevich spiking neuron model [49].

All of these neuron models follow the same basic principle: neurons accept input signals from multiple synapses, sum them, and fire a corresponding output if a threshold is exceeded. The differences between them arise in how the threshold and the bio-physical interactions are modeled. The neuron models discussed here help in mimicking the biological operation of the brain and hence are an integral part of a neural-network model called Spiking Neural Networks (SNNs), which we discuss next. This inter-disciplinary concept of mimicking the brain using advanced neuron models and implementing neural systems is often referred to as neuromorphic engineering or neuromorphic computing.

### 2.2 Spiking Neural Networks (SNNs)

The idea behind Spiking Neural Networks (SNNs) is to emulate the human brain as closely as possible. The brain exhibits low power consumption, fast inference, event-driven processing, continuous learning, and massive parallelism. It is also based on event-based computation, where information is encoded in spikes [50]. Indeed, SNNs were introduced in 1997 to emulate this spike-based method of computation [48]. SNNs utilize asynchronous, event-driven processing to implement neural networks. The inputs to an SNN neuron are referred to as action potentials or spikes (see figure 3), which the neuron receives from its pre-synaptic neuron. These binary spikes can carry information through the network either via rate coding or temporal encoding. Rate coding—also referred to as frequency coding—is the model of neuronal firing which assumes that the information about the stimulus that triggered the neuron to fire can be encoded in the rate at which the neuron fires. Thus, this method of information encoding requires precise calculation of firing rates. Temporal encoding utilizes the temporal resolution or the time between consequent spikes to carry information. For both encoding types, the connection between the neurons is represented by synaptic weights, which influence the input spikes, to create a weighted spike train at each neuron's input. The weighted input spikes affect the membrane potential of the neuron, which refers to the intensity of activation of the neuron. Once the membrane potential exceeds a threshold, the neuron generates a spike (i.e., fires an action potential) to its post-synaptic neuron. This activity is illustrated in figure 3.

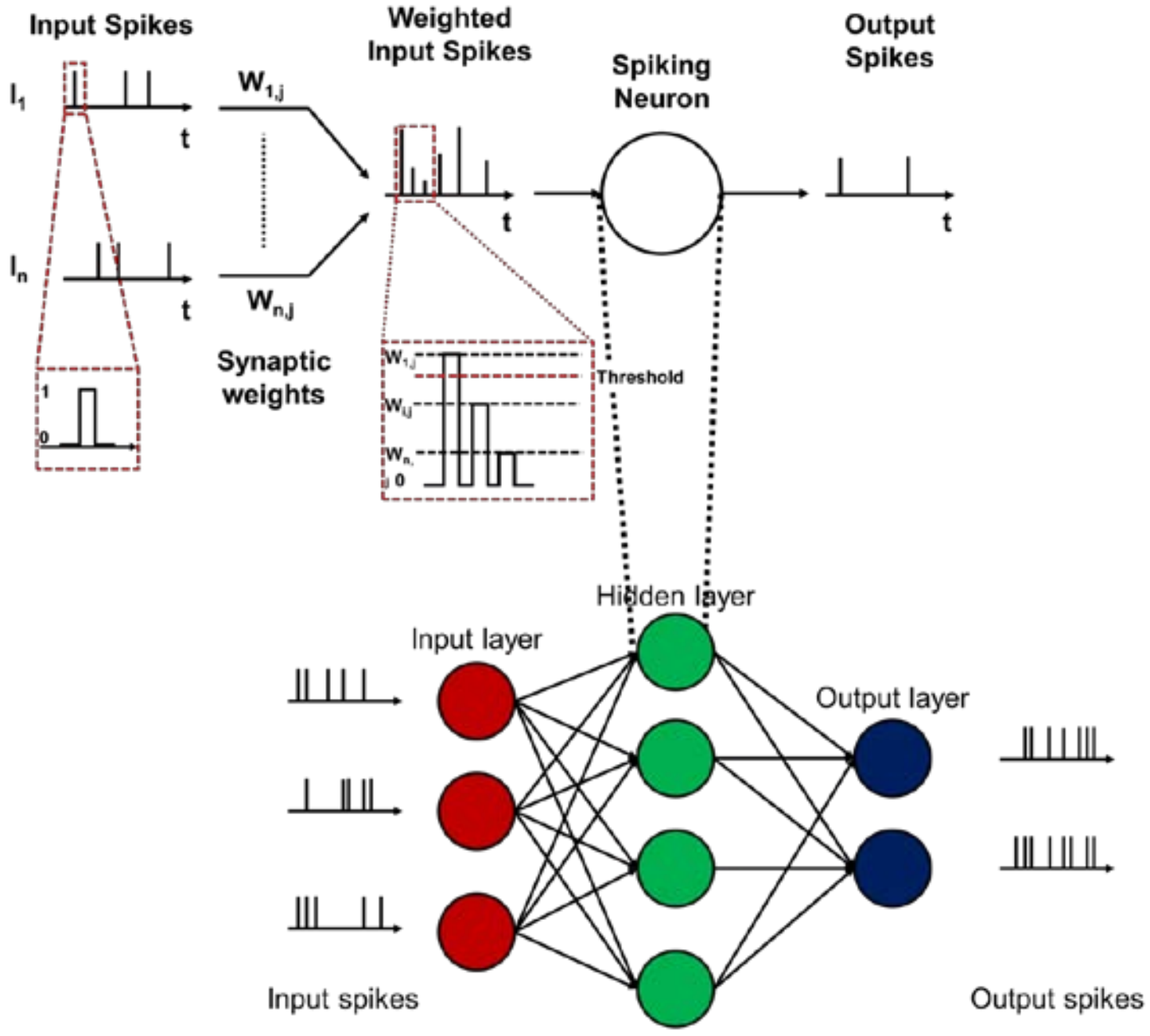

Figure 3: A simple representation of how a spiking neuron, the fundamental unit of an SNN, functions. The spiking neuron shown here can be any of the models mentioned in Section 2.1 in order to implement an SNN.

There has been increased interest in implementing brain-like computation in the last decade [51] to overcome limitations set by conventional Von-Neumann architectures. Supercomputers today can achieve hundreds of peta FLOPS (floating point operations per second) in processing data but at the cost of tens of millions of Watts [52], whereas the human brain achieves this feat at the cost of just 20 Watts [53]. SNN implementations hope to achieve this remarkable level of energy efficiency exhibited by the human brain. To realize this goal, many technologies are actively being explored, including CMOS [54]–[56], new types of transistors [57]–[59], and non-volatile memory [60]–[63]. Employing such technological advances, there have been various SNN accelerator implementations. As an example, *SpiNNaker* [64], from the University of Manchester, was built using ARM processors and implements the Izhikevich neuron model for computational efficiency. It utilizes a Globally Asynchronous, Locally Synchronous (GALS) communication system between the processing cores and Synchronous Dynamic Random Access Memories (SDRAMs) to store the synaptic weight values. *TrueNorth* [65] from IBM contained 5.4 billion transistors while using only 70 mW to operate. The processor is comprised of arrays of low power neurosynaptic processing units, each containing memory, processor, and communication subsystems to mimic neural functions. TrueNorth implements the LIF neuron model in its SNN. *Loihi* [66] from Intel with 128 neuromorphic cores and 130,000 neurons, is another such implementation that exhibited 1000× speed and 10,000× energy efficiency compared to a CPU [67]. Loihi implements a variant of the LIF neuron called "current based synapse (CUBA) LIF neuron."

### 2.3 Artificial Neural Networks (ANNs)

The emergence of the computational model for representing neural activities paved the way for Artificial Neural Networks (ANNs). When compared to SNNs, ANNs are markedly abstract in their approach to implementing brain functions. The weights, which represent the synaptic plasticity, are simple scalars. The neurons utilized are also much simpler, and tasked with accumulation of input-weight products followed by passing the resulting output through a non-linear function. ANNs emulate brain activity by simulating a collection of interconnected neurons, arranged in layers. The simplest representation will have three layers: an input layer, an output layer, and a hidden layer in between these two layers (figure 4(a)). The input layer accepts data from outside the ANN; the hidden layer is where the computation happens; and the output layer is where we can get the results from the neural network. The activation functions also play an important part in simulating intelligence. Mathematically, without appropriate activation functions, the neuron model is a simple linear model, which multiplies and accumulates input-weight products. To introduce non-linearity into the network and make it possible for the model to approximate more complex functions, we need to use appropriate non-linear activation functions, such as *sigmoid*, *Rectified Linear Unit (ReLu)*, and *tanh* to list a few. Utilizing these functions, ANNs are able to learn very complex non-linear relationships between input features.

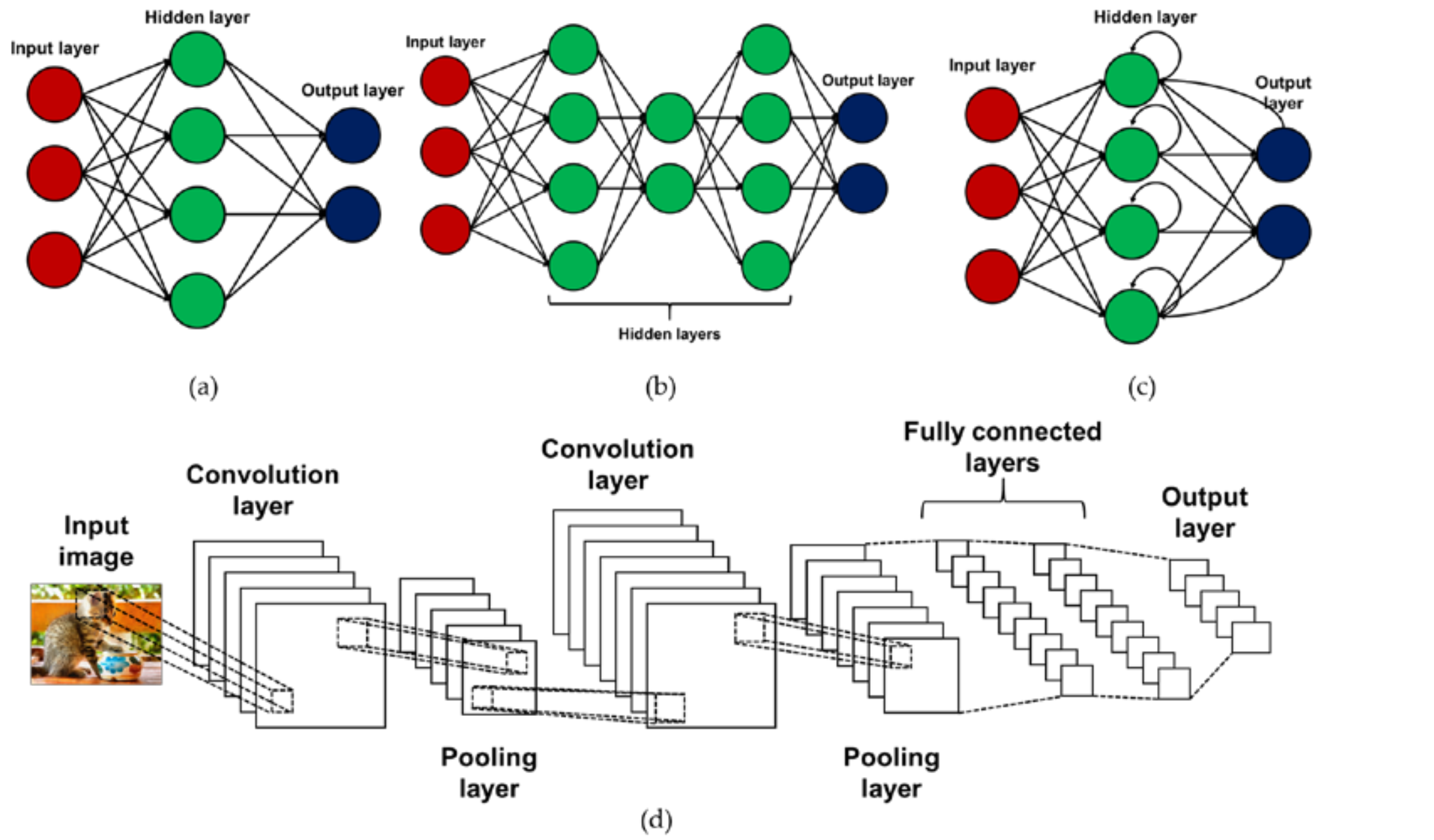

Figure 4: (a) Layered architecture of an ANN; this figure shows a shallow ANN with one hidden layer. (b) As the number of hidden layers in the ANN increase, we achieve DNNs. Note that the layers of an ANN are populated by neurons, with associated weights and biases. (c) A fully connected RNN, which shows the feedback connections in its hidden layer, and simulates memory or saved states in this architecture. (d) Representation of various layers and operations in a CNN.

An important distinction to be made here is between ANNs and traditional ML algorithms such as Support -Vector Machines (SVMs), K-Nearest Neighbours (KNN), Random Forests (RF), etc. What makes ANNs distinct is their ability to handle large quantities of data, with minimal human intervention. Traditional ML algorithms usually require a human expert to provide the necessary rule set on which they operate. Often, assistance for feature extraction from the data is also needed, e.g., for kernel selection in SVMs.

Deep Neural Networks (DNNs) are ANNs with multiple hidden layers (see figure 4(b)) which can utilize the complex interconnectivity among the neurons to compute and efficiently represent very complex non-linear relationships after being trained. During the training phase, input activations traverse a forward path from the input to the hidden layers and finally to the output layer. The error (often called the loss) between the DNN output and the expected output is backpropagated through the model to update the neuron weights and biases, in a manner that reduces the loss. This process is iteratively repeated until the model output (e.g., image class prediction) is as close as possible to the expected outputs, i.e., the loss is minimized. After training, the model can make predictions given an input, in what is referred to as the inference phase. The training phase for DNNs is a time and resource intensive process, compared to the inference phase. The notable learning architectures which utilize DNNs include Multi-Layer Perceptrons (MLPs), Recurrent Neural Networks (RNNs), Deep Boltzmann Machines (DBMs), Stacked Auto-Encoders (SAEs), and Convolutional Neural Networks (CNNs).

MLPs only include feedforward fully connected (FC) layers as shown in figure 4(b), where each neuron in a layer is connected to each neuron in the preceding and following layers. Some model architectures can exhibit temporal dynamic behaviour and possess an internal state or memory because of their network structure. These can be broadly referred to as Recurrent Neural Networks (RNNs). The inherent memory in their structure makes them ideal for recognition tasks such as pattern recognition, handwriting and speech recognition, natural language processing, etc. Research on RNNs began with David Rumehalt in 1986 [68]. As of today, many RNN variants are popular, including Long Short-Term Memory (LSTM) networks, Gated Recurrent Units (GRUs), Continuous Time RNNs (CTRNNs), etc. A simplified RNN is shown in figure 4(c).

CNNs target the processing of 2D or higher dimensional features instead of the 1D ones in MLPs. They are widely used for classification problems in image and video processing. The structure of a CNN is depicted in figure 4(d). A typical CNN contains three types of layers: convolutional (Conv), pooling (Pool), and fully connected (FC). In Conv and Pool layers, there are multiple channels (called feature maps) that extract different local features from the input data. These layers combine the lower-level features from multiple channels of the previous layer, into higher-level features that are passed to the next layer, till the final classification layer where an output prediction is generated. Conv layers have much fewer parameters than FC layers, but involve a high computational footprint due to the many convolution operations that are required between filter weights and the input activations, across all their channels. Pooling layers generate output activations based only on the local receptive field in the corresponding input feature map (e.g., a single “pooled” output from a group of 2x2 inputs). The two widely used variants of pooling layers are max and average pooling, and they produce the maximum or average value of each receptive field, respectively. Lastly, FC layers follow Conv and Pool layers, and act to work as a classifier with the extracted features, similar to how these layers are used in MLPs.

DNNs are beginning to be widely used in real-world applications such as autonomous driving, robotics, and IoT processing. The resource intensity needed for training DNN models was met by the emergence of GPUs that are used for significantly reducing the training time of DNNs, due to the greater data and thread level parallelism supported in GPUs than CPUs. Much like for SNNs, there is growing interest in designing energy-efficient ASIC accelerators for DNNs. Such DNN accelerators, e.g., the Neural Processing Unit (NPU) [69] are designed to accelerate the inference phase, although a few accelerators are also aimed at improving training performance. An example of a DNN accelerator that has been very successful for accelerating both training and inference with DNNs is the TPU [12] from Google. The TPU has dedicated matrix-multiplication units and distributed memory management that makes it ideal for handling the heavy lifting needed to train DNN models and also for inference tasks. TPUs are deployed widely in Google’s data centers. Newer GPU architectures have also adopted similar Tensor cores for DNN acceleration [70]. Researchers have also suggested utilizing non-volatile memory technology and Processing-In-Memory (PIM) for DNN accelerators. PRIME [71] and ISAAC [20] are examples of such accelerators that utilize Resistive Random-Access Memories (ReRAMs) and PIM to accelerate DNN execution.

### 2.4 Reservoir Computing (RC)

RC is a less popular neural-network model than ANNs and SNNs but is covered here briefly because of its amenability to photonics-based implementations and consideration in prior photonics-based designs. RC can be thought of as a type of RNN where only the parameters of the last, non-recurrent output layer (called readout layer) are trained, while all the other parameters are randomly initialized, subject to some condition that essentially prevents chaotic behavior, and then they are left untrained. RC thus represents a type of partially adaptive RNN, which is in contrast with the fully adaptive approach of conventional ANNs and SNNs. The reservoir is comprised of connected non-linear nodes and is a fixed recurrent network as shown in figure 5. This reservoir performs many non-linear operations and the outputs from these are combined into linear combinations to complete a task. The user has little direct access to the reservoir and the output manipulation is restricted to the readout layer. To reach the desired behavior, trained linear classifiers at the readout layer are utilized in a supervised learning framework. The advantage of having such a fixed random network becomes apparent with certain (particularly photonics-based) hardware platforms where the possibility of setting all the internal parameters is not possible.

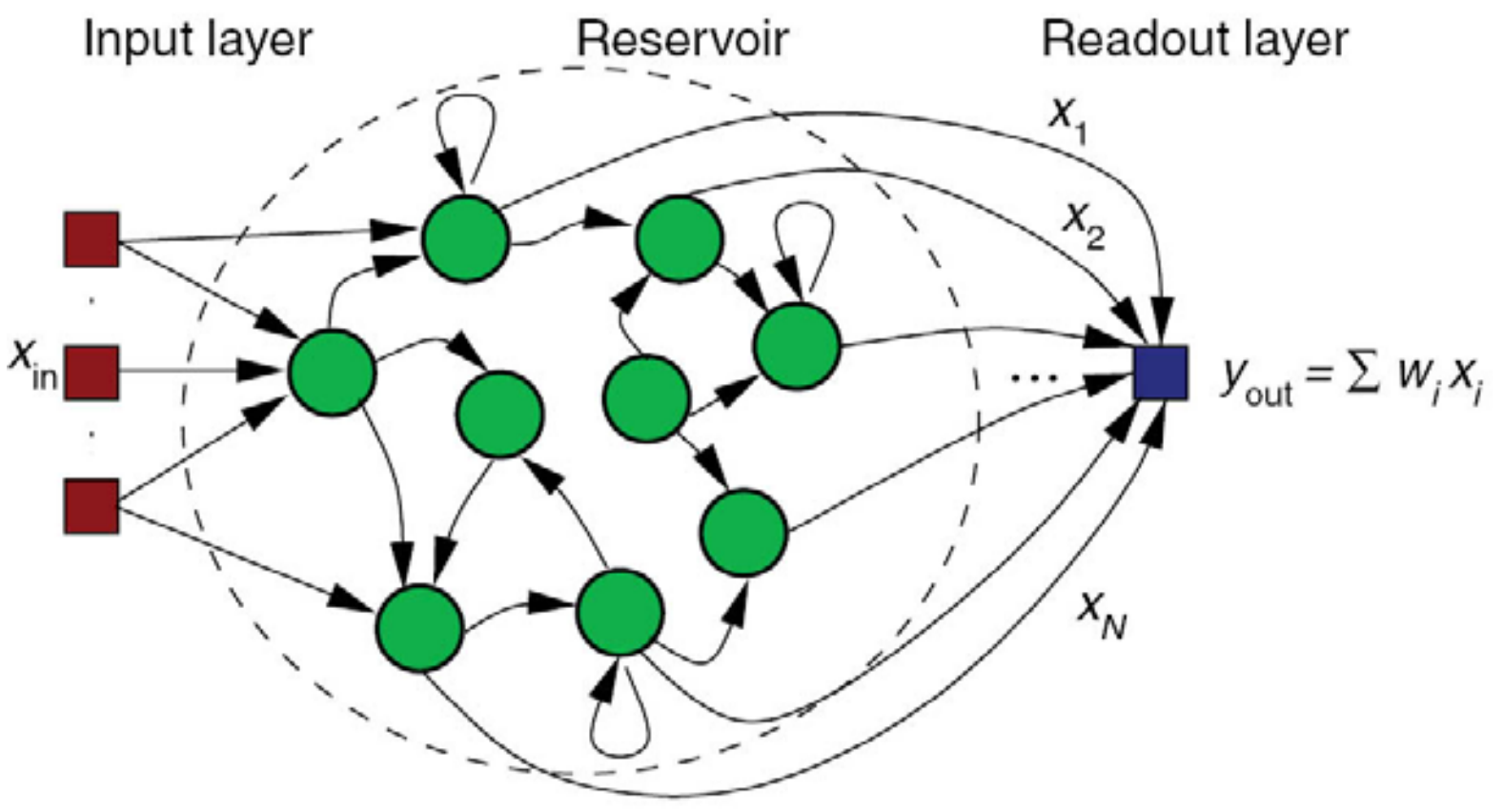

Figure 5: Standard layout of a reservoir computing (RC) architecture with the input layer in red, the reservoir in green (with randomized but fixed connections), and the readout layer in blue where the outputs from the reservoir are consolidated into the desired output.

RC can be utilized to emulate the behavior of conventional ANNs due to its intrinsically parallel nature. Like a neural network, a reservoir often consists of a large number of interconnected non-linear nodes. Therefore, existing hardware implementations of neural networks can be and have been used as reservoirs, as in [72]. However, unlike traditional neural networks, the interconnection weights need not be adaptable or even exactly controllable. In fact, only a global gain scaling is required for weight manipulation in RC. This makes the requirements for reservoir implementations more relaxed and allows for the exploration of technologies that might be less suitable for implementing traditional, fully trainable neural networks. Thus, RC was a popular target for early implementation of all-optical computing and there have been many bench-top models which demonstrated how all-optical reservoir computing can be achieved [73]–[81]. These implementations were often built using telecom equipment (e.g., fiber optical loops, MZIs, lasers, photodetectors and Arrayed Waveguide Gratings (AWGs)) and provided proof of concept validations of the effectiveness of optical reservoir computing.

To implement RC and the other DNN models on a computing chip, silicon photonics is a promising emerging technology candidate. We will now provide an overview of silicon photonics technology (Section 3), followed by in-depth discussions on neuron microarchitectures implemented using this technology (Section 4), and various deep learning architectures built using photonic neurons (Section 5).

## 3 AN OVERVIEW OF SILICON PHOTONICS

Optical communication has been widely employed in communication networks wherever low-cost and high-bandwidth communication at low power consumption and over large distances is required, e.g., in long-haul telecommunication networks. In recent years, silicon photonics has enabled CMOS-compatible integrated photonics and gained widespread adoption in commercial offerings for low-cost optical interconnects in data centers. Optical interconnects are now being aggressively considered at much smaller scales, to connect multiple processing chips at the board level, and even to connect cores within a single computing chip. As the name suggests, silicon photonics employs light, which is guided through the silicon (Si) medium on a CMOS chip, for communication. In a silicon-on-insulator (SOI) fabrication platform, the high refractive index contrast between the waveguide core (silicon) and the waveguide's cladding and substrate (e.g., silicon dioxide) results in guided optical signal propagation through total internal reflection. A single waveguide can be used to carry multiple wavelengths of light simultaneously, each capable of carrying data at high speed and high frequency, and without any interference. This is possible using a technique called Wavelength-Division Multiplexing (WDM). The number of wavelengths in a waveguide is referred to as the WDM degree of the waveguide. The WDM degree can be increased to 64 or beyond, at which point the multiplexing is often referred to as Dense Wavelength-Division Multiplexing (DWDM). For chip-scale communication with silicon photonics, digital data from electronic components (e.g., processor, memory) can be encoded into an optical signal using electronic to optical (E/O) conversion with devices such as microring resonator (MR) modulators, subsequently transmitted over a waveguide with multiple carrier wavelengths, and then detected at a receiver, where optical to electronic (O/E) conversion is performed with devices such as photodetectors (PDs).

There has been growing interest in using silicon photonics for more than just communication. In particular, silicon photonic devices can also be used to perform computation in the optical domain. Together, such light speed communication and computation can significantly accelerate the execution of deep-learning workloads.

While silicon photonic devices face several challenges for robust computation and communication at the chip-scale (e.g., they are sensitive to thermal and fabrication-process variations [82]), they also offer several advantages (e.g., high speed, high bandwidth, and low power) to support inter-neuron communications and implement different neural functions required in photonic neural networks. Such neural functions and their implementations are discussed in the next section. In this section, we review some of the fundamental silicon photonic devices that are employed to implement photonics-based ANNs and SNNs.

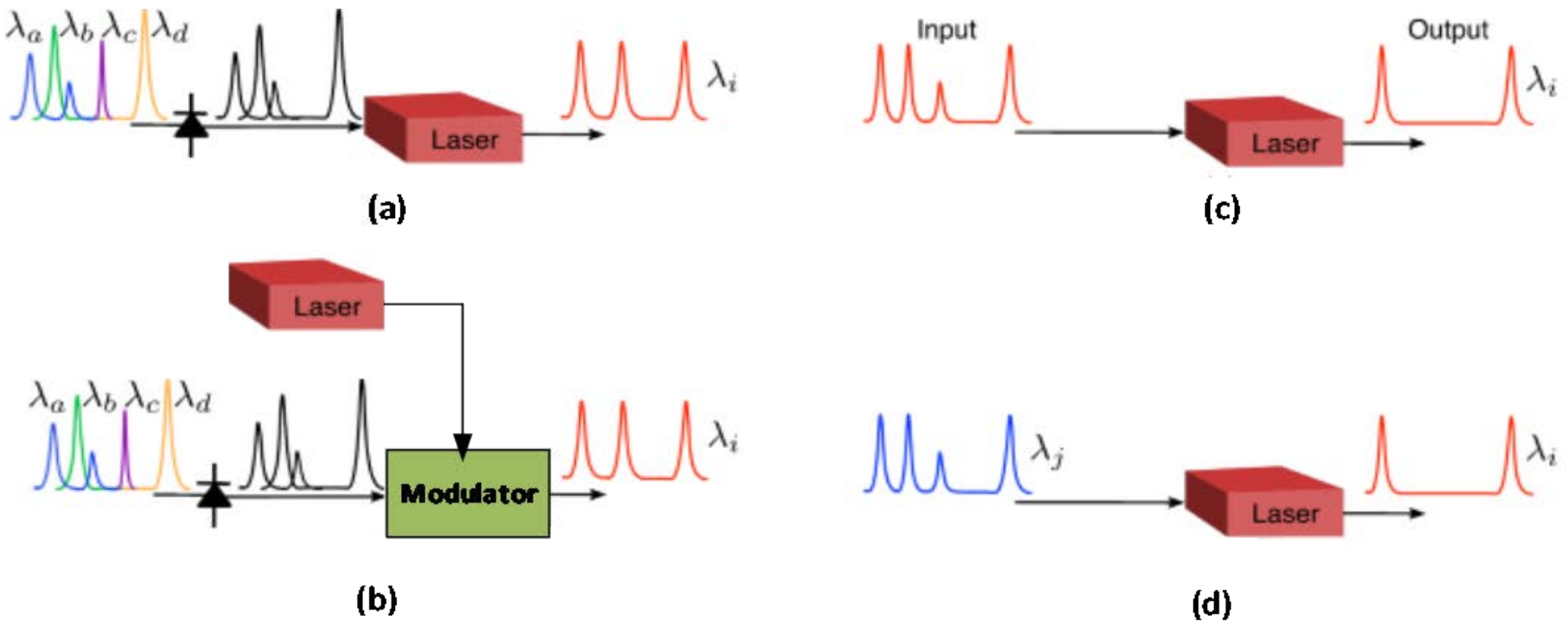

Figure 6: Lasers used in photonic neural networks [83], [84]: (a) Directly modulated laser; (b) Laser connected to a modulator [83], [85]; (c) Coherent laser where the wavelength in the input and output is the same; and (d) Incoherent laser with different wavelengths at its input and output. Note that the signal waveform in black is an electrical signal and the other colored signal waveforms are optical ones. Different colors represent different wavelengths.

### 3.1 Lasers

A laser is a key requirement in optical circuits and neural networks, serving as the light source to support optical communication and computation. Lasers can be either off-chip or on-chip. Although off-chip lasers offer a better light emission efficiency, they necessitate the use of couplers to couple the off-chip optical signal to the chip where such couplers impose high optical power losses. On the other hand, on-chip lasers provide a better integration density and lower optical loss, as there is no need to couple light from an off-chip source. However, on-chip lasers suffer from low emission efficiency and instability against thermal variations [86].

Lasers are used in photonic neural networks to implement different neural functions and requirements in such systems. In directly modulated lasers (see figure 6(a) [87]), the laser itself modulates the data onto an optical signal, while in another arrangement, as shown in figure 6(b), the laser output can be modulated by a modulator which is responsible to modulate the data onto the optical signal. Indeed, employing a laser in conjunction with modulators is common in optical interconnection networks [88], [89]. In photonic neural networks, this laser configuration can be used to design a scalable neural network [83], [85], where an off-chip laser source in combination with modulators can support multiple on-chip inter-neuron communications. In addition to the modulation, lasers can be also used to implement neural activation functions [90], as discussed in section 4, because lasers have shown potential to mimic neural activation functions [91]–[94] where an optical stimulus in the input of the laser can result in an optical output based on an activation function (see figures 6(c) and 6(d)). Thus lasers are used in all photonic neural networks to not only support inter-neuron communication but also, in some cases, to implement different neural functions in the optical domain.

A laser can be implemented in different ways. A Vertical Cavity Surface Emitting Laser (VCSEL) is a semiconductor laser diode with laser beam emission perpendicular to the chip surface, as shown in figure 7(a). Such a feature allows for several VCSELs to be placed in an array to power a large number of optical neurons and hence design scalable neural networks [91], [92], [97]. In addition to the scalability advantage, VCSEL-based neural networks can be realized using both off-chip and on-chip VCSELs [98]. Moreover, VCSELs have shown excitability behaviours of neurons [99], in which the laser emits light when the combination of inputs reaches a threshold. As a result, VCSELs offer scalability, efficiency, and several functionalities required for photonic neural network designs. Microdisk lasers, shown in figure 7(b), are another type of lasers in which a ring resonator is formed by successive total internal reflections inside a circularly shaped waveguide [101]. Compared to VCSELs, microdisk lasers are more area efficient (a laser apparatus radius—radius of microcavity—of a few micron) and deliver a lower threshold current and maximum on-chip optical power [102].

Moreover, microdisk lasers offer a low optical loss [100], and similar to VCSEL arrays, they can be placed in an array of lasers [101] to enable scalable photonic neural network implementations. In addition, microdisk lasers have shown excitability dynamics [103] to support spiking neuron implementations.

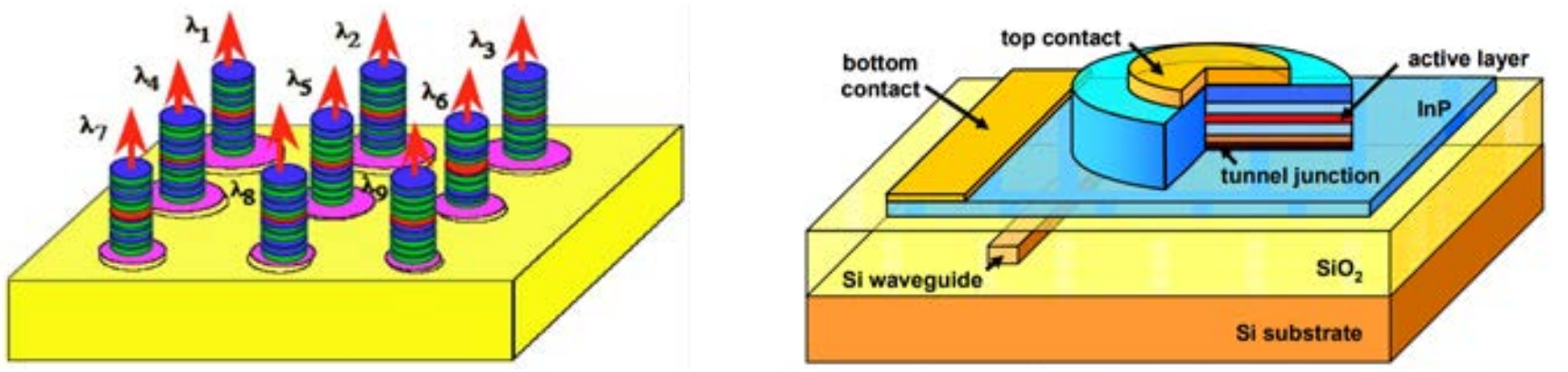

(a) (b)

Figure 7: (a) A VCSEL array [95]. (b) A microdisk laser [96].

## 3.2 Waveguides

A silicon photonic waveguide is analogous to a metallic wire, enabling optical signal transmission and routing in photonic neural networks. As shown in figure 8, waveguides can be classified into ridge and strip waveguides. Ridge waveguides are often employed in active devices and networks as they allow for electrical connections to be made to the waveguide (e.g., through PN junctions) where the characteristics of the optical signal can be actively controlled and altered using electro-optic or thermo-optic effects in silicon [28]. On the other hand, strip waveguides are usually employed in passive devices and networks to passively route optical signals [28]. As discussed earlier, a single waveguide can support simultaneous transmission of multiple optical wavelengths with no interference (using WDM). This allows for ultra-high bandwidth communication, which is of great interest in neural-network designs to support demanding inter-neuron communication.

When an optical signal traverses a waveguide, it experiences some optical loss (i.e., the propagation loss, often characterized in dB/cm) imposed due to, for example, some imperfections in the waveguide structure (e.g., waveguide sidewall roughness). Minimizing such optical loss in silicon photonic waveguides is essential as it limits the scalability of photonic neural networks and substantially degrades the power and energy efficiency in such networks. There have been a lot of efforts to minimize the propagation loss in silicon photonic waveguides and SOI waveguides with propagation losses as low as 0.026 dB/cm have been proposed [104]. In general, this propagation loss in waveguides depends on precise geometry adjustment in these devices, and hence any shape distortion in a waveguide (e.g., angular sidewalls) reduces its transmission efficiency (i.e., increasing the propagation loss). In addition to propagation loss, waveguide bends create optical bending loss where an optical signal will be attenuated due to the mode-mismatch and radiation loss in waveguide bends. This bending loss is proportional to the radius of the waveguide bend.

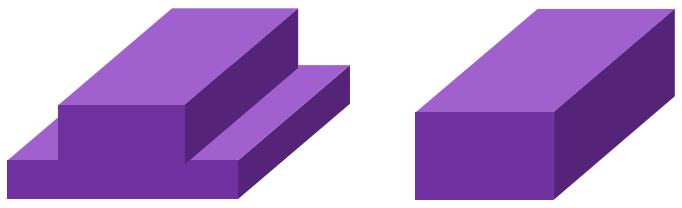

Ridge Strip

Figure 8: Silicon photonic waveguides [28]: a ridge waveguide and a strip waveguide.

## 3.3 Couplers

Silicon photonic couplers, as shown in figure 9, are used to couple an optical signal from an optical fiber (e.g., connected to an off-chip laser source) to an on-chip waveguide due to the significant mismatch between the cross-section of optical fibers (tens of micron) and that of silicon photonic waveguides (hundreds of nanometers). Such mismatch usually imposes some optical loss (i.e., coupling loss) which is considered as a significant portion of optical loss in optical networks employing off-chip lasers. Two major coupling solutions are surface grating coupling and edge coupling. Surface-grating couplers, shown in figure 9(a), are advantageous in terms of a simpler and low cost fabrication process but at the cost of a low coupling efficiency, while edge couplers, shown in figure 9(b), provide a better coupling efficiency but requires a more complex fabrication and packaging process [28]. In edge couplers, as shown in figure 9(a), a tapered waveguide is used to couple light from the fiber to the chip. Surface-grating couplers couple the input light from a fiber to a waveguide using diffractive gratings where a periodic structure splits and diffracts light and eventually couples the light into the waveguide. Diffractive coupling is a common means of optical coupling between VCSELs because of its simple implementation [106], and is also useful to implement photonic reservoir computing [107].

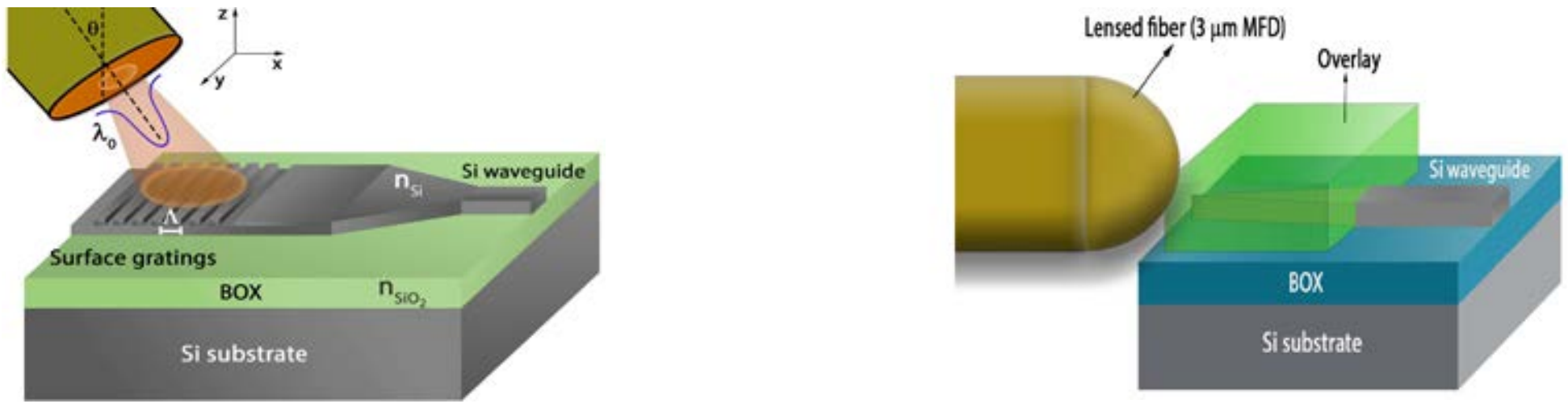

(a) (b)

Figure 9: Couplers [105]: (a) a surface-grating coupler and (b) an edge coupler.

## 3.4 Modulators, Filters, and Switches

Microring Resonators (MRs) are widely employed to design modulators, switches, and optical filters in optical interconnection networks [108], [109]. In addition to such applications in interconnection networks, they are promising devices to implement artificial neural synapses [90], [110], [111] and excitation function of neurons [112], [113], which are further discussed in section 4.

MRs, as shown in figure 10(a), are made with a ring-shaped waveguide in proximity with an input waveguide and a drop waveguide (a.k.a. add-drop filter). When the drop waveguide is missing (e.g., in modulators and some filters), the MR is an all-pass filter (see figure 10(b)). An MR can be in two different states of on- or off-resonance, based on which the optical signal can be switched to different ports. As shown in figure 10(a), when the MR is in the off-state, the input signal is routed to the through port, because the ring is not in resonance with the input optical signal. On the other hand, when the MR is in the on-state, the ring couples the input optical signal and drops it on the drop port. The resonant wavelength of an MR can be tuned to realize various functionalities needed to design optical modulators, switches, and filters. Here, tuning refers to sweeping the resonant wavelength of an MR by leveraging electro-optic or thermo-optic effects of silicon that can alter the optical signal characteristics, and hence the resonant wavelength in the case of an MR. Compared to tuning mechanisms based on electro-optic effects, those based on thermo-optic effects are slower (a few microseconds versus tens of nanoseconds in tuning techniques based on electro-optic effects) but are more power efficient. Figure 10(c) shows an example of an MR-based modulator that is responsible for modulating electronic data onto an optical signal. The modulator can modulate electronic data onto a specific optical wavelength and this modulated optical signal can be filtered with a wavelength-selective MR-based filter at the receiver (see figure 10(b)), and then detected and converted to electronic data through a photodetector.

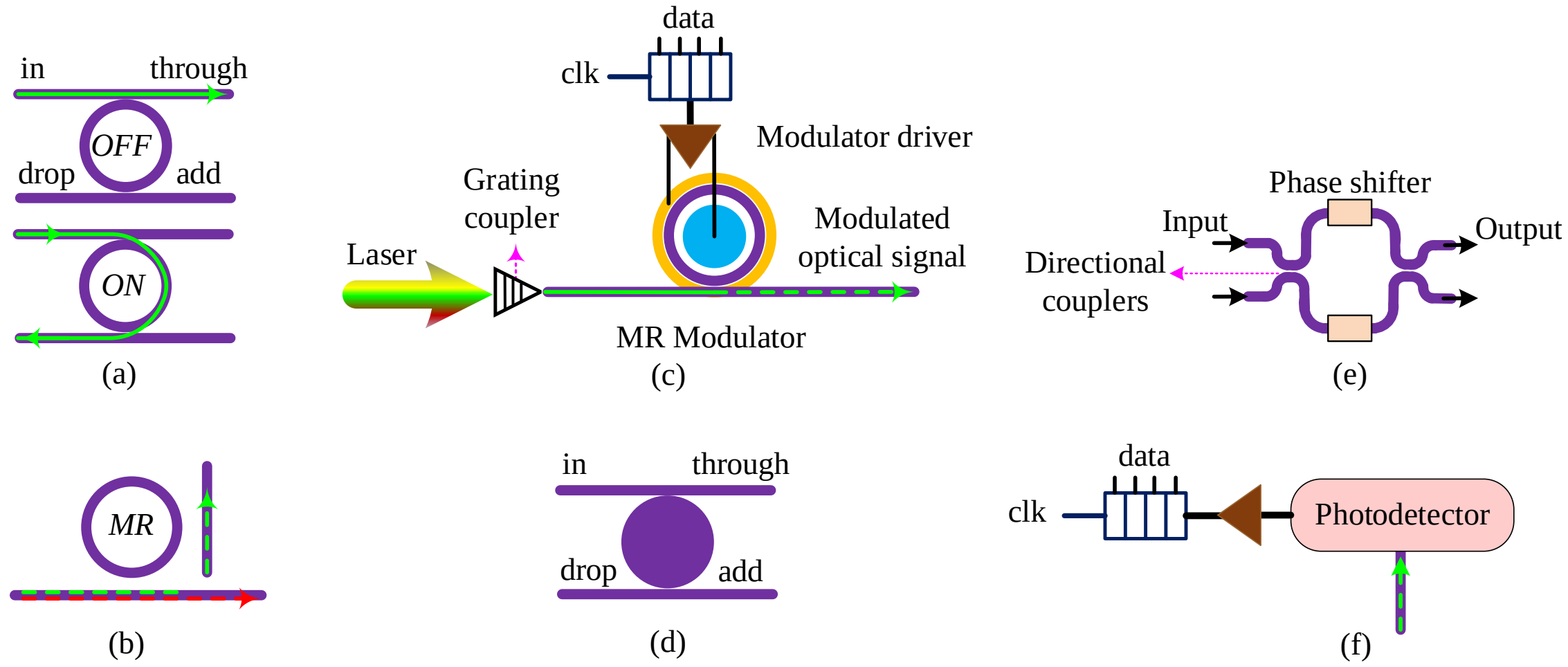

Figure 10: Silicon photonic switching devices [28]: (a) MR add-drop filter/switch; (b) MR all-pass filter; (c) MR modulator; (d) Microdisk resonator; (e) MZI; and (f) Photodetector.

Compared to an MR, a microdisk resonator (see figure 10(d)), which employs a disk instead of the ring structure, offers a better optical confinement to provide a smaller disk size and potentially lower power consumption [114]. MZIs, as shown in figure 10(e), are made of two waveguides with directional couplers and phase shifters. The phase shifters implemented using electro-optic or thermo-optic tuning change the optical phase in one or both arms of the MZI, introducing constructive or destructive interferences at the output to switch an optical signal between the output ports. Similar to MRs, MZIs have been applied to the design of optical modulators, switches, and filters. In comparison to MZIs, MRs have smaller footprint and lower power consumption. On the other hand, MZIs provide high bandwidth and better tolerance to thermal variations.

In summary, MRs, microdisks, and MZIs are widely employed to design modulators, switches, and filters. In photonic neural networks, a set of optical filters, in each of which the optical transmission can be adjusted, can be grouped into a weight bank to support weighting of activation signals as part of a photonic neuron [40].

### 3.5 Photodetectors

Photodetectors (PDs), as shown in figure 10(f), can be used to detect an optical signal and convert it to an electrical one. A small photodetector offers high bandwidth at the cost of low power efficiency. An efficient photodetector provides the desired electrical output with a small optical signal at its input. However, this small optical signal at the input of a photodetector may result in a low bandwidth performance in the photodetector. An optical signal power at the input of a photodetector should be larger than the responsivity of the photodetector, which is defined as the electrical output per optical input. This means that the power of a laser source in an optical link should be large enough to correctly drive a photodetector while considering the sum of different optical losses on the link. In photonic neural network designs, photodetectors not only convert optical signals to electrical ones but also combine (i.e., sum the magnitudes of) several optical signals over different wavelengths [40], [85], [111], which is a useful function in designing a silicon photonic neuron.

### 3.6 Devices based on Phase-change Materials

Devices that utilize phase-change materials (PCM) for tuning are of great interest in silicon photonic circuits to design modulators [115], MZI-based switches [116], and low-loss phase shifters [117]. The main principle in these PCM-based photonic devices is to employ a PCM (e.g., GST: $Ge_2Sb_2Te_5$ used in [115]) to efficiently induce high refractive-index changes for efficient phase tuning. Unlike electro- and thermo-optic tuning, PCM-based tuning is non-volatile: it only requires power for transition between the amorphous and crystalline states in the PCM [117]. This can allow for low overhead tuning of silicon photonic devices, e.g., from on-resonance to off-resonance in a PCM-based MR. In some photonic neural networks, PCM-based devices [118]–[120] are proposed as part of the design of neurons. For example, in the synapse design proposed in [118], several PCMs are placed on a waveguide to control optical transmission in the waveguide and implement the function of a synapse. Moreover, PCMs are useful to implement summation [119] and weighting functions [118], [119], [121].

### 3.7 Other Devices

A Semiconductor Optical Amplifier (SOA) is a device in which a semiconductor is used to add a gain to an optical signal without electro-optical or opto-electrical conversions. SOAs are mainly used to compensate for optical losses in optical communication systems. In photonic neural networks, SOAs can be employed to implement learning functions [122], [123]. However, SOAs suffer from poor coupling efficiency to optical fibers and are sensitive to polarization because of their planar shape [124]. Vertical Cavity Semiconductor Optical Amplifiers (VCSOAs) provide a better coupling efficiency and a lower sensitivity to polarization, and they can also be integrated into 2D arrays [113]. Moreover, based on the proposed learning function implementation in [125], VCSOAs can offer low-power consumption to implement learning functions in neural networks.

A spatial light modulator (SLM) is a device that can be used to change the amplitude, polarization, and phase over the spatial extent of a light beam [126]. Integrated spatial light modulation in silicon photonics can enable all-optical reconfigurable devices with possible applications in testing of optical circuits and reconfigurable multi-port optical filters, splitters, and modulators for data communication [127]. SLMs can be employed in reservoir computing architectures, as described in Section 5.3.

A pillar scatterer is a type of device that can be employed for implementing reservoir computing [34]. These devices can help speed up the classification of biological cells [128], [129]. For example, [129] provided a proof of concept, based on Finite-Difference Time-Domain (FDTD) simulations, of an integrated photonic application of Extreme-Learning Machine (ELM) for fast and label-free classification of biological cells. In this application, a passive optical stage comprising a collection of pillar scatterers embedded in a silicon nitride cladding is used to process the light forward-scattered by a cell when illuminated via a green monochromatic source.

An optical comparator is a common device in the design of analog-to-digital convertors [130]. Optical comparators can be made using MRs, SOAs, and lasers [131]. All-optical comparators are preferred over optoelectronic ones as they can provide higher speed and lower power consumption by avoiding electro-optical conversions [131]. An optical comparator is also useful to implement the max pooling layers in CNNs [132].

Lastly, silicon photonic Arrayed-Waveguide Gratings (AWG) are commonly used as optical (de)multiplexers in WDM systems. These devices are capable of multiplexing many wavelengths into a single optical fiber, thereby considerably increasing the transmission capacity of optical networks. AWGs have been used to implement matrix multiplication [133] and CNNs [134].

## 4 SILICON PHOTONIC NEURON MICROARCHITECTURES

In this section, we review different implementations of silicon photonic neurons that form the building blocks of photonic neural networks. Subsection 4.1 discusses how various functionalities within an individual neuron are implemented using silicon photonic devices. Subsection 4.2 describes two approaches for classifying the implementation of photonic neuron microarchitectures.

### 4.1 Intra-Neuron Functionality Implementation with Silicon Photonic Devices

Artificial neurons are designed to mimic different functions of biological neurons, and can be combined to create a scalable, energy-aware, and high-performance neural network. A high-performance photonic neuron is expected to provide adequate reliability [40], [85], scalability [90], and cascadability [111], [135], [136]. Reliability of a neuron can be improved by either reducing the noise at the output of the neuron or increasing the power of the desired optical signal (i.e., the signal carrying the data being exchanged through inter-neuron communication) to ensure that a neuron is not excited by unwanted noise and only excited by the desired signal. A scalable neuron supports sufficient number of fan-in inputs to enable large-scale networks. Indeed, one of the main factors contributing to the neural network computation power efficiency, in comparison with traditional Von Neumann computing, is the high connectivity inspired by mammalian brains. Cascadability, which directly affects the neural network reliability, is another important factor affecting the performance of a neuron. Cascadability of a neuron design is defined based on the optical signal power of a neuron to drive other neurons. Together, reliability, scalability, and cascadability are important metrics when evaluating the performance of a photonic neuron.

Two types of neurons are widely used in photonic neural networks: conventional (non-spiking) and spiking, as discussed in Section 2. In general, a photonic neuron includes four main functions: weighting, summation, activation, and learning. There are significant differences between the two neuron types when it comes to the learning functionality. In spiking neuron models, the learning function is implemented at the neuron-microarchitecture level: e.g., unsupervised learning (e.g., Spiking Time Dependent Plasticity (STDP)) is often implemented as a part of the spiking neuron to closely mimic the functionality of a biological neuron. For conventional neurons, the learning function is not part of the neuron model and is instead implemented at the neural network architecture level (e.g., with backpropagation learning) rather than at the neuron-microarchitecture level. We discuss the four main neuron functions in the following subsections.

#### *4.1.1 Weighting function*

In biological neural networks, synapses are of great importance because a synapse is a memory for a learning process. Synapses provide weighted connections among neurons where changing the weights is the main function of the learning process (discussed in Subsection 4.1.4). Because of the adaptive weighting in synapses, the weight of a synapse is manipulated (through the learning process) to change the effect of each input. To mimic such dynamic weighting of connections, silicon photonic devices such as MRs can be used to control optical transmission between two neurons. As a result, such devices can implement weighting functions where the transmission can be controlled by a learning feedback over the input and output.

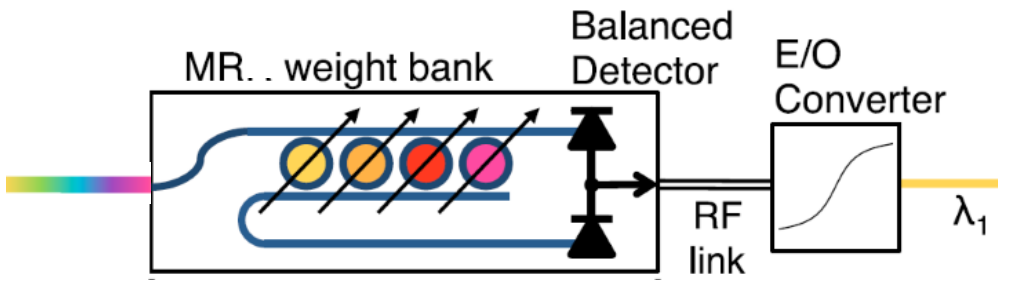

Figure 11: The neuron model employed in [40]. Here, several wavelengths share a waveguide to offer a high bandwidth. Inputs over different wavelengths enter the MR weight bank. Then, weighted inputs are summed using a balanced photodetector and a laser at the final stage converts the electrical summation signal to optical spikes (E/O converter can be a modulator [83] instead of a laser).

An MR is one the key devices used in the design of a weighting function. MRs can be placed in arrays to offer a bank of dynamic filters (see MR weight bank in figure 11) on the input connections of a post-synaptic neuron [90], [110], [111]. Each MR in an MR weight bank has an assigned weight value. When an optical signal passes an MR in an MR weight bank, the MR can alter the optical signal power proportional to its weight value. The weighted optical signals are then sent to a photodetector to perform the summation function, which is discussed in Subsection 4.1.2. Multiple wavelengths can be used in an MR weight bank with the WDM paradigm, to support high bandwidth communication and provide great scalability [110], [111]. However, the number of wavelengths that can be used in an MR weight bank is limited by cross-weight penalty [40] where the channel spacing—the frequency space between two consecutive optical channels/wavelengths—should guaranty the desired tuned weight for each synapse. Reducing the channel spacing (by increasing WDM degree) in an MR weight bank increases undesired effects (e.g., inter-channel crosstalk) on the spike-to-noise ratio, which can result in an undesired weight tuning. Such cross-weight effects can be improved at the cost of increasing the optical signal power to improve spike-to-noise ratio and, therefore, neuron reliability. In [40], the authors proposed an analytical model to design an MR weight bank while considering the channel spacing and power efficiency.

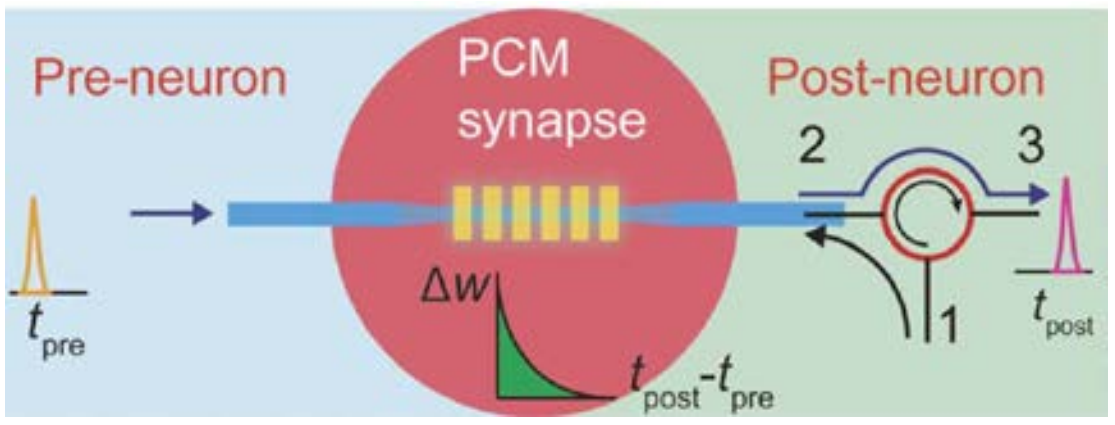

Figure 12: A PCM-based synapse proposed in [118]. The synapse is based on placing several PCM islands (yellow) on a tapered waveguide (blue) to control the optical transmission between the pre-synaptic and post-synaptic neurons. Weighting pulses to change the weight of synapse, optically, is applied via port 1.

Devices that utilize Phase-Change Materials (PCMs) for tuning can also be used to implement weighting functionality. In [118], $Ge_2Sb_2Te_5$ (GST), which is a PCM, is used to efficiently change optical transmission of a waveguide to implement a photonic synapse (weighting function) in a spiking neuron. In this design, shown in figure 12, multiple PCM pieces/dopants, which are called a PCM island, are placed on the waveguide. In comparison with using a single PCM island, this design is improved by using several PCM islands on the waveguide for each synapse that helps realize a more efficient optical-transmission change in the waveguide. The results also show that using a tapered waveguide structure in combination with PCM islands is more efficient than using a standard, non-tapered waveguide. The experiments in [118] confirm that each weight of the PCM-based synapse can be obtained using a predefined number of input optical pulses. Therefore, an accurate and all-optical weight tuning can be achieved by employing several PCM islands with a tapered waveguide. However, such an all-optical synapse suffers from low operation speed due to the photo-structural transformation process, which influences movement of atoms or ions because the speed of atoms and ions is much lower than that of a photon [137].

*Excitatory and inhibitory functions:* In addition to synapse weight, the type of neurotransmitters plays a significant role in a biological neuron. A neurotransmitter is a chemical to transmit information over a synapse to the receptors of a post-synaptic neuron [138]. Based on the type of neurotransmitters, weighted inputs can increase or decrease the membrane potential of a post-synaptic neuron. If a weighted input increases the membrane potential, the weighted input is excitatory (i.e., it encourages neuron to excite). On the other hand, an inhibitory weighted input decreases the membrane potential (i.e., discouraging the neuron to excite). In an artificial neuron, this corresponds to considering a positive or negative weight for each synapse. Therefore, a weighted input can increase or decrease membrane potential of the post-synaptic neuron to support excitatory and inhibitory functions, respectively. Several silicon photonic devices have been investigated to implement excitatory and inhibitory functions which are necessary in neural network designs. Excitatory function of distributed-feedback (DFB) lasers [93], MRs [112], and VCSELs [97] have been studied. Moreover, [133] analyzed the inhibitory functionality of VSCELs. To enable an efficient photonic neural network, both excitatory and inhibitory functions are required in the neuron design. For example, [91] proposed a neuron model in which a VCSEL is used to realize both excitatory and inhibitory functions of a neuron by injecting orthogonally polarized and parallelly-polarized fields at the same time. In addition to excitatory and inhibitory functions, injecting the two fields makes the neuron more reliable in the presence of noise and provides a faster response of the VCSEL. The proposed neuron design in [90], which is an opto-electronic neuron to support WDM, also provides both excitatory and inhibitory functions. The neuron uses two MR-based filters to represent positive and negative weights for excitatory and inhibitory functions, respectively. Furthermore, [140] proposes the experimental implementation and analysis of summation with excitatory and inhibitory functions in the opto-electronic neuron proposed in [90]. However, in this opto-electronic neuron, two wavelengths are required to enable both excitatory and inhibitory functions in the summation. The proposed neuron in [141] employs a modulation technique, which is based on using two push-pull MZIs and a phase shifter, to realize both positive and negative weights over a single wavelength. However, in the laser-based excitatory and inhibitory functions described above, active devices are used. Photonic neurons employing active devices, in which the light is generated by the device itself, suffer from integration challenges as it is costly to fabricate them with a standard CMOS process. To this end, the proposed neuron in [113] provides both inhibitory and excitatory functions by using MRs, which are passive devices.

#### *4.1.2 Summation function*

As described in Section 2, in a biological neuron, the soma or body of the neuron is responsible to combine (i.e., sum) inputs of a neuron so that the post-synaptic neuron can be excited with the aggregated input spikes. Similarly, in a conventional (non-spiking) artificial neuron model, a summation function integrates all the neuron inputs and forwards the result to an activation function. Summation over inputs is quite important because it directly affects the neuron scalability. A scalable neuron can effectively integrate a large number of inputs to enable a large fan-in and hence a large-scale photonic neural network. To design such a scalable neuron, neurons proposed in [90] and [111] combine inputs on different wavelengths to provide a compatible design with WDM. In such implementations, the neuron employs a photodetector to combine inputs transmitted on multiple wavelengths (see the photodetector in figure 11). As a result, there is a need for signal conversion (i.e., optical-to-electrical and electrical-to-optical) in such opto-electronic neurons where such conversion imposes some power losses, hence degrading the neuron performance. Alternatively, the summation function can be implemented in a photonic neuron using an all-optical approach. For instance, the summation functionality in micropillar-semiconductor lasers based on an integrated saturable absorber is investigated in [142]. Results show that the micropillar laser is able to combine spiking stimuli and excite an activation function. Moreover, DFB lasers [93] and VCSELs [91], [97] are also used to implement the summation function in photonic neurons. The proposed neuron in [112] employs MRs to implement the summation function. Nevertheless, all-optical approaches do not support WDM, hence cannot provide high interconnectivity to realize scalable photonic neural networks.

#### *4.1.3 Activation functions*

The activation function of a neuron can be linear or non-linear. In [141], a linear neuron model is presented to support linear additions and subtractions with both positive and negative weights to realize excitatory and inhibitory functionalities. The proposed linear neuron in [141] can support non-linear *sigmoid* and *ReLU* activation functions to be added to the base linear neuron. In [113], to implement a non-linear activation function, non-linearity effects in MRs are leveraged to realize a low-power neuron.

There are differences in the activation functions in spiking and conventional neuron models. In spiking neurons, an activation function defines the spiking time of the neuron based on the aggregated input spikes. Therefore, the optical signal propagation time depends on the activation function in this event-driven approach. Alternatively, in conventional neurons used in ANNs, an optical signal is propagated from the input to the output in predefined times. In [143], SOAs are used to emulate a *sigmoid* activation function for conventional neurons. An activation function can also be implemented by a photonic laser with electrical control signals in opto-electronic neurons (see figure 11). In the neuron used in [90], which is called Broadcast-and-Weight (B&W), an excitable laser and a photodetector are used to mimic the excitation function of artificial neurons. In particular, the activation function is implemented using an excitable laser and can fire when the summation signal, provided by the photodetector, reaches a threshold. The decision of firing a spike, realized by the activation function, triggers a spiking optical signal in the laser. The B&W approach has also been used in conventional neurons. For example, in [144] an MZI-based neuron is used in which the activation function is a binary function of +1 or -1 (symbolic decision function)

VSCELs have also shown great potential to implement an activation function, due to their relatively small footprint [145], low-power consumption [135], capability for 2D or 3D integration in arrays [99], low manufacturing costs [135], [145], and efficiency in coupling to optical fibers [135], [147]. In [97], the spiking behavior of VCSEL-based neurons, which is electrically controlled, is investigated. In [146], the spiking behavior of VCSELs with both parallel and orthogonal polarized optical stimuli is explored, and results show that VCSELs are able to produce controllable spikes required to enable ultra-fast optical neural networks. In [135], the excitation behavior of a VCSEL is studied. To investigate cascadability of VCSELs while considering their excitation behavior, two VCSELs—a transmitter VSCEL and a receiver VCSEL—are considered in [135]. Results show that controllable spikes (using an external control signal) are propagated from the first VCSEL to the second one, confirming cascadability of VCSELs and that they can be used as an excitation device in photonic neural networks. However, after firing a spike using a VCSEL, an inherent relaxation oscillation can occur which deteriorates reliability and speed of photonic neurons [91], [148]. The majority of VCSEL-based photonic neurons [99], [139], [146], [149] do not support excitatory and inhibitory activations at the same time. However, in [91] and [148], both excitatory and inhibitory functions are realized in VCSELs by employing double polarized injections, i.e., orthogonally and parallelly polarized injections. Silicon photonic devices are often designed for a single-polarization operation, hence employing double-polarized VCSELs is a challenge.

PCMs can also be used in the design of photonic devices to realize activation functions. The all-optical neuron proposed in [120], shown in figure 13, employs PCMs in an MR structure to realize excitation behavior of spiking neurons. Figure 13(a) shows the schematic of the proposed neuron, figure 13(b) shows its main components, and figure 13 (c) shows the circuit design of the proposed neuron. As shown in the figures, PCM is not only used to implement the weighting function but it is also placed on the MR to mimic an activation function. The transmission response of the PCM on the MR (shown in figure 13(b)-IV) was used to emulate the *ReLU* activation function.

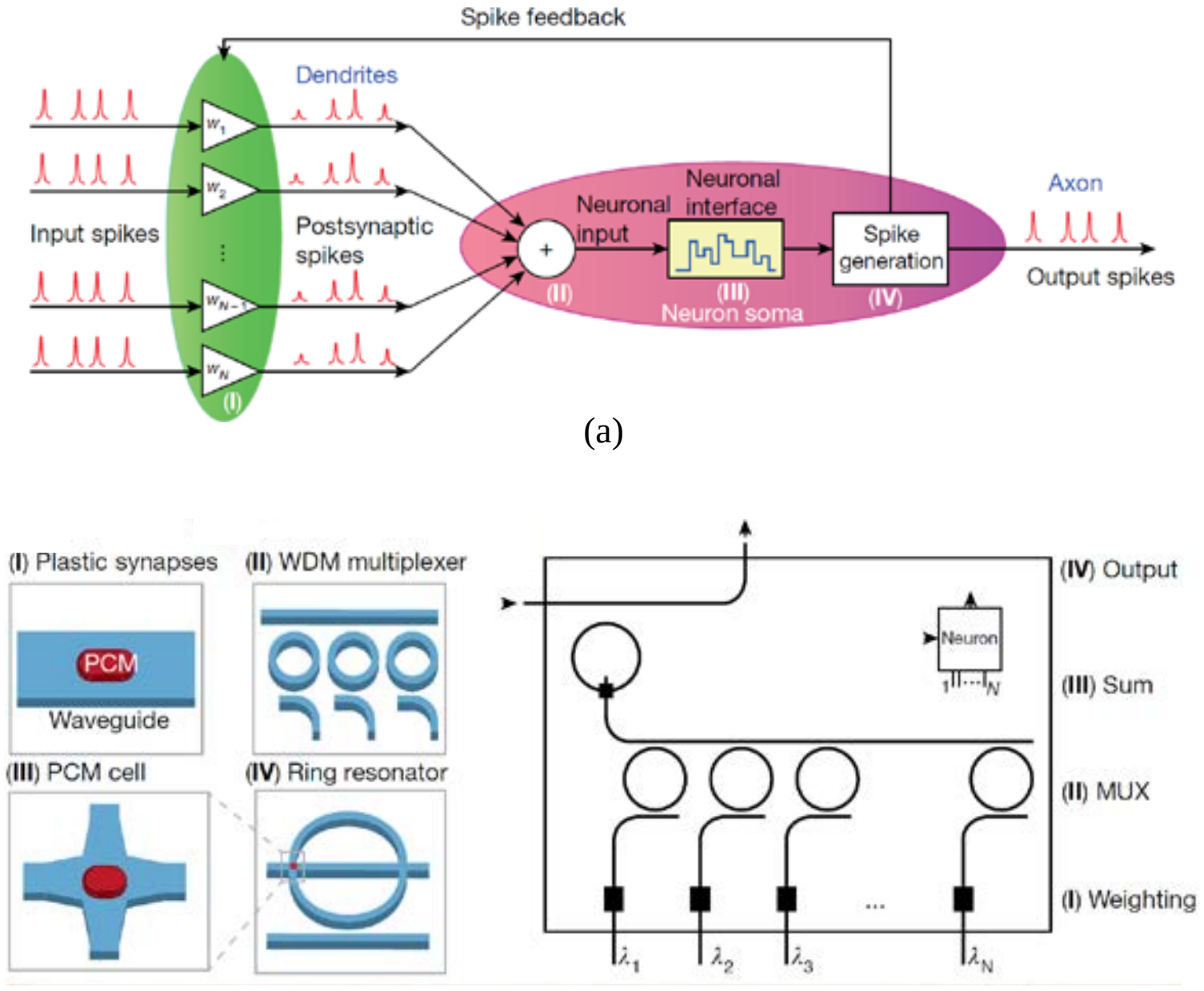

(a)

(b) (c)

Figure 13: PCM-based neuron proposed in [120]: (a) schematic of the neuron model, (b) main components of the neuron, and (c) photonic circuit of the neuron.

### *4.1.4 STDP learning function*

In spiking neurons, Spiking Time Dependent Plasticity (STDP) learning is usually employed to closely mimic a biological neuron, while in conventional artificial neurons, the learning function is implemented at the neural-network-architecture level and usually in the electronic domain. In spiking neurons, the STDP learning function updates weights based on pre-synaptic and post-synaptic spikes to help gradually decrease the neural network error, which corresponds to the difference between the desired and the actual output. In the STDP learning process, the strength of connections (i.e., synapses) is adjusted based on the spiking time of pre-synaptic and post-synaptic neurons. Changes in weight synapse are based on the output and input spiking time. According to the spiking time of pre-synaptic neuron and post-synaptic neuron, the weight can be increased or decreased to implement a learning function for a photonic neuron. A synapse's weight increases, which is called "potentiation," if the pre-synaptic spike occurs right before the post-synaptic spike. On the other hand, a synapse's weight decreases when the pre-synaptic spike misses the excitation of the post-synaptic neuron, i.e., the post-synaptic neuron fires as a result of the spikes received from the other pre-synaptic neurons.

In [122], STDP is implemented using an SOA and an Electro-Absorption Modulator (EAM), which can be deployed in high-speed (picosecond timescale) neural network computation. Also, [123] discusses photonic implementation of STDP and its application in both supervised and unsupervised learning. Employing a single SOA device to implement STDP learning improves the neuron scalability to realize supervised and unsupervised learning algorithms in large-scale neural networks. However, employing SOAs to implement STDP imposes high-power consumption (e.g., in comparison with using passive devices). In [150], STDP learning algorithm is implemented based on using passive MRs, which is suitable to design large-scale photonic neural networks with low-power consumption. In [125], a VCSOA is employed to improve the high-power consumption in STDP learning implementations based on SOAs.

### *4.1.5 Summary*

The state-of-the-art neural functions and their implementation using silicon photonic devices discussed in this section are summarized in Table.1.

Table 1: A summary of neural functions and their implementation using silicon photonic devices

| | | Implementation | Advantages | Disadvantages | References |
|---|---|---|---|---|---|
| **Weighting function** | | MR filters | Compatible with WDM | Not all-optical (require signal conversions) | [40], [110] |
| | | Several PCM islands on a tapered waveguide | All optical weight update | Low operation speed | [118] |
| **Summation function** | | Photodiode | Compatible with WDM | Low scalability | [90] |
| | | Micropillar laser | Both summation and excitation in laser | Does not support WDM and inhibitory function | [142] |
| **Activation function** | **Support excitatory and inhibitory functions** | VCSEL by double polarized injection | Support both excitatory and inhibitory functions; VCSEL advantages (low power, small footprint, implementation in arrays for large-scale designs) | Require double polarized injection | [91], [148] |
| | | MR | Use passive devices compatible with standard CMOS technology; support both excitatory and inhibitory functions | Sensitive to temperature and fabrication-process variations; has a limited cascadability | [112], [113] |
| | | VCSEL | Low power, small footprint, implementation in arrays for large-scale designs | Uses active laser devices that consume high power compared to passive ones [112], [113] | [97], [99], [146], [135], [147] |
| | | PCM and MR | Power efficient due to the use passive devices | low cascadability | [120] |
| **STDP Learning function** | | Semiconductor optical amplifier (SOA) | Scalable | Power inefficient | [122], [123] |
| | | Vertical-cavity semiconductor optical amplifier (VCSOA) | Power efficient in comparison with SOA-based STDP learning | Employing active devices, which are power hungry | [125] |
| | | MR | Power efficient because of implementation with passive device – does not require neurons to spike at different wavelengths | Only support unsupervised STDP learning | [150] |

## 4.2 Classifications of Silicon-Photonic Neuron Microarchitecture Implementations

Neuron implementations with silicon photonics can be classified in two ways: 1) all-optical versus opto-electronic neurons, and 2) coherent versus noncoherent neurons. In the following subsections, we discuss these two neuron implementation classification approaches in detail.

### *4.2.1 All-optical versus opto-electronic neurons*

In opto-electronic neuron designs, there is a need for electrical-to-optical and optical-to-electrical conversions. In such neurons, weighted inputs are typically summed/combined using a photodetector to control a laser [90]. Therefore, optical inputs should be converted to electrical signals and the electrical output of the photodetector should then be converted to an optical signal using a laser (see figure 11). Because of such conversions from optical to electrical and electrical to optical domains, the neuron is also called an O/E/O neuron. Compared with all-optical neurons, O/E/O neurons are power inefficient due to the power losses enforced in the required conversions. Moreover, due to the analog nature of intra-neuron communication [85] in both electrical and optical domains, the photodetector and the modulator laser are susceptible to noise. A noise analysis for opto-electronic neurons is presented in [85]. To compensate for the noise, the power of the modulator or electric transimpedance gain should be increased by adding a transimpedance amplifier (TIA) [85]. However, increasing the modulator power and adding a TIA both result in power consumption overhead. Conventional O/E/O neurons [40], [85], [111] often employ a directly modulated laser for excitation and spiking which in turn necessitates the placement of the photonic devices and the laser on the same chip. Consequently, such neurons suffer from thermal issues and variations caused by the on-chip laser. To this end, Modulator Neuron [83], [85] employs a modulator instead of a directly modulated laser. Therefore, the neurons can use an off-chip laser as the light source to compensate for the thermal issues.

In all-optical neurons, there is no need for electro-optical conversion during intra-neuron communication, i.e., all the devices within the artificial neuron support optical signal communication [113], [120]. For example, [120] proposed an all-optical spiking neuron, shown in figure 13, including an STDP learning implementation. Moreover, [113] proposed an all-optical neuron based on passive devices (in which the light is not generated by the device itself). The models presented in [113] suggest that MRs provide fast and power efficient excitatory and inhibitory functions in photonic neurons. Besides inhibitory and excitatory functions, the proposed model shows refractory behavior, which is an important functionality in a neural network implementation. Moreover, because the proposed neuron employs passive MR devices, it can be easily implemented with standard CMOS technology. However, such all-optical neurons lack high cascadability to support a large neural network. In addition, as we discussed in Section 4.1.1, all optical synapses (as a part of all-optical neurons) suffer from low-speed operation to implement weighting functions.

#### *4.2.2 Coherent and noncoherent neurons*

Based on the wavelength of operation in neurons, neuron implementations can be classified as coherent or noncoherent [151]. Coherent neurons manipulate the electrical field phase and amplitude with a single wavelength. Noncoherent neurons, such as those that employ the B&W photonic neuron configuration discussed earlier, manipulate optical signal power and rely on multiple wavelengths. The coherent neurons proposed in [32] and [141] employ MZIs and are power efficient as they require a single wavelength. However, MZIs impose a high area overhead and thus the design cannot be extended to support large-scale neural networks. Moreover, MZIs in coherent neurons require phase shifters in which the tuning error is inevitable. This tuning error can be propagated and magnified in the neural network, reducing the network reliability. The use of microdisk lasers was investigated in [113] to design a coherent neuron in which excitatory and inhibitory functions are realized by controlling optical phases. However, adjusting optical phases, which can be done using a microheater [113], adds a new challenge. In addition to phase-control challenges, coherent neurons operate at a single wavelength and are unable to distinguish between different wavelengths. Consequently, a neural network based on coherent neurons does not support reconfigurability [84] and WDM, resulting in a low-bandwidth performance. On the other hand, noncoherent neurons can operate with multiple wavelengths and support WDM in which several wavelengths share a waveguide to offer a high connectivity with lower number of waveguides. However, the dependency between the input and output wavelengths in photonic ANNs using noncoherent neurons necessitates wavelength conversions [84]. Such conversions can require high-power consumption overheads [152]. Moreover, noncoherent neurons also suffer from inter-channel crosstalk, which can reduce reliability [40].

#### *4.2.3 Summary*

Table 2 summarizes the state-of-the-art all-optical, opto-electronic, coherent, and noncoherent neuron microarchitecture implementation approaches.

Table 2: A summary of some proposed neuron microarchitectures using silicon photonic devices

| | | Implementation | Advantages | Disadvantages | References |
|---|---|---|---|---|---|
| O/E/O neuron | Noncoherent | Use weight bank for synapse, photodetector for summation, and laser for spiking | Use WDM to offer high bandwidth | Not all optical – consumes power in electro-optical and opto-electrical conversions; requires wavelength conversions | [90] |
| | | Use modulator instead of direct modulation by laser | Use WDM to offer high bandwidth; support on-chip neurons with off-chip lasers | Not all optical; high power consumption because of employing couplers | [83], [85] |
| All-optical neuron | Noncoherent | Use MR weight banks for synapse and PCM for excitation function | Supports both unsupervised and supervised learning; no need for electro-optical or opto-electrical conversion | Cross-weight penalty [40] for weight tuning; low cascadability | [120] |
| | Coherent | MR (require an all-optical synapse: synapse [118] is suggested by the paper) | Use passive devices; no need for electro-optical or opto-electrical conversion | Optical learning is not included; low cascadability | [113] |
| | | Splitters and MZIs | Higher reliability and low power overhead in comparison with using several wavelengths | Low bandwidth because of using one wavelength; area inefficient; exact splitting ratios are hard to achieve after fabrication due to variations; susceptible to noise in phase and splitting ratios [153] | [141] |
| | | MZIs | Higher reliability and low power overhead in comparison with using several wavelengths | Low bandwidth because of using one wavelength; area inefficient and hence not scalable; susceptible to phase noise | [32] |

# 5 SILICON PHOTONIC NEURAL NETWORK ARCHITECTURES

At the architecture level, prior work focuses on implementing different types of neural network models (discussed in Section 2): ANNs (MLPs, CNNs, RNNs), SNNs, and RC. The overarching innovation of a particular work is governed by the basic optical devices used in the architecture and the fundamental principles, such as optical resonance and optical interference, that govern those devices. These principles ultimately have the biggest impact on the performance, power, and reliability centric design decisions, and also the inherent limitations of the architecture built using them. Thus, devices and the driving principles behind them drive the innovations required to realize the architectures with silicon photonics technology. Hence, our classification in this section will be based on the primary photonic principles used to construct the neural network architectures.

## 5.1 Optical Resonance based Network Architectures

Optical resonance based neural network implementations usually rely on the wavelength specificity of MRs or microdisks, which leads to the utilization of WDM-based implementations where multiple wavelengths are utilized in a waveguide. These architectures are noncoherent architectures and utilize the noncoherent neuron microarchitectures discussed in Section 4.2. Most of these architectures utilize or build on the B&W protocol, illustrated in figure 14, for setting and updating the weights as it was demonstrated in [154] to have isomorphism to Continuous Time Recurrent Neural Networks (CTRNNs). The feedback loops, which are characteristic of RNNs, can be emulated by MRs when they reach optical bistability. Under favourable conditions pertaining to the resonant material and incident transmission intensity, the output transmission of the resonator can enter a hysteresis cycle, with two stable transmission levels. This is referred to as optical bistability of resonators. The work in [154] also suggested using Mach–Zehnder Modulators (MZMs) to generate the *sigmoid* activation function. This specific work provided the proof-of-concept that B&W-based MR architectures can be used to implement neural networks and that they can yield better performance over traditional CPU-based CTRNNs. For benchmarking, they considered a Lorenz Attractor [155] simulation application and reported a 294× acceleration with their photonic architecture compared to CPU-based simulations.

The B&W protocol is a multi-wavelength analog networking protocol in which multiple all-photonic neuron outputs are multiplexed and distributed to all-neuron inputs. These architectures tend to make use of the parallelism that is inherent in photonic architectures, employing multiple wavelengths to transfer data in parallel using WDM. Different wavelengths in a waveguide represent the input signals to the neuron. Weights are reconfigured by tuning the MRs, so that the characteristics of a specific wavelength are modified. MR weight banks comprise of tuneable MRs which can be tuned to drain energy from their resonant wavelength so that intensity of the wavelengths reflect the weights or the kernel values. The change in intensity is read using photodetectors (PDs) and summed to obtain the output values from the weight bank. This process was described in detail in Subsections 4.1.1 and 4.1.2. The obvious advantage of this approach is the utilization of the well-studied and mature MR technology to implement photonic neural networks, which makes the hardware implementation and integration easier. However, an issue this protocol can face is the number of MRs needed to implement it for real-world applications and depending on the feature map and the kernel size for CNNs this can become exorbitant. The research utilizing this protocol tries to work around this issue.

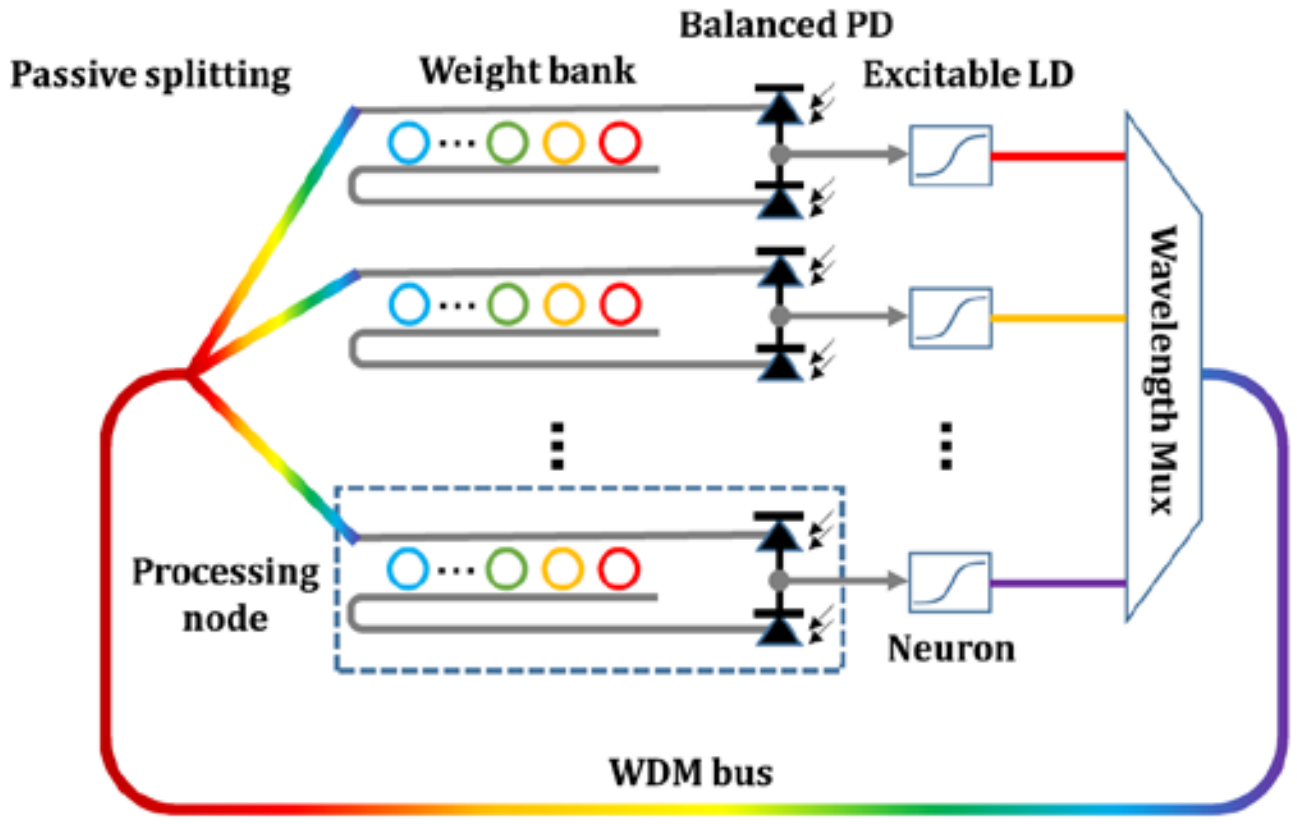

Figure 14: The Broadcast-and-Weight (B&W) protocol as illustrated in [37].

The authors in [156] utilize the B&W protocol with an MZM to implement the *sigmoid* non-linearity as in [154] to propose a CNN accelerator, dubbed Photonic CNN Accelerator (PCNNA). PCNNA implements one CNN layer and reuses that layer sequentially, with varying kernel sizes to implement the whole CNN. The input feature map and the required kernels are loaded from an off-chip DRAM. The results from the execution of individual layers are fed back into the memory. This is a sequential execution of kernels using an optical core, which runs at a higher clock frequency than its electrical components. The optical core mentioned here is comprised of the weighting MR banks and Digital to Analog Converters (DACs) that feed data into the MR banks and the laser diodes (LDs). The authors argue that because CNNs use kernels with the same dimensions per layer, they share the same receptive fields, and hence convolution computations for different kernels can be performed in parallel. They demonstrated the effectiveness of this work by implementing *AlexNet*, which is a deep CNN architecture with eight layers, five convolution layers and three fully connected layers. The authors showed how their filter-based approach to implementing *AlexNet* had substantially fewer number of MRs than an approach which does not consider any optimizations for implementing *AlexNet* (they claimed a reduction in the number of MRs from one-billion range to 100,000 range).

Another architecture which utilizes photonic weight banks for implementing CNNs is described in [157]. The authors have described an architecture which implements the entirety of the CNN layers using connected convolution units which are comprised of weight banks, where the tuned MRs assume the kernel values by using phase tuning to manipulate the energy in their resonant wavelengths. The architecture was tested using the *MNIST* dataset [158] and was shown to have better execution time than GPU-based classification, with the AMD Vega FE, AMD M125, NVIDIA Tesla P100 and NVIDIA GTX 1080 Ti GPUs. However, they do not consider any optimization methods on the model to reduce the MR count required to represent it, and they report a very high 100W power utilization for a 1024 MR modulator array in their proposed architecture.

A CNN accelerator implemented using MRs and memristors, is described in [132]. In this work, weights are fed into the MR-weight bank through memristors, which in turn gets their weight values from off-chip memories via SRAM buffers. The architecture includes individual layers needed for CNN implementation. The convolution layer is comprised of the memristor-based photonic weight bank. The activation layer (*ReLU* layer) is built using SOAs. The work also uses an all-optical analog comparator, proposed in [130], to implement maxpooling layer.

These three layers form a single feature-extraction layer. Two feature-extraction layers are interconnected using an interface layer, which demodulates the output from the previous maxpool layer, generates the corresponding electronic voltage values, and then feeds them into the memristors of the next feature extraction layer. This work focused on recognition of handwritten digits, using the *MNIST* dataset and shows better execution time against the FPGA-based Caffeine accelerator [159] and the memristor based ISAAC accelerator [20] on various benchmarks. The architecture was further extended by the authors in [209].

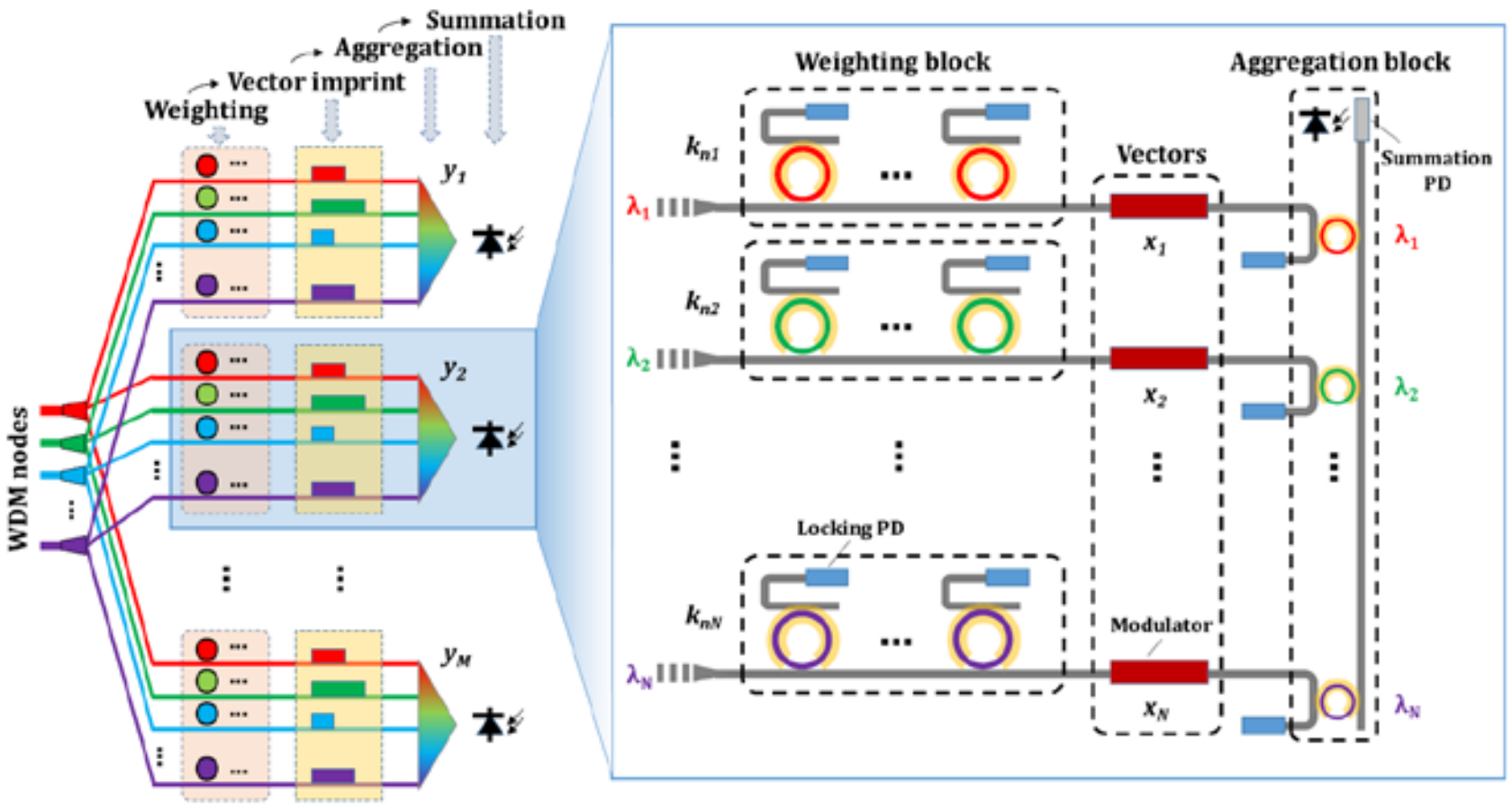

Figure 15: The hitless weight and aggregate architecture for optical matrix-vector multipliers (OMMs), from [37]. The architecture aims to avoid the thermal crosstalk-based weight corruption that can occur in the B&W protocol-based architectures.

A variation of the B&W protocol for MLP implementations was explored in [37] where the authors described the "Hitless weight-and-aggregate" architecture. This method to accumulate weight values from the weight banks was devised to overcome the possible corruption of weight values from thermal variations and thermal crosstalk. The proposed "Hitless weight-and-aggregate" architecture for MR-weight banks separates the wavelengths and weighs them in a parallel manner instead of the cascaded approach proposed in [154]. The process of updating the input matrix is simplified to counteract the delay induced by updating the weights in the weight bank. This is done by encoding the kernel directly into the input matrix of the Optical Matrix Multiply (OMM) unit, which is illustrated in figure 15. The control of the Hitless MR bank architecture is given to an FPGA, which use PDs and ADCs to obtain the summed signals from the OMM. The OMMs were implemented using MRs but the vector storage was implemented using MZIs. Given that the main issue with B&W is the large MR count, it is unclear if the modified architecture can address that issue, as the work does not elaborate on the MR count, even with the input matrix minimization approach discussed in the paper.

An MR-based neural network accelerator designed to be resilient to on-chip fabrication process variations (FPV) and thermal variations was proposed in [160]. The proposed Crosslight accelerator utilized FPV resilient MR designs coupled with thermal eigen decomposition based tuning approach and intelligent ring placement to combat the on-chip variations mentioned. These device and circuit level considerations helped increase the weight resolution achievable to 16 bits for a single MR. Also, in this work the weight matrices are decomposed to component vectors, which effectively converts matrix multiplication to a collection of vector dot product operations. The hardware design considered the splitter losses and waveguide propagation losses to design vector granularity aware, compact vector dot product (VDP) units. The VDPs operate based on the B&W protocol, and the partial sums generated are added photonically before passing to the electronic control unit. The Crosslight architecture was shown to have 9.5× lower energy per bit and 15.9× higher performance (frames per second) per watt on average than DEAP-CNN [157] and Holylight [161] photonic accelerators.

MRs are prominently used to implement the B&W protocol, but an architecture which considered microdisks over MRs for its lower area and power consumption is described in [161]. This work explores designing and implementing photonic Matrix-Vector Multipliers (MVMs), adders, and shifters, which are the fundamental computing components for CNN inference, using microdisks (figure 16). The MVM (figure 16(a)) uses the transmissivity of the microdisk array to represent the elements of one matrix and the input power into the microdisk array from the LD array to represent the other matrix. The output power from the microdisk array is product of the two matrices. The electro-optic full adder (figure 16(b)) utilizes CMOS logic gates to calculate the propagate bit ($P_n$) and the generate bit ($G_n$) needed for the full adder, and microdisks to calculate the sum and carry. The $P_n$ and $G_n$ values are used to modulate the microdisks. By modulating the light intensity, one of the optical combiners at the output implements an XOR gate while the other generates an OR gate, thereby implementing all the necessary operations for the sum and carry operations. The authors also propose a binary shifter using microdisks as shown in figure 16(c). The shifting operation is performed by configuring the on/off states of the microdisk crossing switches. The authors also describe simplified CNN models called power of two quantized CNN (P2Q-CNN) models to avoid reliance on ADCs and boost CNN inference accuracy with negligible drop in accuracy (below 1%). This alternative architecture uses a photonic adder and shifter combination instead of the MVM. For testing the architecture, the authors used benchmarks based on *MNIST* and *ImageNet* datasets. They compared this architecture to a ReRAM-based PIM accelerator ISAAC [20], and showed 13× better performance-per-watt.

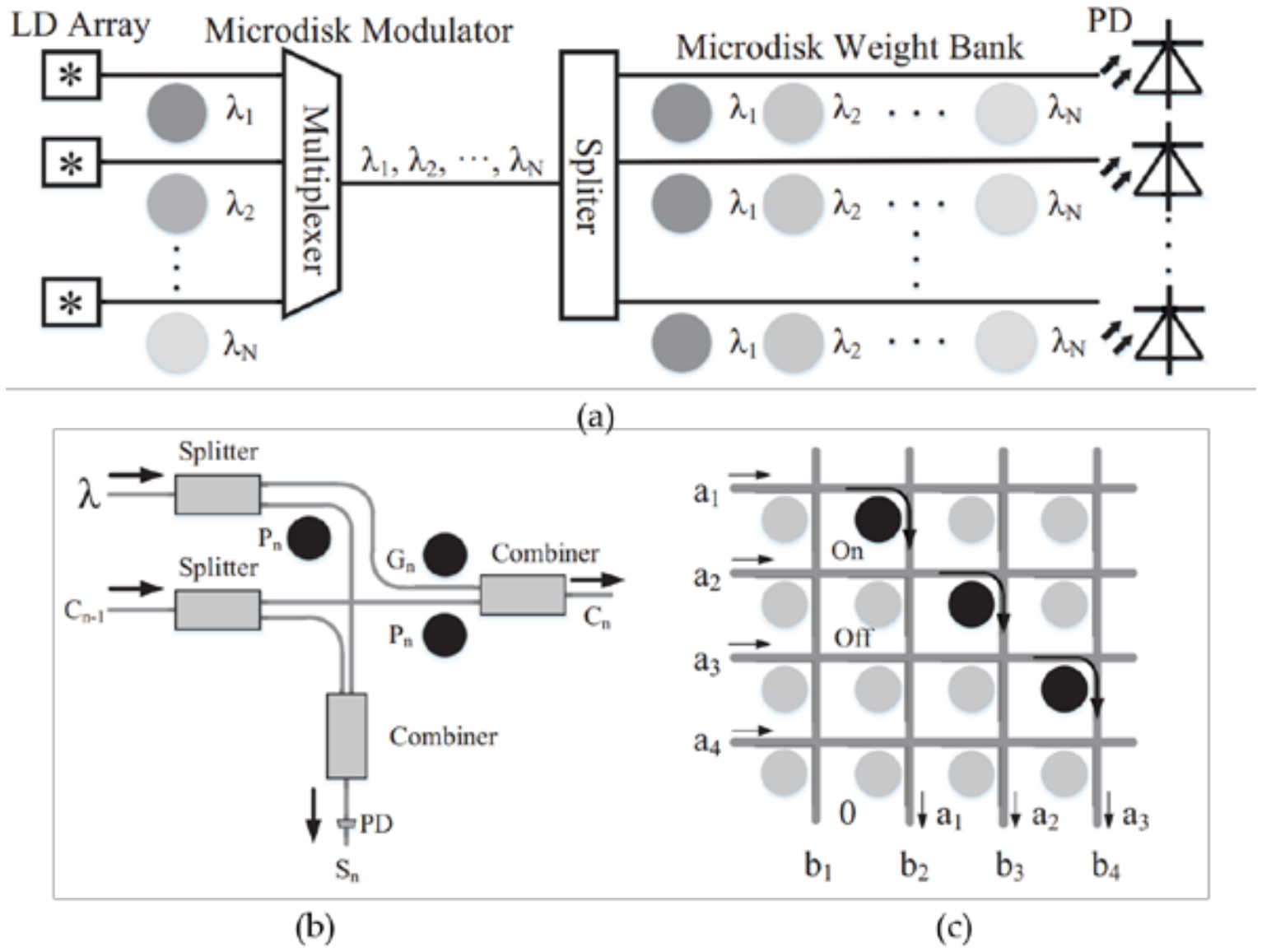

Figure 16: (a) Microdisk-based on-chip matrix-vector multiplier (MVM), (b) electro-optic full adder; (c) photonic bit shifter, from [161].

A recent work [162] combined MRs and MZIs to design the basic Optical Multiply and Accumulate (OMAC) unit, which is used in an accelerator for CNNs, called PIXEL. The work describes two versions of the accelerator: a hybrid version that multiplies optically and accumulates electrically, and a fully optical version that multiples and accumulates optically. The hybrid version uses MRs to implement an AND function with the MRs controlled using a synapse-lane, with shift-accumulation being done electrically. The bitwise AND operation along with shift-accumulate is used as an alternative to the opto-electrical MAC operation. The all-optical shift and ADD design uses MZIs to perform low-latency, low-power shift-accumulate operation optically and by cascading MZIs together. Synchronization of the signals from AND output is achieved with the help of propagation delay in MZI arms. Large dimensional MZIs and 6mm waveguide arms between these MZIs are used, to induce propagation delay in optical signals. The output signal from these interconnected MZIs effectively bit shifts the input. The proposed architecture has register files for filter weight storage in each OMAC. The OMACs are arranged in a grid, and neuron outputs are passed through photonic interconnects in both x- and y-dimensions. The synapses are pre-loaded into the OMAC and the proposed design assumes timed firing of the neurons to implement the MAC functionality. The hybrid and all-optical approaches were compared against an all-electrical architecture via simulations for the ResNet, GoogleNet, and ZFNet models. The all-optical approach shows better energy efficiency than the all-electrical approach and is comparable to the hybrid approach in this regard. The hybrid approach which relies only on MRs occupies significantly smaller area than the all-optical version of the architecture which uses both MRs and MZIs.

The B&W protocol has also been used for SNNs as discussed in [90]. The suggested architecture utilizes laser neurons in conjunction with two different MR weight banks to interact with the WDM signals. The weight banks are used to represent excitatory and inhibitory weights. The weights are accumulated using a balanced PD pair before being used to excite a laser neuron. One of the PDs in the pair accepts the signal power from the excitatory weight bank and the other accepts power from the inhibitory weight bank. A short wire is incorporated to perform a subtraction operation, thereby considering the values from the inhibitory bank as negative values. The combination of the weight banks, the PDs, and the LD acting as the firing mechanism, simulates a basic spiking neuron and is called a processing network node (PNN) in this work. The WDM signals are transmitted between these PNNs using a broadcast loop (BL). Multiple broadcast loops can be connected together in hierarchical manner via interfacing PNNs. Interfacing PNNs are PNNs which are tasked with the purpose of accepting output values from one BL and passing it to another, essentially acting as an optical router for the signals. The authors of [162] explored this strategy to allow for spectrum reuse and improve parallel processing. The work does not provide an experiment section to demonstrate the capabilities of the proposed architecture. Rather this work explored the feasibility of the B&W protocol based spiking neural networks. The key observation in the work involves how utilizing the hierarchical broadcast loop architecture would allow for better spatial layout freedom than other conventional hardware neuromorphic systems.

There are other instances where MRs are used to implement SNNs other than through the B&W protocol. For example, in [163] MRs are employed to implement STDP on a chip. The authors of [163] incorporated $Ge_2Sb_5Te_5$ (GST), a popular and well-studied PCM material [164]–[167] on top of the ring waveguide in the ring resonator. This allowed the control of light propagation through the ports by merely changing the state of the GST. In this case, the PCM and its different phases act as the memory in the synapse. The authors of [163] also discussed potential integration of the integrate-and-fire neuron using MRs and GST in an SNN framework consisting of bipolar weights (weights with positive and negative values). The positive and negative weighted sums are computed using two separate dot-product engines and input to two different MRs. The bidirectional integrating action of the two ports of the MR is leveraged to calculate the effective membrane potential under the action of the bipolar weighted sums. Output spikes are generated when the effective membrane potential of the neuron crosses a threshold. Upon receiving the dot product stimulus, the neurons integrate their membrane potential at that time-step. The work tested this architecture using the *MNIST* data set and assumed that the dot product engine has perfect operation. The work highlights the viability of PCM-based STDP in neuromorphic architectures and showcased this by demonstrating faster read/write operations and low energy consumption for the photonic architecture when compared to an electronic counterpart. In their simulations, a testing accuracy of 98.3% was achieved using this architecture. A related work [168] discussed how MRs can be used to incorporate spike delays into a photonic SNN.

A few works have also proposed MR-based photonic reservoir computing (RC) architectures, such as the 5x5 MR reservoir for high-bit-rate digital pattern classification in [169]. In this work, the reservoir was formed by randomly interconnected MRs. The simulated architecture was able to achieve a classification error of only 0.5% while offering bit rates up to 160 Gbps for eight-bit-length digital words for bit-pattern recognition on a custom dataset. The authors in [170] explored a 4x4 swirl topology-based reservoir design which utilizes MRs. The work also demonstrated basic Boolean operations. In such architectures the nodes are comprised of non-linear elements (MRs) and are part of the recurrence of the network, which is a departure from the original swirl topology introduced in [171]. This architecture has been widely used in photonic RC research. The swirl in the data paths allows for sufficient mixing of the input signals/weight matrix. Traditionally, reservoir architectures set their nodes at near instability for proper operation of the reservoir to ensure that they have sufficient memory of past inputs and respond well to new inputs. The MRs in [170] are set to this operating point after detailed analysis of MR stability in operation and resonance at various input power values, as well as temperature induced optical detuning from resonance.

In summary, noncoherent neuron microarchitectures that use MRs are one of the most prolifically used components to implement photonic neural network architectures. These noncoherent architectures that use MRs span SNNs, ANNs (MLPs, CNNs), and RCs. The majority of the SNN, MLP, and CNN architectures utilize the B&W protocol for propagating weights through the network. Some efforts have identified several issues with this protocol, such as heterodyne crosstalk corrupting the weight values and the increasing large number of MRs needed for the implementation of larger networks, especially when larger WDM degrees are utilized [37]. Other works have suggested reusing implemented layers as in [156], and also methods to reduce energy consumption and increase speed of operation by reducing the involvement of electronic components [161]. Moreover, there have been suggestions to use microdisks instead of MRs to further increase the integration density in chips as in [161]. MRs have also been used for weight propagation in SNNs using the B&W protocol as in [90] and for implementing STDP in photonic SNNs [163], [172]. They also appear in RC to create the nodes in reservoirs that are comprised of randomly interconnected nodes [170].

### 5.2 Optical Interferometry based Network Architectures

Optical interferometry based neural network implementations usually rely on the manipulation of the electrical field phase and amplitude of a single optical wavelength. These are coherent architectures and utilize the coherent neuron microarchitectures discussed in Section 4.2. Coherent architectures rely extensively on MZIs and have been widely employed in MLP implementations. MZIs have less commonly been used to play the part of intensity modulators in some SNN implementations. The large number of nodes needed in the reservoir and the large area requirement of MZIs make them not very popular in RC implementations. MZI-based coherent architectures utilize universal linear meshes of MZIs to implement the required matrix multiplications needed by neural networks. The weights are controlled by controlling the phase and amplitude of optical signals which is done by implanting attenuators and phase shifters on the MZI arms. This was demonstrated in [173] where 2×2 beam splitters and phase shifters in the form of an MZI was programmed to enable independent control of amplitude and phase of light for a set of optical channels.

The work in [174] fabricates and demonstrates an MZI-based 4x4 optical matrix multiplier. Here, the architecture is constructed based on the premise that an ideal NxN multiport reconfigurable MZI-based interferometer represents a special unitary (SU) group of degree N, SU(*N*), which is comprised of n MZIs with N optical channels forming a unitary transformation matrix. In [174], the structure is made of an SU(*N=4*) section followed by a diagonal matrix multiplication (DMM) section (figure 17(a)). The DMM can be extended depending on the application and can form a complete SVD through cascading. This 4x4 optical matrix multiplier was used for implementing a single-layered neural network. The performance of the neural network was evaluated by tasking it to classify 50 data samples of a synthetic linearly separable multivariate Gaussian dataset, for which it was able to achieve a 72% accuracy. The work in [175] describes another matrix SVD-based mesh implementation, which can implement arbitrary non-unitary matrices using MZIs. The Singular Value Decomposition (SVD) based methodology is used to perform decomposition of matrices to unitary matrices, and these simplified matrices are implemented on-chip. More specifically, SVD is the process by which a matrix can be decomposed into three matrices, two unitary matrices V and U, and a diagonal matrix comprised of non-zero singular values Σ. The SVD process can be implemented in an MZI mesh by using a diagonal matrix which implements the amplitude and phase while the universal unitary matrices follow the designs as proposed by [173] and [176]. The final architecture of this approach is shown in figure 17(b).

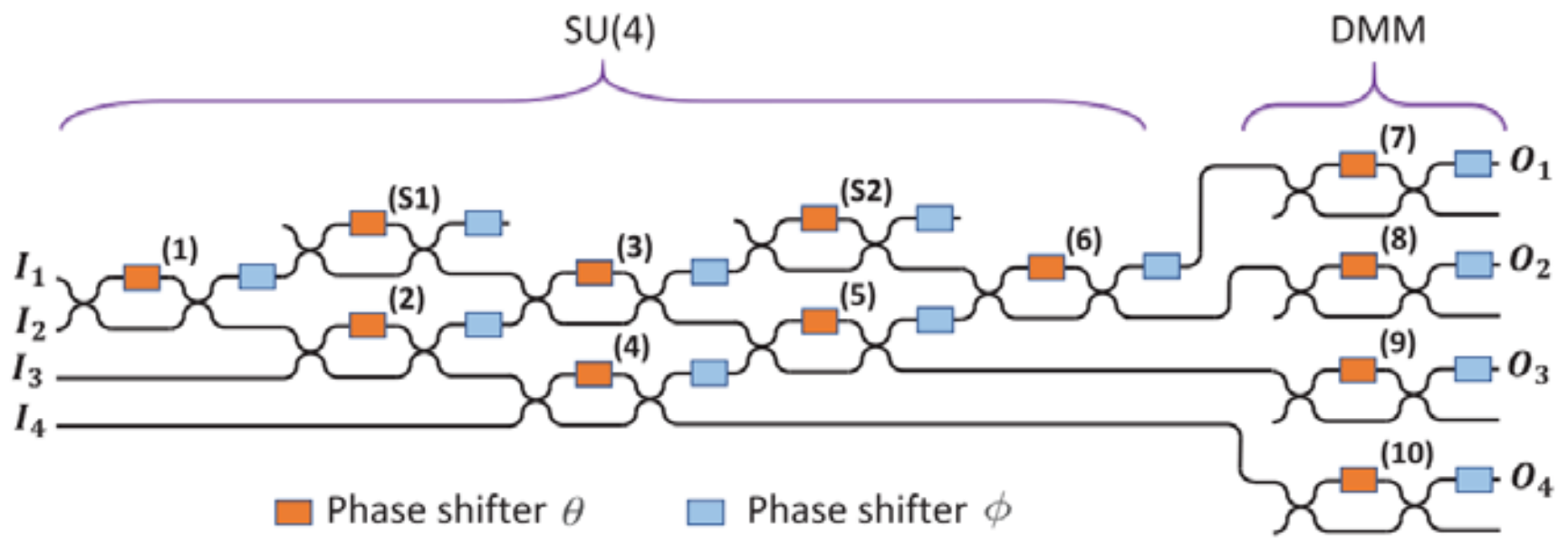

(a)

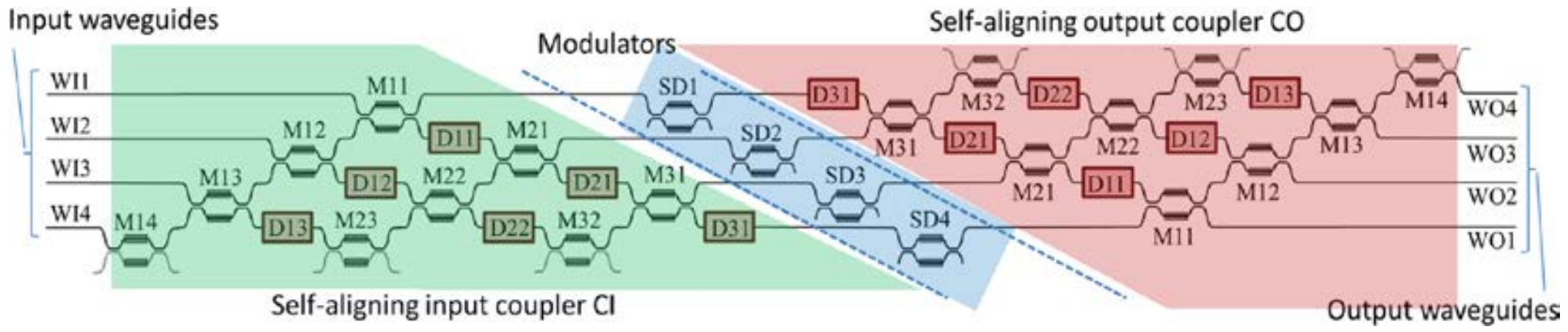

(b)

Figure 17: (a) The 4 × 4 MZI-based reconfigurable linear optical processor from [174]; SU is special unitary group and DMM is the Diagonal matrix multiplication unit; (b) The universal linear mesh for MZI based Singular Value Decomposition (SVD) as described in [175]. The matrices involved in SVD implemented: V in green, diagonal matrix Σ in blue and $U^T$ in red.

The authors in [32] proposed an architecture that utilizes an SVD-based approach for implementing the necessary matrix calculations (figure 18), where vectors were encoded in the intensity and phase of light and then fed into each layer of the network. SVD is used to decompose the matrices to be multiplied into unitary matrices which can be encoded into the MZI mesh. The SVD operation and encoding signals for the MZIs were generated using a digital computer. Once the matrices are encoded into the MZI mesh, matrix multiplication between them can be performed by allowing the optical signal to pass passively through the mesh. The key advantages of SVD is the reduced complexity of operation and reduction in dimensionality which helps with reduced cost of operation of the DNN model at hand. Each layer of this proposed model is comprised of an Optical Interference Unit (OIU) and an Optical Non-linearity Unit (ONU). In this work, ONU functionality was implemented with digital electronics while the OIU was implemented in a photonic integrated circuit, which performed optical matrix multiplications using the SVD approach as described in [175]. The work in [32] discussed how to use such an architecture for vowel recognition. They also utilized forward propagation with finite difference method instead of backpropagation to train the architecture. The architecture was able to achieve 76.7% accuracy in classifying vowels, which is lower than the 91.7% achieved by the same architecture implemented on a conventional 64-bit digital computer. The authors attribute the lower accuracy to the limited computational resolution (24 bits as opposed to the 64 bits of the conventional computer) of the optical neural network.

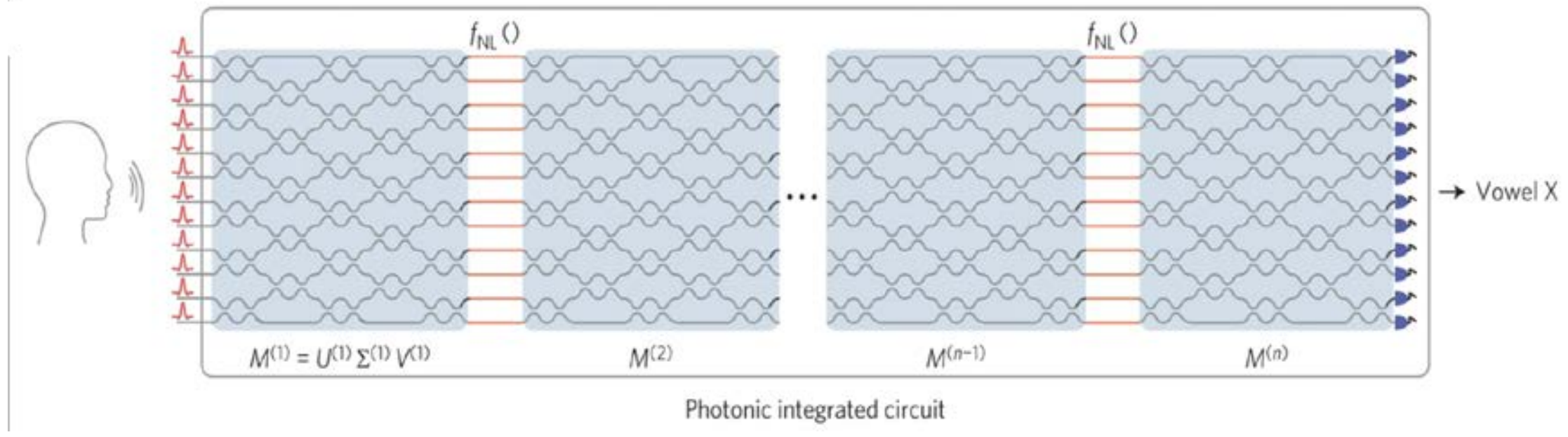

Figure 18: The architecture from [32] that utilizes SVD to implement the matrix multiplications for vowel recognition using a Photonic Integrated Circuit (PIC).

MZIs have been noted to have much larger footprint (which can be up to a few millimeters [162] or tens to hundreds of micrometers) than their counterparts (e.g., MRs) and as [132] noted, this large footprint in combination with accumulation of phase errors throughout an MZI-based mesh can limit the scalability of neural networks built with MZIs. There has been research, such as in [177], focusing on reducing the overall area consumption by these architectures, whether it be by utilizing methods to prune the weight matrices represented by MZI meshes or by utilizing other devices in tandem.

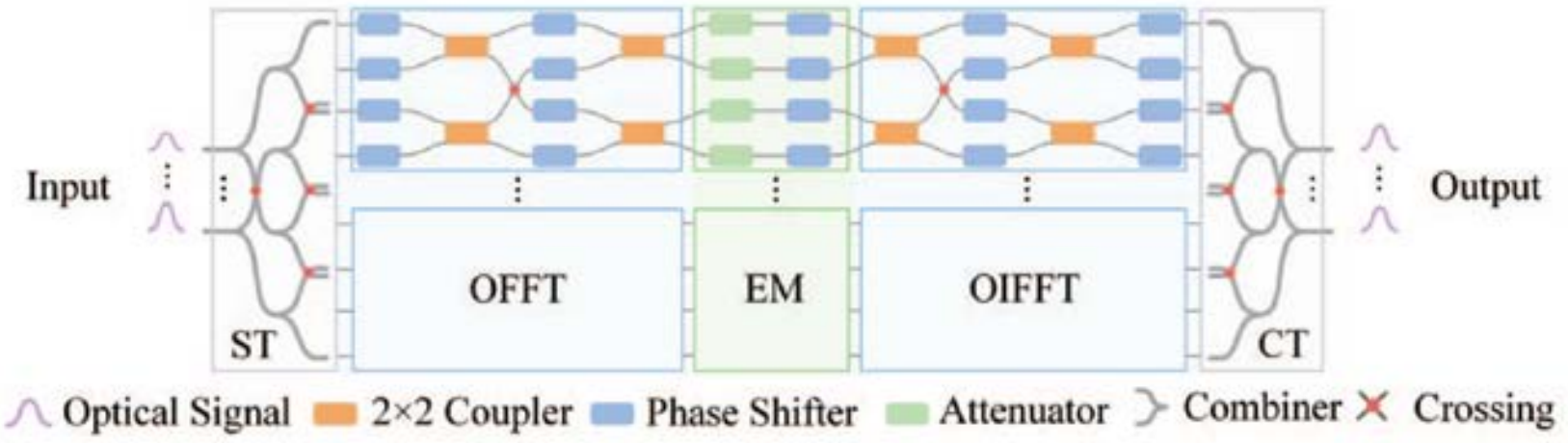

Figure 19: FFT-IFFT based photonic Structured Neural Network architecture described in [177]. The FFT based analyses of the model aims to simplify the model and thereby reduce energy and area consumption of the MZI meshes.

The work in [177] demonstrated a Fast Fourier Transform (FFT) based methodology to reduce area and energy footprint of MZI meshes used to implement MLPs. This is achieved by sparsifying the network through reducing the overall number of weights used, thereby compressing the neural network. The proposed architecture is based on structured neural networks with circulant matrix representation. Structured neural networks are a class of neural networks that are specially designed for computational complexity reduction, whose weight matrices are regularized using the composition of structured sub-matrices [178]. Structured neural networks utilize circulant weight matrices, which can be efficiently calculated using FFT and Inverse FFT (IFFT). The weight matrices are further pruned using Group Lasso regularization [179], and these operations can be implemented in MZI meshes using cascaded attenuator/amplifiers and phase shifters (figure 19). The authors of [177] adapted this methodology because of the difficulty in pruning SVD-based architectures. The architecture was tested using the *MNIST* data set against SVD-based architectures to show how effective their method is in reducing the overall area consumption of MZI meshes. The results indicate the architecture was able to achieve close to 98.5% testing accuracy while substantially reducing the overall area consumption.

The authors of [144] proposed a coherent MZI-based binary neural network implementation with weights restricted to +1 or -1. The activation function is a symbolic decision function which binarizes any real number mapped to it to +1 or -1. The weights of binarization are encoded onto the MZI by shifting voltages on the internal and external phase shifters on MZI arms. The real and imaginary parts of the two-way polarized In-phase and Quadrature component (IQ) modulated optical signal are used for training the model in simulation. The input to the model is the real and imaginary part of the signal, while the output is the prediction of the input position by the neural network. This work tested the architecture, comprised of seven hidden layers, for classifying to nearest neighbours the constellation formed by real and imaginary parts of a 100 GHz DP-QPSK signal. Close to 100% accuracy in classification was achieved for high Signal-to-Noise Ratio (SNR) input signals, while an accuracy close to 90% was achieved for low SNR signals.

As mentioned earlier, MZI-based architectures are typically coherent and utilize only a single wavelength. But MZIs have been used to implement WDM-based or noncoherent architectures as well. For instance, the work in [133] that demonstrated a photonic matrix multiplication accelerator using MZIs and Arrayed Waveguide Grating (AWG)-Multimode Interferences (MMIs). A single unit of AWG-MMI coupler balanced detector can successfully perform matrix multiplication by using WDM and coherent homodyne detection scheme. MZIs are utilized as intensity modulators, which feed into the multiplier (figure 20). The work in [143] demonstrated an all-optical WDM RNN utilizing an SOA-MZI as an activation unit incorporated into the feedback delay loop. In order to emulate a fully functional Gated-Recurrent-Unit (GRU), the authors integrated a gating mechanism (the SOA-MZI) to allow for agile reconfiguration of forget functions within a GRU. The SOA are embedded into the arms of the MZI, and act as cross gain modulation wavelength converters. An RNN was constructed using this unit and tested using a four-input WDM. The utility of the RNN was tested by running a finance forecasting benchmark application using the FI-2010 dataset. The gated optical RNN was able to achieve a higher F1 score (41.85%) than the optical and regular RNNs.

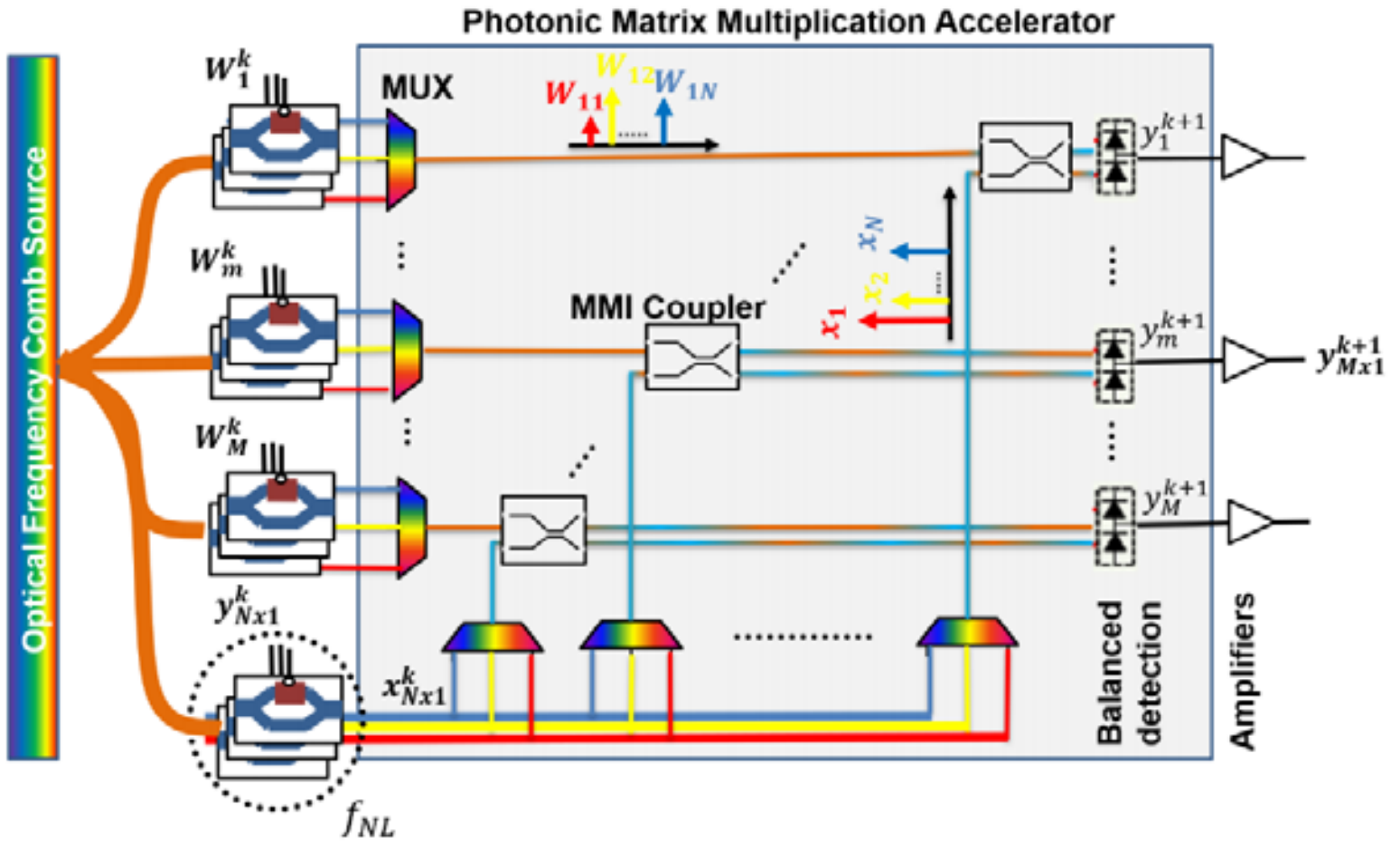

Figure 20: Cascadable analog feed-forward artificial neural network structure with photonic matrix-vector multiplier circuit and Mach–Zehnder modulator non-linearity, as depicted in [133].

In summary, coherent neuron microarchitectures that make use of interferometer devices such as MZIs have been widely used in photonic neural network architectures because of their ability to effectively represent matrices for neural network operations, but at the cost of a larger area overhead than MRs and susceptibility to phase-noise corruption. The basic principle of MZIs for neural network operations relies on phase and amplitude tuning of a passing optical wavelength, which can be easily achieved by integrating phase and amplitude tuners in MZI arms. MZIs are typically arranged in a mesh configuration in the works that use them, with SVD also being used to efficiently represent matrices. A reduction in MZI area footprint and phase-noise corruption is attempted via regularization approaches (e.g., [32], [144]) or by utilizing niche neural network models to reduce the overall MZI count ([177]). Usually, architectures which make use of MZIs use coherent principles to function, but MZIs have also been used in noncoherent approaches which make use of WDM, e.g., the RNN implementation in [133] which utilizes SOAs and MZIs in combination. They are not found to be used in any RC based architectures, probably because of the large area requirement an MZI based implementation would require to realize the large number of non-linear nodes in an RC implementation.

### 5.3 Diffractive Optics based Network Architectures

On-chip diffractive optics have also also used for implementing photonic neural network architectures. The obvious advantage of using these techniques is implementing the necessary functionalities passively, by leveraging the physics of diffractive optics. This is different from the MRs and MZIs as they are used as active devices which need active tuning (as in the case of MRs) or control mechanisms for phase control (as in the case of MZIs).

Various architectures discuss MLP implementations by integrating on-chip diffractive optics. These usually utilize AWG/Star couplers along with polarization controllers and SOAs to achieve the various functionalities needed to implement the neural network. The architecture described in [180] describes one such implementation where the AWGs are utilized to reduce noise and increase resolution of the accumulated weight values, demonstrating neuromorphic weighted addition operations in an 8x8 InP cross-connect. The weights are multiplied onto the optical signals by tuning the gain of SOAs. The output signals are then combined to accumulate the results and the weighted addition operation is executed by using PDs to obtain the resultant opto-current. A highly precise 4-bit precision multiplication and accumulation operation is achieved with an error of less than 0.2 in this system. The authors claim that this system can be scaled to form viable photonic DNNs.

The work in [134] explored a combination of AWGs (figure 21(a)) and MZIs to implement a CNN (figure 21(b)). The free-space propagation in the AWG was utilized to mimic an approximate Discrete Fourier Transform operation (DFT). Cascading two DFT operations with a phase and amplitude mask in between them was used to represent a convolution operation. The pooling layer was implemented as a low pass filter which only passes low frequency components of the DFT. The filter was implemented with three AWGs with a phase and amplitude mask between the first two. Lastly, the fully connected layer was implemented as an MZI mesh with tuneable attenuators/amplifiers in its arms. The MZI mesh implements SVD to represent the unitary matrix obtained from the DFT operations. The authors used the Cooley-Tukey FFT Algorithm [181] to reduce the number of MZIs used and thus reduce the implementation footprint. The Cooley-Tukey FFT algorithm utilizes a composite of DFTs to generate an approximation of the continuous FFT, and is extremely popular in FFT based applications. This work also explored how noise in the masks applied to the AWG outputs will affect the accuracy of the architecture for the MNIST dataset classification problem. The work explored how different noise sources would impact the test accuracy of the architecture, by considering Gaussian amplitude, phase, and complex noise addition to the AWG matrix. The architecture was shown to be resilient to noise, once it was trained with noisy input signals. By retraining the output layer with noise, the architecture was claimed to have substantial recovery in accuracy even with severely noisy inputs.

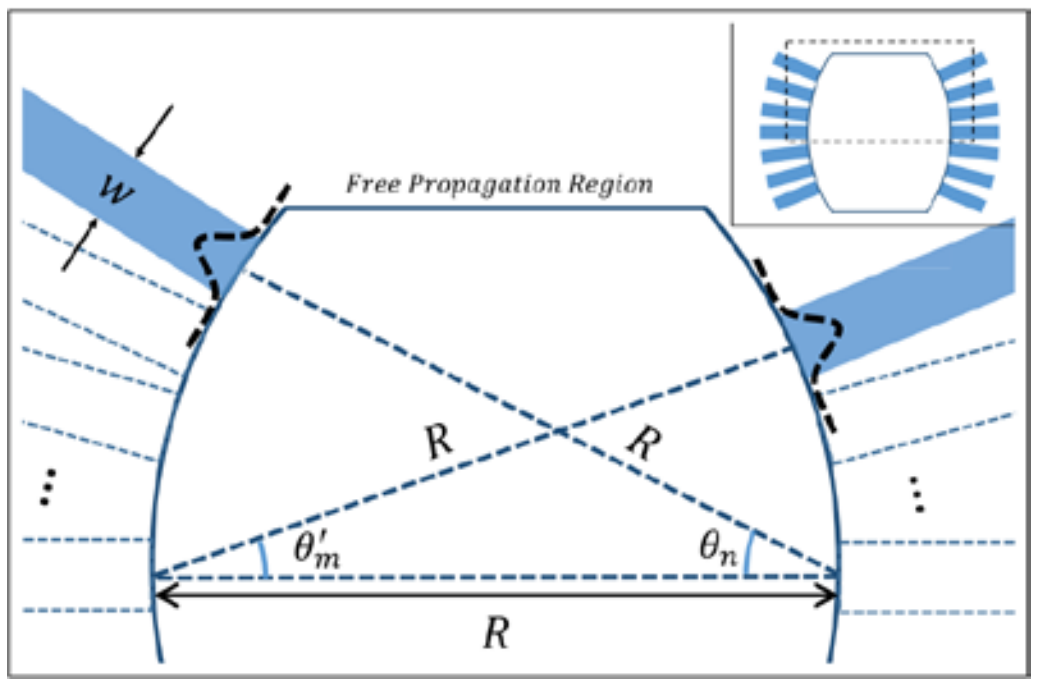

(a)

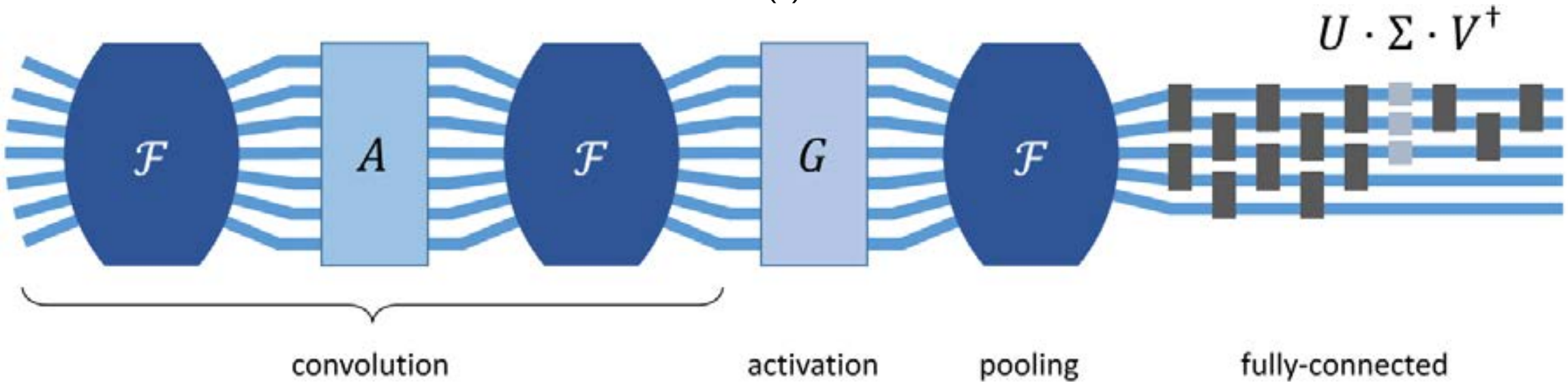

(b)

Figure 21: (a) Schematic of an N×M star coupler or an AWG. $R$ is the radius of the confocal circles that make up the free-space propagation region. $\theta_n$ is the angle of the n[th] input waveguide, $\theta_m$ is the angle of the m[th] output waveguide. $w$ is the waveguide mode width parameter. (b) An AWG based CNN implementation, which utilizes the fact that optical signals passing through the free propagation region of a star coupler undergoes Discrete Fourier Transform (DFT). $A$ and $G$ are filter masks [134].

The work in [34] discussed a reservoir based on the 4x4 swirl topology, with its readout layer comprised of non-linear optical modulators (figure 22). The reservoir utilized is of passive elements as in [170]. The notable approaches adapted here include a demonstration of using 4-Pulse-Amplitude Modulation (4-PAM) in a reservoir computing setting, where Boolean operations like XOR are the benchmark. In addition, the authors presented an RC architecture which uses pillar silica scatterers with cavity as the passive element in the reservoir. For experiments and simulations, the authors scaled their reservoir up to 20x20 nodes. They simulated this architecture, shown in figure 23, using FDTD simulations. The architecture also demonstrated classification capabilities by being trained to identify cancer cells from normal cells. The performance of this label-less classification was compared to previous work [129] that used pillar scatterers without cavity, which caused the work in [129] to use lower wavelength waves (UV) for the reservoir. The approach of using UV for this task was found to be impractical due to the high cost of UV lasers and the possible damage to the cells that UVs can cause. The new architecture in [34] based on pillar scatterers with cavity was reported to have achieved comparable accuracy to the approach in [129].

The on-chip diffractive mechanism of utilizing VCSELs to form a diffractively couple VCSEL array was used to form a reservoir in [182]. This work proposed to set weights using a spatial modulator. The architecture was tasked with header recognition and was able to recognize up to 5-bit headers. The work in [183] describes a large-scale system which employs diffractive mechanics in its readout layer, which is all-optical and is made of digital micromirrors. But the non-linearity is implemented in the electrical domain which severely limits the update rate to 5Hz. This work showcased an architecture with 2025 non-linear nodes, realized as a pixel in a Spatial Light Modulator (SLM). The SLM would display the current state of the reservoir as a speckle pattern which can be read using a camera and the next state needed for the reservoir is calculated and is encoded into the SLM. The architecture was tasked to predict the next step in the non-linear Mackey-Glass chaotic time series [184], with normalized mean-square error (NMSE) as the criteria to evaluate the performance of the architecture. The architecture was shown to achieve an NMSE value of 0.013 for the prediction task. Another reservoir architecture which utilized SLM is described in [185]. This architecture also utilized an SLM-based reservoir and was operated at 640 Hz, which the authors attribute to the superior SLM equipment. This architecture was also benchmarked using a Mackey-Glass chaotic time series prediction. The architecture has up to 16385 nodes, again as pixels in the SLM, and was reported to have an NMSE value below 0.3.

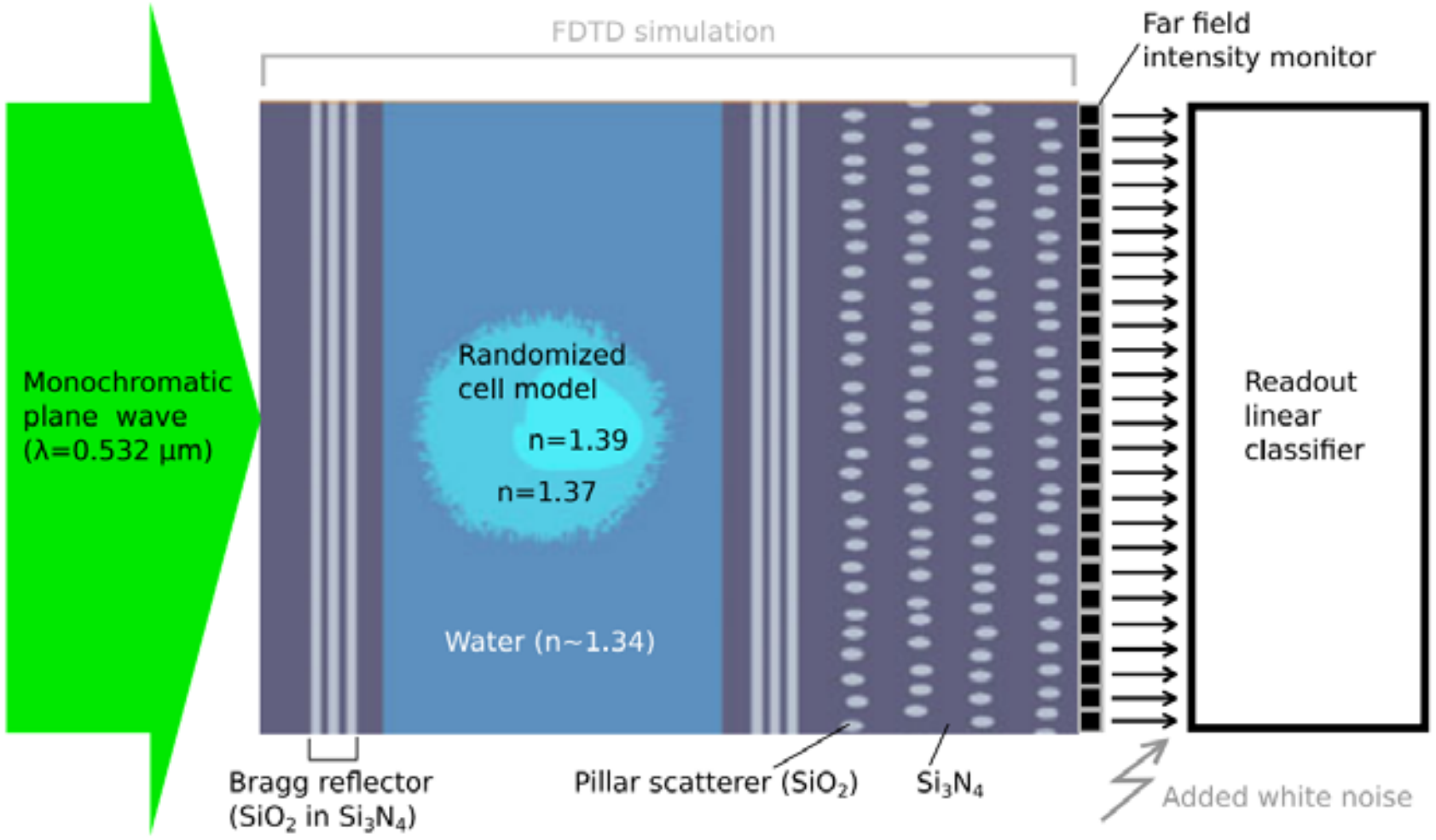

Figure 22: Detailed diagram from [34] depicting the FDTD simulation of their pillar scatterer based architecture showing the various components simulated.

In summary, diffractive optics have been used to implement MLP, CNN, RNN, and RC architectures. These implementations use a diverse set of devices, such as on-chip AWGs, passive elements like pillar scatterers, diffractively coupled VCSELs, and SLMs. These architectures tend to utilize the passive properties of optical devices to achieve the necessary functionalities, like the passive DFT transformation of light waves as they pass through AWGs or using SLMs to form huge reservoirs. Often, architectures which use diffractive optics also utilize other devices such as MZIs, SOAs, and VCSELs. However, the requirement of specifically designed devices (SLMs, micromirrors, scatterers) prevents programmability and implementation of compact and scalable implementations. As a result, diffractive optics based implementations are not as popular for chip-scale neural network acceleration.

### 5.4 Optical Amplification and Lasing based Architectures

Here we discuss photonic neural network architectures that utilize SOAs and VCSELs. These are prominently SNN implementations to realize STDP in the network. STDP is believed to be a fundamental plasticity mechanism in the synapses of the human brain [186]–[188]. This is how weights are assigned to synapses in the brain, depending on temporal relationships between pre-synaptic and post-synaptic spikes. The weight associated with the synapse is increased if the pre-synaptic spike appears before the post-synaptic spike and is decreased otherwise. This technique is usually implemented in photonic neuromorphic architectures by utilizing in-plane Semiconductor Optical Amplifiers (SOAs).

In [189], a DNN is implemented and experimentally verified with SOAs. The bias and the activation functions are implemented via digital electronics. The value of the bias is added to the data after detection. The work used a *tanh* activation function. This architecture implements one neuron operation by biasing up to six SOAs: one SOA as pre-amplifier, one SOA to select the input vector, and four SOAs acting as intensity modulators to represent weights. To represent the operation of a layer requires a total of four weighted additions, which are performed by biasing 21 SOAs: one pre-amplifier SOA, four SOAs for selecting the input vectors, and 16 SOAs acting as weights. AWGs were also used in the architecture, for multiplexing/demultiplexing wavelengths. The resulting neural network that was proposed had three layers of neurons. Fisher's Iris flower classification was used to test the accuracy of the architecture, on which a prediction accuracy of 85.8% was achieved in simulations when compared to 95% accuracy with digital electronics.

The work in [190] demonstrates optical STDP driven supervised learning utilizing an SOA and electro-absorption modulator (EAM). The linear combination of the gain depletion effects in SOA and absorption saturation in EAM is used to implement the effects of STDP. A teacher spike sample, which represents the expected spike train output, was used to train a pulse processing device, with the photonic STDP automatically tuning its gain so that the pulse processor matches the teacher spike sample. Following this model, the authors of [191] implemented reward-based reinforcement learning enabled by a photonic STDP unit constructed with two SOAs. This was an emulation of the biological behavior of STDP synapses and how the brain learns via principles of reinforcement learning. Here, a new modulatory element was introduced by varying the current injection into the SOA and used to emulate the reward function necessary for reinforcement learning. The work experimentally demonstrated how a reward function is tuned by the photonic STDP, depending on the reinforcement. The work in [123] demonstrated a photonic STDP module towards supervised learning and unsupervised pattern recognition based on a single SOA. The proposed setup demonstrated for the first time a generalized Hebbian algorithm [192] for synaptic modification, called Activity-Dependent Synaptic Plasticity (ADSP) in neuroscience. The SNN is photonic but the computation of correlation between post-synaptic and pre-synaptic signals were calculated using a CPU, along with calculating the update rule and controlling the SOA-based weight bank.

In [193], vertical-cavity SOAs (VCSOAs) [194]–[196] along with VCSELs were used to form photonic SNNs. VCSOAs are considered as they are VCSELs operating below their lasing threshold, thus providing ease of integration with the VCSELs, due to their structural similarities, and low power utilization. The authors based this implementation of VCSOA-based STDP on their previous work in [125], which introduced a theoretical and mathematical model to achieve photonic STDP using VCSOAs. The SNN was tested by tasking it to recognize arbitrary spike patterns. The results show the post-synaptic spike timing converging to the spike timing of the input spike train through supervised learning.

The authors of [197] discussed a fully connected photonic SNN consisting of excitable VCSELs with an embedded saturable absorber to implement spike sequence learning via supervised training. The authors incorporated photonic STDP into a classical remote supervised method (ReSuMe) algorithm to implement supervised training of the SNN. The work in [145] introduced fast VCSEL-neuron systems for neuromorphic photonic applications in two different architectures, namely a single VCSEL-neuron subject to delayed optical feedback, and two mutually coupled VCSEL-neurons. This emulated the operation of biological retinal neuronal circuits. The mutually coupled VCSEL-neurons were used to emulate the connection between bipolar cells and retinal ganglion cells in the eye, with a VCSEL-neuron representing the photoreceptors. By using these VCSEL-neurons, the study successfully emulated ON and OFF type neuronal circuitry in the eye.

In [198], coupled SOAs were used to form a reservoir. But employing active elements such as SOAs would make the reservoir architecture power inefficient, even though using the active elements greatly reduces the architecture footprint. They circumvented these issues by utilizing passive elements in an RC architecture in [170]. The work in [170] demonstrated a reservoir using only passive elements: waveguides, splitters, and combiners were the only components used in the reservoir. The reservois was realized as a 16-node square-mesh network with multiple feedback loops. In the architecture in [170], the required non-linearity is no longer within the reservoir and is implemented at the readout layer using PDs. The output from each node itself is a linear superposition of the complex amplitudes of the input waveguides of that node. At the readout layer, the complex amplitudes of the reservoir nodes are converted into real-valued power levels, which are then used as inputs for a linear classifier. The architecture from [170] was fabricated to perform basic Boolean tasks and header recognition for up to 5-bit headers using the proposed architecture. The authors also demonstrated the architecture's ability to recognize spoken digits and reported a minimum Word Error Rate (WER) of 4.5% for their coherent SOA based reservoir.

In [199], electrically modulated silicon-nano lasers (SNLs) are used as the reservoir layer of their RC architecture. The SNL's delay loop is used to generate the virtual nodes and time multiplexing is utilized to form the reservoir. The weights are set using a random weight matrix introduced via the input layer, while the weighting and linear summation takes place at the output layer. Weight optimization is done by minimizing the least-square error between the current and target weights. The task set to test the architecture was to predict the next step in the Santa-Fe chaotic time series [200]. The performance of the architecture was evaluated by calculating the normalized mean-square error (NMSE) between the predicted and target values. The feedback rate of the SNL was fine tuned to test the architecture and its performance. An NMSE of 0.0359 was reported for the feedback rate of 10 $ns^{-1}$.

In summary, the discussions in this section pertain mostly to the implementation of SNNs using SOAs and lasers. Different works listed in this section focused on STDP implementations for SNNs, for the most part, utilizing SOAs and VCSELs to achieve on-chip synaptic plasticity. There are also some RC architectures geared for machine-learning applications, which also use SOAs and lasers. We discussed a reservoir constructed from SOAs implementing the non-linear nodes and an all-passive photonic element reservoir. The passive element-based implementation was explored to circumvent the power and speed constraints of active elements such as SOAs on the reservoir. We also discussed a recent work where on-chip nano lasers called SNLs were used to form an RC architecture, where the delay loops of the laser were used to form virtual nodes which operated using time multiplexing.

### 5.5 Summary

The literature concerning photonic neural network architectures is vast, and so are the techniques and devices used to realize these architectures. In this section, we reviewed different architectures and divided the literature into resonator-based implementations, interferometer-based implementations, diffractive-optics-based implementations, and optical amplification/lasing based implementations. We have provided a summary of the literature on architectures covered as part of Section 5 in Table 3. The table has the references to the works (first column); the devices prominently used in the architecture (second column); a brief summary of the application(s) considered as part of experiments (third column); whether the work featured fabricated results or simulations or both, or none (fourth column); and the significant results provided in the paper (fifth column). A "—" in the table represents information that is not provided.

Table 3: Summary of prior work on photonic neural network architectures

| Reference | Devices utilized | Application | Fabricated (F) or Simulated (S) | Results achieved |
|---|---|---|---|---|
| [154] | MRs | Lorenz Attractor simulation to benchmark against a traditional CPU based CTRNN. | F | Reports 294× acceleration in simulation over traditional CPU based CTRNN. |
| [156] | MRs and MZMs | *AlexNet* CNN model | S | Claims 5 orders of magnitude faster speeds than fully electrical implementations. |
| [157] | MRs | MNIST classification using CNNs. | S | Faster when compared to GPU based implementations (2.8 to 1.4 times faster) and 0.75 times the power consumption. |
| [132] | MRs and SOAs. Weight fed into MR banks using memristor arrays. | Various benchmarks including MNIST tested on photonic CNNs. | S | Reduction in operation cost when compared to GPU based implementations, with up to 25× better computational efficiency. |
| [37] | MRs and MZIs | MNIST classification using MLPs. | S | Higher than 95% accuracy achieved at 14 bit resolution and custom MLP with 2048 neurons in hidden layer, for |
| | | | | both types of weights. Non-negative weights give lower accuracy. |
| [160] | MRs | CNNs acceleration. Image classification and one-shot learning acceleration demonstrated. | F + S | 9.5× lower energy per bit and 15.9× higher performance (frames per second) per watt on average than DEAP-CNN [157] and Holylight [161]. |
| [161] | Microdisks | Image classification with CNNs. | S | 13× better performance per Watt than ISAAC. |
| [162] | MRs and MZIs | Image classification with CNNs. | S | All optical design consumes only 5.1% the energy needed by all electrical accelerator, while being 31.9% faster. |
| [90] | MRs | This was an exercise to prove the feasibility of B&W based SNNs. No application-based experiment was conducted in this work. | – | – |
| [163] | GST embedded MRs | MNIST classification with MLPs. | S | 98.06% accuracy. |
| [169] | MRs | High-bit-rate digital pattern classification using RC. | S | Classification error of 0.5% at 160 Gbps for 8-bit-length digital words. |
| [170] | MRs | Demonstration of Boolean operations using RC. Detailed analysis of XOR operations. | S | Demonstrated XOR operations at an error rate of 0.1. Also explored the relationship between error rate and input power modulation and optical detuning. |
| [173] | MZIs | Mathematical discussion of phase and amplitude control for unitary operator representation, using MZIs. | – | – |
| [174] | MZIs | Single layer neural network using the 4x4 optical processor described in the work, set to classifying data samples | F + S | Demonstrated an accuracy of 72% in classification of data samples. |
| [175] | MZIs | Mathematical and theoretical discussion of MZI based unitary matrix representation, and consequently, how a universal linear device may be fashioned. No application-based testing done. | – | – |
| [176] | MZIs | Mathematical and theoretical discussion of MZI based unitary matrix representation, with added discussion into error and loss tolerance of such a device. | – | – |
| [144] | MZIs | Binary neural network set to nearest neighbor classification of a constellation formed from 100 GHz DP-QPSK signal. | S | Close to 100% accuracy in classification achieved for high SNR signal, while accuracy close to 90% was achieved for low SNR signal. |
| [32] | MZIs | Photonic DNN for vowel recognition | F + S | Achieved 76.7% accuracy in vowel recognition. Lower accuracy attributed to limited resolution (24 bits). |
| [177] | MZIs | MNIST dataset classification using Structured Neural Network. | S | 98.5% accuracy. |
| [133] | MZMs and MMIs | Analog feed-forward ANN with photonic MVM and MZM non-linearity demonstrated using a 2-by-1 vector dot-product experiment. Energy efficient binary multiplication demonstrated in simulation. | S | – |
| [143] | SOA-MZIs | RNN benchmarked using a finance forecasting application utilizing FI-2010 dataset | S | Gated optical RNN achieved an F1 score of 41.85% |
| [180] | AWGs and SOAs | Demonstration of precise 4-bit multiplication and accumulation operation | F + S | Error less than 0.2 |
| [134] | AWG and MZIs | MNIST classification with CNN architecture. CNN implemented using Cooley-Tukey FFT algorithm, with AWGs used to implement DFT photonically. | S | Various noise sources (amplitude, phase and linear noises) and their combinations introduced to the CNN; 99.6% accuracy for 14280 parameter CNN. |
| [34] | Pillar silica scatteres | XOR computation and label-less classification of cancer cell images from healthy cells | S | 20×20 node reservoir achieves symbol error rate below 5%. |
| [129] | Laser diodes | Label-less classification of cancer cell images from healthy cells | S | – |
| [182] | Diffractively coupled VCSELs | Demonstrated header recognition up to 5-bit headers. | S | – |
| [183] | SLM | Mackey-Glass chaotic time series prediction | S | Achieves an error of 0.013 for the prediction task. |
| [185] | SLM | Mackey-glass chaotic time series prediction | S | Reports NMSE below 0.3 for time series prediction task. |
| [189] | SOAs and AWGs | DNN implementation. Tested on Fisher's Iris classification. | F + S | Prediction accuracy of 85.8% achieved. |
| [190] | SOA and EAM | Experimental demonstration of photonic STDP and its utilization for supervised learning. | S | – |
| [191] | SOAs | Theoretical discussion and experimental demonstration of photonic STDP implementation using feedback signals. Demonstrated STDP used for reward based reinforcement-learning demonstration. | S | – |
| [123] | SOAs | Supervised and unsupervised pattern recognition. Demonstrated Hebbian algorithm for synaptic modification | S | – |
| [193] | VCSOA and VCSELs | SNN for learning and recognizing arbitrary spike patterns | S | – |
| [197] | VCSEL-SAs | SNN for learning and recognizing arbitrary spike patterns | S | – |
| [145] | VCSELs | SNN to simulate biological retinal neuronal circuitry. Simulated the ON and OFF stages of the retinal neuron circuitry. | S | – |
| [198] | SOA | Spoken digit recognition using RC. | S | The work reports a minimum Word Error Rate (WER) of 4.5% for their coherent SOA based reservoir |
| [171] | Passive photonic elements | Successful recognition of up to 5-bit headers and spoken digit recognition using RC. | S | Reports error rate "very close to" 0% |
| [199] | SNLs | Santa-Fe chaotic time series prediction | S | NMSE of 0.0359 obtained while the SNL is tuned to a feedback rate of $10ns^{-1}$ |

## 6 CHALLENGES AND OPPORTUNITIES

State-of-the-art silicon photonic devices have shown great promise to implement artificial neurons. Deep learning architectures built with photonic neurons support high parallelism in transmitting and processing weights by utilizing WDM, fast execution time, and low energy expenditure. However, there are several outstanding challenges to efficiently implement different neuron functions with silicon photonic devices, as well as to achieve high reliability, scalability, and cascadability in architecture implementations. Here we summarize challenges and opportunities for future research needed to overcome these challenges.

- **Coherency Challenges:** On-chip interferometers (e.g., MZIs) have been used extensively in photonic neural network architectures due to their ability to effectively represent matrices for neural network operations. The main issues with MZIs are the large area requirement and phase-noise corruption in MZI meshes. Due to thermal and fabrication-process variations in MZIs, the phase values can deviate from their target values, which can impact the inference accuracy of the neural network employing them. Recent efforts [201], [202] explore how to avoid these issues by factoring in these issues at the training phase and tuning the photonic neural network while considering variations. Further research is required to more efficiently overcome the area and noise limitations of these coherent architectures.
- **Noncoherency Challenges:** The broadcast-and-weight (B&W) protocol is widely employed for implementing photonic neural network architectures. Some efforts have recognized the possible issues with this protocol, such as heterodyne crosstalk corrupting the weight values and the very large number of MRs needed for implementing larger networks, especially when DWDM is utilized. Some recent efforts have attempted to address these limitations. The authors in [37] suggested a new architecture which utilizes parallel arrangements of MRs as opposed to cascading them, though they utilize the area inefficient MZIs for vector storage. The parallel "hitless" arrangement of MRs was used to reduce thermal crosstalk between adjoining MRs and to achieve better weight resolution. Other works have suggested reusing implemented layers, e.g., [156], and also methods to reduce energy consumption and increase speed of operation by reducing the involvement of electronic components [161]. There have also been suggestions to use microdisks instead of MRs to further increase integration density on chips [162]. But as noted in [203], all variants of noncoherent architectures can suffer from low throughput as the electronic components, such as the memory, may not be running at as high a frequency as their photonic components. Further research is needed to overcome these limitations of noncoherent architectures.
- **Variations and Reliability:** Many silicon photonic devices (e.g., MRs, MZIs) are susceptible to design time and runtime variations. Fabrication-process [204] and thermal crosstalk [89] as well as device aging [205] can considerably impact the reliability and performance in photonic neural networks by introducing undesirable crosstalk noise, optical phase shifts, resonance drifts (e.g., in MRs), tuning overheads, and photodetection current mismatches. For example, experimental studies have shown that the resonant wavelength in MRs can shift by 4.79 nm within a wafer due to inevitable fabrication-process variations [206], and deviate as large as 0.1 nm/K [207] due to runtime thermal variations. Moreover, silicon photonic devices intrinsically suffer from optical loss that degrade the energy efficiency, reliability, and scalability of photonic neural networks [208]. Also, the finite-encoding precision on phase settings (e.g., in coherent networks) adds extra uncertainty to the weight values obtained during network training, when mapped onto phase shifters as phase angles. A recent study [153] on the impact of uncertainties—due to fabrication-process and thermal variations—in photonic neural networks shows a significant 70% reduction in a photonic neural network inferencing accuracy. Therefore, further research is needed to improve the reliability in silicon photonic devices.
- **Power and Energy:** O/E/O neurons are power hungry because electro-optical and opto-electrical conversions consume considerable power. Moreover, O/E/O requires wavelength conversions to implement a large-scale neural network, which also consume extra power. Therefore, O/E/O might not be a good choice to achieve high power efficiency. All-optical neurons can achieve better power efficiency but at the cost of lower-speed operation (which can increase energy consumption) and reduced cascadability (which makes it difficult to implement complex functionalities). Off-chip lasers consume a significant portion of overall power in photonic neural networks. While such lasers are less susceptible to thermal variations than on-chip lasers, they incur extra optical-power loss due to the need to couple the off-chip light source to on-chip devices through coupling structure (e.g., grating couplers). Moreover, coping with variations (as discussed in the previous bullet) requires power and energy overheads to achieve reliability via spatial, temporal, or information redundancy. As power is such a significant design constraint in today's computing chips, there is thus an urgent need for new research to achieve power and energy efficient implementations of photonic neural networks.

- **Electronic Controllers:** The design of a photonic neural network would be unrealistic without considering its electronic controller challenges. Photonic neural networks require an electronic controller to manage (i.e., tune and control) and orchestrate photonic devices in the network (e.g., MR tuning and supervised learning control). Moreover, the controller should detect and mitigate runtime bias (e.g., due to thermal crosstalk) and maintain correct operation of optical neurons. However, electronic controllers impose high latencies, and there is a frequency mismatch between the electronic controller and the optical network. Therefore, more research is needed towards the implementation of high speed electronic controllers for photonic neural networks.
- **Backpropagation Training:** Almost all photonic neural network architectures in prior work focus on inference acceleration. There is a need to explore photonic architectures that can efficiently support neural network training. This is particularly challenging because training (e.g., via backpropagation) requires a backward flow of information from the output layers towards the input layers which would require additional waveguides, signals, and processing components to calculate gradients and update weight values. Some recent efforts have begun to explore such architectures, e.g., [209] which proposed a hybrid memristor+photonics based accelerator that also supports backpropagation. More research is needed towards the design of low overhead backpropagation support with photonics.
- **Resolution:** Weight resolution plays a crucial role in deep learning accelerator architectures. For inference acceleration, it is desirable to have higher resolution for better accuracy. Most prior works on photonic neural networks achieve very low resolution, such as the work in [157], which achieves 6–7 bits of resolution, and that in [37], which achieves 14 bits of weight resolution. Some proposed architectures tackle lower resolution by dividing weight representation among multiple devices, such as [161], or by utilizing bitwise parallelization of weight matrix operations as in [162], to achieve 16 bits of weight resolution. The work in [32] manages to achieve 24 bits of weight resolution using MZI meshes, but scalability of such an architecture is questionable because of large area consumption of MZIs. The work in [32] could not achieve higher resolution than 24 bits due to phase encoding noise in MZI phase shifters. The main challenges in achieving good resolution in photonic architectures stem from crosstalk noise, photodetector sensitivity, and photodetector noise (shot noise). While the work in [37] presented a detailed analysis on how thermal crosstalk impacts photonic sensitivity to weight values, even inter- and intra-channel crosstalk can affect the achievable resolution. Research is thus needed to achieve effective photonic crosstalk mitigation, phase noise correction, and noise resilient photodetection, to achieve better resolution in photonic deep learning accelerators.
- **Scalability:** Many of the works discussed in this survey have focused on implementing small neural networks [32], [37], [90], [163] to highlight the effectiveness of silicon photonic acceleration. Other works focus on accelerating matrix vector multiplication and reusing it over multiple layers of the deep learning model, such as in [157], [161], and [162]. A major issue that plagues implementations of large-scale networks using silicon photonic devices is area consumption, given that the basic components in a photonic neural network architecture can span micrometers in dimension. Also, the losses related to propagation and crosstalk accumulates over larger architectures involving very large device counts, and the power consumption can reach very high values [156]. MZI meshes, such as those presented in [32], [175], and [180], face severe issues related to area consumption (MZIs being much larger than MRs or microdisks) and phase noise, limiting their scalability. To reduce scalability concerns, some works consider a simplified version of the neural network model in hardware by utilizing regression techniques [177] and efficient matrix convolution calculation using FFT techniques [134], [178]. In order to realize scalable photonic accelerator designs, research is needed into 1) new model compression approaches for reducing silicon photonic hardware complexity, and 2) noise resilient, low loss, and compact silicon photonic devices that can support high cascadability to realize large neural networks.

## 7 REFERENCES

[1] G. E. Hinton, S. Osindero, and Y.-W. Teh, “A fast learning algorithm for deep belief nets,” Neural computation, vol. 18, no. 7, pp. 1527–1554, 2006

[2] V. K. Kukkala, J. Tunnell, S. Pasricha, and T. Bradley, “Advanced driver-assistance systems: A path toward autonomous vehicles,” IEEE Consumer Electronics Magazine, vol. 7, no. 5, pp. 18–25, 2018

[3] S. Levine, P. Pastor, A. Krizhevsky, J. Ibarz, and D. Quillen, “Learning hand-eye coordination for robotic grasping with deep learning and large-scale data collection,” The International Journal of Robotics Research, vol. 37, no. 4-5, pp. 421–436, 2018

[4] F. Monti, F. Frasca, D. Eynard, D. Mannion, and M. M. Bronstein, “Fake news detection on social media using geometric deep learning,” arXiv:1902.06673 [cs], Feb. 2019

[5] S. Lalmuanawma, J. Hussain, and L. Chhakchhuak, “Applications of machine learning and artificial intelligence for covid-19 (sars-cov-2) pandemic: A review,” Chaos, Solitons & Fractals, p. 110059, 2020

[6] K. Kukkala, S. V. Thiruloga, and S. Pasricha, “Indra: Intrusion detection using recurrent autoencoders in automotive embedded systems,” arXiv preprint arXiv:2007.08795, 2020

[7] J. Gu, G. Neubig, K. Cho, and V. O. Li, “Learning to translate in real-time with neural machine translation,” arXiv preprint arXiv:1610.00388, 2016

[8] W. S. McCulloch and W. Pitts, “A logical calculus of the ideas immanent in nervous activity,” The bulletin of mathematical biophysics, vol. 5, no. 4, pp. 115–133, 1943

[9] F. Rosenblatt, “The perceptron, a perceiving and recognizing automaton”, Report 85-460-1, Cornell Aeronautical Laboratory, 1957

[10] A. Merolla, J. V. Arthur, R. Alvarez-Icaza, A. S. Cassidy, J. Sawada, F. Akopyan, B. L. Jackson, N. Imam, C. Guo, Y. Nakamuraet al., “A million spiking-neuron integrated circuit with a scalable communication network and interface,” Science, vol. 345, no. 6197, pp. 668–673, 2014

[11] M. Davies, N. Srinivasa, T.-H. Lin, G. Chinya, Y. Cao, S. H. Choday, G. Dimou, P. Joshi, N. Imam, S. Jainet al., “Loihi: A neuromorphic manycore processor with on-chip learning,” IEEE Micro, vol. 38, no. 1,pp. 82–99, 2018

[12] P. Jouppi, C. Young, N. Patil, D. Patterson, G. Agrawal, R. Bajwa, S. Bates, S. Bhatia, N. Boden, A. Borcherset al., “In-datacenter performance analysis of a tensor processing unit,” Proceedings of the 44th Annual International Symposium on Computer Architecture, 2017

[13] V. Gokhale, J. Jin, A. Dundar, B. Martini, and E. Culurciello, “A 240 G-ops/s Mobile Coprocessor for Deep Neural Networks,” CVPR Workshop, 2014

[14] Z. Du, R. Fasthuber, T. Chen, P. Ienne, L. Li, T. Luo, X. Feng, Y. Chen, and O. Temam, “ShiDianNao: Shifting Vision Processing Closer to the Sensor,” International Symposium on Computer Architecture (ISCA), 2015

[15] C. Zhang, P. Li, G. Sun, Y. Guan, B. Xiao, and J. Cong, “Optimizing FPGA-based Accelerator Design for Deep Convolutional Neural Networks,” FPGA, 2015

[16] Y.-H. Chen, T.-J. Yang, J. Emer, and V. Sze, “Eyeriss v2: A flexible accelerator for emerging deep neural networks on mobile devices,” IEEE Journal on Emerging and Selected Topics in Circuits and Systems (JETCAS), vol. 9, no. 2, pp. 292-308, June 2019

[17] A. Parashar, M. Rhu, A. Mukkara, A. Puglielli, R. Venkatesan, B. Khailany, J. Emer, S. W. Keckler, and W. J. Dally, “SCNN: An accelerator for compressed-sparse convolutional neural networks,” International Symposium on Computer Architecture (ISCA), 2017

[18] S. Markidis, S. W. Der Chien, E. Laure, I. B. Peng, and J. S. Vetter, “Nvidia tensor core programmability, performance & precision,” pp. 522–531, 2018

[19] A. Sodani, R. Gramunt, J. Corbal, H.-S. Kim, K. Vinod, S. Chinthamani, S. Hutsell, R. Agarwal, and Y.-C. Liu, “Knights landing: Second-generation intel xeon phi product,” IEEE micro, vol. 36, no. 2, pp. 34–46, 2016

[20] A. Shafiee, A. Nag, N. Muralimanohar, R. Balasubramonian, J. P.Strachan, M. Hu, R. S. Williams, and V. Srikumar, “Isaac: A convolutional neural network accelerator with in-situ analog arithmetic in crossbars,” ACM SIGARCH Computer Architecture News, vol. 44, no. 3, pp. 14–26, 2016

[21] L. Song, X. Qian, H. Li, and Y. Chen, “Pipelayer: A pipelined reram-based accelerator for deep learning,” 2017 IEEE International Symposium on High Performance Computer Architecture (HPCA), pp. 541–552, 2017

[22] H. Tsai, S. Ambrogio, P. Narayanan, R. M. Shelby, and G. W. Burr, “Recent progress in analog memory-based accelerators for deep learning,” Journal of Physics D: Applied Physics, vol. 51, no. 28, p. 283001, 2018

[23] A. Amravati, S. B. Nasir, S. Thangadurai, I. Yoon, and A. Raychowdhury, “A 55nm time-domain mixed-signal neuromorphic accelerator with stochastic synapses and embedded reinforcement learning for autonomous micro-robots,” pp. 124–126, 2018

[24] H. Valavi, P. J. Ramadge, E. Nestler, and N. Verma, “A mixed-signal binarized convolutional-neural-network accelerator integrating dense weight storage and multiplication for reduced data movement,” pp. 141–142,2018

[25] A. Amaravati, S. B. Nasir, J. Ting, I. Yoon, and A. Raychowdhury, “A55-nm, 1.0–0.4 v, 1.25-pj/mac time-domain mixed-signal neuromorphic accelerator with stochastic synapses for reinforcement learning in autonomous mobile robots,” IEEE Journal of Solid-State Circuits, vol. 54,no. 1, pp. 75–87, 2018

[26] M. M. Waldrop, “The chips are down for Moore’s law,” Nature News, vol.530, no. 7589, p. 144, 2016

[27] S. Pasricha, and N. Dutt. “On-Chip Communication Architectures”, Morgan Kauffman, ISBN 978-0-12-373892-9, Apr 2008

[28] S. Pasricha, M. Nikdast, “A Survey of Silicon Photonics for Energy Efficient Manycore Computing” to appear, IEEE Design and Test, 2020

[29] L. Chrostowski, H. Shoman, M. Hammood, H. Yun, J. Jhoja, E. Luan,S. Lin, A. Mistry, D. Witt, N. A. Jaegeret al., “Silicon photonic circuit design using rapid prototyping foundry process design kits,” IEEE J. Sel. Top. Quantum Electron., vol. 25, no. 5, pp. 1–26, 2019

[30] D. A. Miller, “Silicon photonics: Meshing optics with applications,” Nature Photonics, vol. 11, no. 7, pp. 403–404, 2017

[31] A. R. Totovi ́c, G. Dabos, N. Passalis, A. Tefas, and N. Pleros, “Femtojoule per mac neuromorphic photonics: An energy and technology roadmap,” IEEE J. Sel. Top. Quantum Electron., vol. 26, no. 5,pp. 1–15, 2020

[32] Y. Shen, N. C. Harris, S. Skirlo, M. Prabhu, T. Baehr-Jones, M. Hochberg, X. Sun, S. Zhao, H. Larochelle, D. Englundet al., “Deep learning with coherent nanophotonic circuits,” Nature Photonics, vol. 11, no. 7, p. 441, 2017

[33] G. Van der Sande, D. Brunner, and M. C. Soriano, “Advances in photonic reservoir computing,” Nanophotonics, vol. 6, no. 3, pp. 561–576, 2017

[34] A. Katumba, M. Freiberger, F. Laporte, A. Lugnan, S. Sackesyn, C. Ma,J. Dambre, and P. Bienstman, “Neuromorphic computing based on silicon photonics and reservoir computing,” IEEE J. Sel. Top. Quantum Electron., vol. 24, no. 6, pp. 1–10, 2018

[35] G. Tanaka, T. Yamane, J. B. H ́eroux, R. Nakane, N. Kanazawa, S. Takeda, H. Numata, D. Nakano, and A. Hirose, “Recent advances in physical reservoir computing: A review,” Neural Networks, vol. 115, pp. 100–123, 2019

[36] A. N. Tait, M. A. Nahmias, Y. Tian, B. J. Shastri, and P. R. Prucnal, “Photonic neuromorphic signal processing and computing,'' in *Nanophotonic Information Physics*. Berlin, Germany: Springer, 2014

[37] Q. Cheng, J. Kwon, M. Glick, M. Bahadori, L. P. Carloni, and K. Bergman, “Silicon photonics codesign for deep learning,” Proceedings of the IEEE, 2020

[38] T. F. de Lima, H. Peng, A. N. Tait, M. A. Nahmias, H. B. Miller, B. J. Shastri, and P. R. Prucnal, “Machine learning with neuromorphic photonics,'' *J. Lightw. Technol.*, vol. 37, no. 5, pp. 1515_1534, Mar. 1, 2019

[39] L. De Marinis, M. Cococcioni, P. Castoldi, and N. Andriolli, “Photonic neural networks: A survey,” IEEE Access, vol. 7, pp. 175 827–175 841,2019

[40] A. N. Tait, A. X. Wu, T. F. De Lima, E. Zhou, B. J. Shastri, M. A. Nahmias, and P. R. Prucnal, “Microring weight banks,” IEEE J. Sel. Top. Quantum Electron., vol. 22, no. 6, pp. 312–325, 2016

[41] F. A. Azevedo, L. R. Carvalho, L. T. Grinberg, J. M. Farfel, R. E. Ferretti, R. E. Leite, W. J. Filho, R. Lent, and S. Herculano-Houzel, “Equal numbers of neuronal and nonneuronal cells make the human brain an isometrically scaled-up primate brain,” Journal of Comparative Neurology, vol. 513, no. 5, pp. 532–541, 2009

[42] T. Takeuchi, A. J. Duszkiewicz, and R. G. Morris, “The synaptic plasticity and memory hypothesis: encoding, storage and persistence,” Philosophical Transactions of the Royal Society B: Biological Sciences, vol. 369, no. 1633, p. 20130288, 2014
[43] A.L Hodgkin, A.F. Huxley, "A quantitative description of membrane current and its application to conduction and excitation in nerve", *The Journal of Physiology*, Aug 1952
[44] A. Borisyuk, “Morris–lecar model,” Encyclopedia of Computational Neuroscience. Springer, 2015
[45] S. Binczak, S. Jacquir, J.-M. Bilbault, V. B. Kazantsev, and V. I. Nekorkin, “Experimental study of electrical fitzhugh–nagumo neurons with modified excitability,” Neural Networks, vol. 19, no. 5, pp. 684– 693, 2006
[46] M. Hayati, M. Nouri, D. Abbott, and S. Haghiri, “Digital multiplierless realization of two-coupled biological hindmarsh–rose neuron model,” IEEE Transactions on Circuits and Systems II: Express Briefs, vol. 63, no. 5, pp. 463–467, May 2016
[47] L.F. Abbott, "Lapique's introduction of the integrate-and-fire model neuron (1907)" (PDF). Brain Research Bulletin, May 1999
[48] W. Maass. “Networks of spiking neurons: The third generation of neural network models”, Neural Networks 10, 1997
[49] E. M. Izhikevich, “Simple model of spiking neurons,” IEEE Transactions on neural networks, vol. 14, no. 6, pp. 1569–1572, 2003
[50] C. A. Runyan, E. Piasini, S. Panzeri, and C. D. Harvey, “Distinct timescales of population coding across cortex,” Nature, vol. 548, pp. 92–96,Jul. 2017
[51] J. Hines, “Stepping up to summit,” Comput. Sci. Eng., vol. 20, no. 2, pp. 78–82, 2018
[52] R. J. Douglas and K. A. C. Martin, “Recurrent neuronal circuits in the neocortex,” Current Biol., vol. 17, no. 13, pp. 496–500, 2004
[53] J. Hasler and B. Marr, “Finding a roadmap to achieve large neuromorphic hardware systems,” Front. Neurosci., vol. 7, Sep. 2013
[54] C.Mead, “Neuromorphic electronic systems,” Proc. IEEE, vol. 78, no. 10, pp. 1629–1636, Oct. 1990
[55] D. Tank and J. J. Hopfield, “Simple ‘neural’ optimization networks: An A/D converter, signal decision circuit, and a linear programming circuit,” IEEE Trans. Circuits Syst., vol. 33, no. 5, pp. 533–541, May 1986
[56] I. Sourikopoulos, S. Hedayat, C. Loyez, F. Danneville, V. Hoel, E. Mercier, and A. Cappy, “A 4-fj/spike artificial neuron in 65 nm CMOS technology,” Frontiers in neuroscience, vol. 11, p. 123, 2017
[57] J. Shi, S. D. Ha, Y. Zhou, F. Schoofs, and S. Ramanathan, “A correlated nickelate synaptic transistor,” Nature Commun., vol. 4, Oct. 2013
[58] W. Xu, S. Y. Min, H. Hwang, and T. W. Lee, “Organic core-sheath nanowire artificial synapses with femtojoule energy consumption,” Sci. Advances, vol. 2, no. 6, Jun. 2016
[59] J. Zhu, Y. Yang, R. Jia, Z. Liang, W. Zhu, Z. U. Rehman, L. Bao,X. Zhang, Y. Cai, L. Songet al., “Ion gated synaptic transistors based on 2d van der waals crystals with tunable diffusive dynamics,” Advanced Materials, vol. 30, no. 21, p. 1800195, 2018
[60] M. Prezioso, F. Merrikh-Bayat, B. Hoskins, G. C. Adam, K. K. Likharev, and D. B. Strukov, “Training and operation of an integrated neuromorphic network based on metal-oxide memristors,” Nature, vol. 521, no. 7550,pp. 61–64, 2015
[61] S. Park, M. Chu, J. Kim, J. Noh, M. Jeon, B. H. Lee, H. Hwang, B. Lee, and B.-g. Lee, “Electronic system with memristive synapses for pattern recognition,” Scientific reports, vol. 5, p. 10123, 2015
[62] I. Boybat, M. Le Gallo, S. Nandakumar, T. Moraitis, T. Parnell, T. Tuma,B. Rajendran, Y. Leblebici, A. Sebastian, and E. Eleftheriou, “Neuro-morphic computing with multi-memristive synapses,” Nature communications, vol. 9, no. 1, pp. 1–12, 2018
[63] S. Hu, G. Qiao, Y. Liu, L. Rong, Q. Yu, and Y. Liu, “An improved memristor model connecting plastic synapse and nonlinear spiking neuron,” Journal of Physics D: Applied Physics, vol. 52, no. 27, p. 275402, 2019
[64] X. Jin, S. B. Furber, and J. V. Woods, “Efficient modelling of spiking neural networks on a scalable chip multiprocessor,” pp. 2812–2819, 2008
[65] Introducing a Brain-inspired Computer, [Online]. Available: https://www.research.ibm.com/articles/brain-chip.shtml.
[66] Beyond Today’s AI, [Online]. Available: https://www.intel.com/content/www/us/en/research/neuromorphic-computing.html.
[67] S. Moore, “Intels neuromorphic system hits 8 million neurons, 100million coming by 2020,” IEEE Spectrum, vol. 15, 2019
[68] D. E. Rumelhart, G. E. Hinton, and R. J. Williams, “Learning representations by back-propagating errors,” nature, vol. 323, no. 6088, pp.533–536, 1986
[69] H. Esmaeilzadeh, A. Sampson, L. Ceze, and D. Burger, “Neural acceleration for general-purpose approximate programs,” pp. 449–460, 2012
[70] NVIDIA Corp., NVIDIA A100 Tensor Core GPU Architecture, Whitepaper, 2020
[71] P. Chi, S. Li, C. Xu, T. Zhang, J. Zhao, Y. Liu, Y. Wang, and Y. Xie, “Prime: A novel processing-in-memory architecture for neural network computation in reram-based main memory,” ACM SIGARCH Computer Architecture News, vol. 44, no. 3, pp. 27–39, 2016
[72] D. Verstraeten, S. Xavier-de-Souza, B. Schrauwen, J. A. K. Suykens, D. Stroobandt, and J. Vandewalle, “Pattern classification with CNNs as reservoirs,” Proc. Int. Symp. Nonlin. Theory Appl., Budapest, Hungary, pp. 101–104, 2008
[73] Y. Paquot, F. Duport, A. Smerieri, J. Dambre, B. Schrauwen, M. Hael-terman, and S. Massar, “Optoelectronic reservoir computing,” Scientific reports, vol. 2, p. 287, 2012
[74] R. Martinenghi, S. Rybalko, M. Jacquot, Y. K. Chembo, and L. Larger, “Photonic nonlinear transient computing with multiple-delay wavelength dynamics,” Phys. Rev. Lett., vol. 108, Jun. 2012
[75] L. Larger, M. C. Soriano, D. Brunner, L. Appeltant, J. M. Gutiˊerrez, L. Pesquera, C. R. Mirasso, and I. Fischer, “Photonic information processing beyond turing: an optoelectronic implementation of reservoir computing,” Optics express, vol. 20, no. 3, pp. 3241–3249, 2012
[76] F. Duport, B. Schneider,A. Smerieri, M. Haelterman, and S.Massar, “All optical reservoir computing,” Opt. Express, vol. 20, no. 20, pp. 22783–22795, Sep. 2012
[77] D. Brunner, M. C. Soriano, C. R. Mirasso, and I. Fischer, “Parallel photonic information processing at gigabyte per second data rates using transient states,” Nature Commun., vol. 4, 2013
[78] K. Hicke, M. A. Escalona-Morˊan, D. Brunner, M. C. Soriano, I. Fischer, and C. R. Mirasso, “Information processing using transient dynamics of semiconductor lasers subject to delayed feedback,” IEEE J. Sel. Top. Quantum Electron., vol. 19, no. 4, pp. 1 501 610–1 501 610, 2013
[79] M. C. Soriano, S. Ortˊın, D. Brunner, L. Larger, C. R. Mirasso, I. Fischer, and L. Pesquera, “Optoelectronic reservoir computing: tackling noise-induced performance degradation,” Optics express, vol. 21, no. 1, pp.12–20, 2013
[80] S. Ortin, M. C. Soriano, L. Pesquera, D. Brunner, D. San-Martin, I. Fischer, C. Mirasso, and J. Gutierrez, “A unified framework for reservoir computing and extreme learning machines based on a single time-delayed neuron,” Scientific reports, vol. 5, p. 14945, 2015
[81] F. Duport, A. Smerieri, A. Akrout, M. Haelterman, and S. Massar, “Fully analogue photonic reservoir computer,” Sci. Rep., vol. 6, 2016
[82] M. Nikdast, G. Nicolescu, J. Trajkovic, and O. Liboiron-Ladouceur, “Deeper: Enhancing performance and reliability in chip-scale optical interconnection networks,” Proceedings of the 2018 on Great Lakes Symposium on VLSI, pp. 63–68, 2018
[83] A. N. Tait et al., “Silicon photonic modulator neuron,” Physical Review Applied, vol. 11, no. 6, p. 064043, June 2019
[84] P. R. Prucnal, B. J. Shastri, T. F. de Lima, M. A. Nahmias, and A. N. Tait, “Recent progress in semiconductor excitable lasers for photonic spike processing”, Adv. Opt. Photonics, vol. 8, no. 2, pp. 228–299, Jun. 2016

[85] T. F. de Lima et al., “Noise analysis of photonic modulator neurons,” IEEE J. Sel. Top. Quantum Electron., vol. 26, no. 1, pp. 1–9, Jan-Feb 2020
[86] D. Liang and J. E. Bowers, “Recent progress in lasers on silicon,” Nat. Photonics, vol. 4, no. 8, Art. no. 8, Aug. 2010
[87] N. H. Zhu et al., “Directly modulated semiconductor lasers,” IEEE J. Sel. Top. Quantum Electron., vol. 24, no. 1,pp. 1–19, Jan-Feb 2017
[88] D. Vantrease et al., “Corona: System implications of emerging nanophotonic technology,” ACM SIGARCH Computer Architecture News, vol. 36, no. 3, pp. 153–164, 2008
[89] A. Mirza, S. M. Avari, E. Taheri, S. Pasricha, and M. Nikdast, “Opportunities for Cross-Layer Design in High-Performance Computing Systems with Integrated Silicon Photonic Networks,” 2020 Design, Automation Test in Europe Conference Exhibition (DATE), 2020
[90] A. N. Tait, M. A. Nahmias, B. J. Shastri, and P. R. Prucnal, “Broadcast and Weight: An Integrated Network For Scalable Photonic Spike Processing,” JLT, vol. 32, no. 21, pp. 4029–4041, 2014
[91] S. Xiang, Y. Zhang, X. Guo, A. Wen, and Y. Hao, “Photonic generation of neuron-like dynamics using vcsels subject to double polarized optical injection,” Journal of Lightwave Technology, vol. 36, no. 19, pp. 4227–4234, 2018
[92] Z. W. Song, S. Y. Xiang, Z. X. Ren, S. H. Wang, A. J. Wen, and Y. Hao, “Photonic spiking neural network based on excitable vcsels-sa for sound azimuth detection,” Optics Express, vol. 28, no. 2, pp. 1561–1573, 2020
[93] I. Aldaya, C. Gosset, C. Wang, G. Campuzano, F. Grillot, and G. Cas-tanon, “Periodic and aperiodic pulse generation using optically injected dfb laser,” Electronics Letters, vol. 51, no. 3, pp. 280–282, 2015
[94] G. Sarantoglou, M. Skontranis, and C. Mesaritakis, “All optical integrate and fire neuromorphic node based on single section quantum dot laser,” IEEE J. Sel. Top. Quantum Electron., vol. 26, no. 5,pp. 1–10, 2019
[95] F. Koyama, “Recent Advances of VCSEL Photonics,” JLT, vol. 24, no. 12, pp. 4502–4513, 2006
[96] J. Van Campenhout, P. Rojo-Romeo, P. Regreny, C. Seassal, D. Van Thourhout, S. Verstuyft, L. Di Cioccio, J.-M. Fedeli, C. Lagahe, and R. Baets, "Electrically pumped InP-based microdisk lasers integrated with a nanophotonic silicon-on-insulator waveguide circuit," Optics express, vol. 15, 2007
[97] J. Robertson, E. Wade, and A. Hurtado, “Electrically controlled neuron-like spiking regimes in vertical-cavity surface-emitting lasers at ultrafast rates,” IEEE J. Sel. Top. Quantum Electron., vol. 25,no. 6, pp. 1–7, 2019
[98] J. Robertson, M. Hejda, J. Bueno, and A. Hurtado, “Ultrafast optical integration and pattern classification for neuromorphic photonics based on spiking vcsel neurons,” Scientific reports, vol. 10, no. 1, pp. 1–8, 2020
[99] A. Hurtado, K. Schires, I. Henning, and M. Adams, “Investigation of vertical cavity surface emitting laser dynamics for neuromorphic photonic systems,” Appl. Phys. Lett, vol. 100, no. 10, p. 103703, 2012
[100] A. Levi, “Microdisk lasers,” Solid-state electronics, vol. 37, no. 4-6, pp.1297–1302, 1994
[101] L. Mahler, A. Tredicucci, F. Beltram, C. Walther, J. Faist, B. Witzigmann, H. E. Beere, and D. A. Ritchie, “Vertically emitting microdisk lasers,” Nature Photonics, vol. 3, no. 1, pp. 46–49, 2009
[102] Yisu Yang, Gligor Djogo, Moez Haque, Peter R. Herman, and Joyce K. S. Poon, "Integration of an O-band VCSEL on silicon photonics with polarization maintenance and waveguide coupling," Opt. Express 25, 2017
[103] K. Alexander, T. Van Vaerenbergh, M. Fiers, P. Mechet, J. Dambre, and P. Bienstman, “Excitability in optically injected microdisk lasers with phase controlled excitatory and inhibitory response”, Opt. Express, vol. 21, no. 22, pp. 26182–26191, 2013
[104] M. Tran, D. Huang, T. Komljenovic, J. Peters, A. Malik, and J. Bowers, “Ultra-Low-Loss Silicon Waveguides for Heterogeneously Integrated Silicon/III-V Photonics”, Appl. Sci., vol. 8, no. 7, pp. 1139-, 2018
[105] S. Nambiar, S. Purnima, and S. K. Selvaraja, "Grating-assisted fiber to chip coupling for SOI photonic circuits," Applied Sciences, vol. 8, 2018
[106] D. F. Siriani and K. D. Choquette, “Coherent Coupling of Vertical-Cavity Surface-Emitting Laser Arrays”, Semiconductors and Semimetals, vol. 86, pp. 226-264, 2012
[107] S. Maktoobi et al., “Diffractive Coupling for Photonic Networks: How Big Can We Go?”, IEEE J. Sel. Top. Quantum Electron., vol. 26, no. 1, pp. 1–8, Jan. 2020
[108] X. Wu et al., “UNION: A Unified Inter/Intrachip Optical Network for Chip Multiprocessors”, IEEE Trans. Very Large Scale Integr. VLSI Syst., vol. 22, no. 5, pp. 1082–1095, May 2014
[109] H. Shabani, A. Roohi, A. Reza, M. Reshadi, N. Bagherzadeh, and R. F. DeMara, “Loss-Aware Switch Design and Non-Blocking Detection Algorithm for Intra-Chip Scale Photonic Interconnection Networks”, IEEE Trans. Comput., vol. 65, no. 6, pp. 1789–1801, Jun. 2016
[110] A. N. Tait, T. F. de Lima, M. A. Nahmias, B. J. Shastri, and P. R. Prucnal, “Microring Weight Banks for Neuromorphic Silicon Photonics”, 2018 Conference on Lasers and Electro-Optics (CLEO), May 2018
[111] A. N. Tait, T. F. de Lima, M. A. Nahmias, B. J. Shastri, and P. R. Prucnal, “Multi-channel control for microring weight banks”, Opt. Express, vol. 24, no. 8, pp. 8895–8906, 2016
[112] T. V. Vaerenbergh et al., “Cascadable excitability in microrings,” Opt. Express, vol. 20, no. 18, pp. 20292–20308, Aug. 2012
[113] J. Xiang, A. Torchy, X. Guo, and Y. Su, “All-Optical Spiking Neuron Based on Passive Microresonator”, JLT, pp. 1–1, 2020
[114] Z. Ying et al., “Comparison of microrings and microdisks for high-speed optical modulation in silicon photonics,” Appl. Phys. Lett., vol. 112, no. 11, p. 111108, Mar. 2018
[115] Z. Yu, J. Zheng, P. Xu, W. Zhang, and Y. Wu, “Ultracompact Electro-Optical Modulator-Based Ge2Sb2Te5 on Silicon,” IEEE Photonics Technol. Lett., vol. 30, no. 3, pp. 250–253, Feb. 2018
[116] P. Xu, J. Zheng, J. Doylend, and A. Majumdar, “Non-Volatile Integrated-Silicon-Photonic Switches using Phase-Change Materials”, 2019 Asia Communications and Photonics Conference (ACP), Nov. 2019
[117] N. Dhingra, J. Song, G. J. Saxena, E. K. Sharma, and B. M. A. Rahman, “Design of a Compact Low-Loss Phase Shifter Based on Optical Phase Change Material”, IEEE Photonics Technol. Lett., vol. 31, no. 21, pp. 1757–1760, Nov. 2019
[118] Z. Cheng, C. Ríos, W. H. P. Pernice, C. D. Wright, and H. Bhaskaran, “On-chip photonic synapse,” Sci. Adv., vol. 3, no. 9, p. e1700160, Sep. 2017
[119] C. D. Wright, Y. Liu, K. I. Kohary, M. M. Aziz, and R. J. Hicken, “Arithmetic and Biologically-Inspired Computing Using Phase-Change Materials”, Adv. Mater., vol. 23, no. 30, pp. 3408–3413, 2011
[120] J. Feldmann, N. Youngblood, C. D. Wright, H. Bhaskaran, and W. H. P. Pernice, “All-optical spiking neurosynaptic networks with self-learning capabilities”, Nature, vol. 569, no. 7755, Art. no. 7755, May 2019
[121] S. Kim et al., “NVM neuromorphic core with 64k-cell (256-by-256) phase change memory synaptic array with on-chip neuron circuits for continuous in-situ learning”, 2015 IEEE International Electron Devices Meeting (IEDM), p. 17.1.1-17.1.4, Dec. 2015
[122] M. P. Fok, Y. Tian, D. Rosenbluth, and P. R. Prucnal, “Pulse lead/lag timing detection for adaptive feedback and control based on optical spike-timing-dependent plasticity,” Opt. Lett., vol. 38, no. 4, pp. 419–421, 2013
[123] R. Toole et al., “Photonic Implementation of Spike-Timing-Dependent Plasticity and Learning Algorithms of Biological Neural Systems”, JLT, vol. 34, no. 2, pp. 470–476, Jan. 2016

[124] F. Marino and S. Balle, “Experimental study of a broad area vertical-cavity semiconductor optical amplifier”, Opt. Commun., vol. 231, no. 1, pp. 325–330, Feb. 2004
[125] S. Xiang et al., “Numerical Implementation of Wavelength-Dependent Photonic Spike Timing Dependent Plasticity Based on VCSOA”, IEEE J. Quantum Electron., vol. 54, no. 6, pp. 1–7, Dec. 2018
[126] S. R. Restaino, “Introduction to Liquid Crystals for Optical Design and Engineering”, 2015, PDF ISBN: 9781628416619, Print ISBN: 9781628418071
[127] R. Bruck et al., “All-optical spatial light modulator for reconfigurable silicon photonic circuits,” Optica, vol. 3, no. 4, pp. 396–402, Apr. 2016
[128] A. Lugnan et al., “Photonic neuromorphic information processing and reservoir computing,” APL Photonics, vol. 5, no. 2, p. 020901, Feb. 2020
[129] A. Lugnan, J. Dambre, and P. Bienstman, “Integrated pillar scatterers for speeding up classification of cell holograms,” Opt. Express, vol. 25, no. 24, pp. 30526–30538, Nov. 2017
[130] P. Li et al, “All-optical Analog Comparator”, Nature Comm., 2016
[131] Aikawa, Yohei. "Ultracompact optical comparator for 4-bit QPSK-modulated signal based on silicon photonic waveguide." IEEE Photonics Journal VOL. 11, no. 3, pp. 1-10, 2019
[132] D. Dang, J. Dass, R. Mahapatra, “ConvLight: A Convolutional Accelerator with Memristor integrated Photonic Computing”, IEEE 24th International Conference on High Performance Computing (HiPC), 2017
[133] M. B. On, H. Lu, H. Chen, R. Proietti and S. J. Ben Yoo, "Wavelength-Space Domain High-Throughput Artificial Neural Networks by Parallel Photoelectric Matrix Multiplier," Optical Fiber Communications Conference and Exhibition (OFC), 2020
[134] J. R. Ong, C. C. Ooi, T. Y. L. Ang, S. T. Lim and C. E. Png, "Photonic Convolutional Neural Networks Using Integrated Diffractive Optics," IEEE J. Sel. Top. Quantum Electron, vol. 26, no. 5, pp. 1-8, Sept.-Oct. 2020
[135] S. Y. Xiang et al., “Cascadable Neuron-Like Spiking Dynamics in Coupled VCSELs Subject to Orthogonally Polarized Optical Pulse Injection”, IEEE J. Sel. Top. Quantum Electron., vol. 23, no. 6, pp. 1–7, 2017
[136] Z. Zhang, Z. Wu, D. Lu, G. Xia, and T. Deng, “Controllable spiking dynamics in cascaded VCSEL-SA photonic neurons”, Nonlinear Dyn., vol. 99, no. 2, pp. 1103–1114, 2019
[137] X. Zhuge, J. Wang, and F. Zhuge, “Photonic Synapses for Ultrahigh-Speed Neuromorphic Computing”, Phys. Status Solidi RRL – Rapid Res. Lett., vol. 13, no. 9, p. 1900082, 2019
[138] U. H. Lodish et al., “Molecular Cell Biology”, Macmillan, 2008.
[139] J. Robertson, T. Deng, J. Javaloyes, and A. Hurtado, “Controlled inhibition of spiking dynamics in VCSELs for neuromorphic photonics: theory and experiments”, Opt. Lett., vol. 42, no. 8, pp. 1560–1563, 2017
[140] A. N. Tait, J. Chang, B. J. Shastri, M. A. Nahmias, and P. R. Prucnal, “Demonstration of WDM weighted addition for principal component analysis”, Opt. Express, vol. 23, no. 10, pp. 12758–12765, 2015
[141] G. M. Alexandris et al., “Neuromorphic photonics with coherent linear neurons using dual-IQ modulation cells”, J. Lightw. Technol., vol. 38, no. 4, pp. 811–819, Feb. 2020
[142] F. Selmi, R. Braive, G. Beaudoin, I. Sagnes, R. Kuszelewicz, and S. Barbay, “Temporal summation in a neuromimetic micropillar laser”, Opt. Lett., vol. 40, no. 23, pp. 5690–5693, Dec. 2015
[143] G. M. Alexandris et al., "All-Optical WDM Recurrent Neural Networks With Gating," IEEE J. Sel. Top. Quantum Electron., vol. 26, no. 5, pp. 1-7, Sept.-Oct. 2020
[144] G. M. Alexandris et al., “Neuromorphic photonics with coherent linear neurons using dual-IQ modulation cells”, J. Lightw. Technol., vol. 38, no. 4, pp. 811–819, Feb. 2020
[145] J. Robertson, E. Wade, Y. Kopp, J. Bueno, and A. Hurtado, “Toward Neuromorphic Photonic Networks of Ultrafast Spiking Laser Neurons”, IEEE J. Sel. Top. Quantum Electron., vol. 26, no. 1, pp. 1–15, Jan. 2020
[146] A. Hurtado and J. Javaloyes, “Controllable spiking patterns in long-wavelength vertical cavity surface emitting lasers for neuromorphic photonics systems”, Appl. Phys. Lett., vol. 107, no. 24, 2015
[147] T. Deng, J. Robertson, and A. Hurtado, “Controlled Propagation of Spiking Dynamics in Vertical-Cavity Surface-Emitting Lasers: Towards Neuromorphic Photonic Networks”, IEEE J. Sel. Top. Quantum Electron., vol. 23, no. 6, pp. 1–8, Nov. 2017
[148] A. Hurtado, I. D. Henning, and M. J. Adams, “Optical neuron using polarisation switching in a 1550nm-VCSEL”, Opt. Express, vol. 18, no. 24, pp. 25170–25176, Nov. 2010
[149] T. Deng et al., “Stable Propagation of Inhibited Spiking Dynamics in Vertical-Cavity Surface-Emitting Lasers for Neuromorphic Photonic Networks”, IEEE Access, vol. 6, pp. 67951–67958, 2018
[150] C. Mesaritakis, M. Skontranis, G. Sarantoglou, and A. Bogris, “Micro-Ring-Resonator Based Passive Photonic Spike-Time-Dependent-Plasticity Scheme for Unsupervised Learning in Optical Neural Networks”, Optical Fiber Communications Conference and Exhibition (OFC), Mar. 2020
[151] B. J. Shastri, A. N. Tait, T. F. de Lima, M. A. Nahmias, H.-T. Peng, and P. R. Prucnal, “Principles of Neuromorphic Photonics”, ArXiv180100016 Phys., pp. 1–37, 2018
[152] Y. Zhao, D. Lombardo, J. Mathews, and I. Agha, “Low control-power wavelength conversion on a silicon chip”, Opt. Lett., vol. 41, no. 15, pp. 3651–3654, Aug. 2016
[153] S. Banerjee, M. Nikdast, and K. Chakrabarty, “Modeling Silicon-Photonic Neural Networks under Uncertainties,” IEEE/ACM Design, Automation and Test in Europe (DATE) Conference and Exhibition, 2021
[154] A. N. Tait et al., “Neuromorphic photonic networks using silicon photonic weight banks”, Sci. Rep., vol. 7, no. 1, Art. no. 1, Aug. 2017
[155] E. N. Lorenz, “Deterministic nonperiodic flow”, Journal of Atmospheric Sciences, 1963
[156] A. Mehrabian, Y. Al-Kabani, V. J. Sorger, and T. El-Ghazawi, “PCNNA: A Photonic Convolutional Neural Network Accelerator”, 31st IEEE International System-on-Chip Conference (SOCC), 2018
[157] V. Bangari et al., “Digital Electronics and Analog Photonics for Convolutional Neural Networks (DEAP-CNNs)”, IEEE J. Sel. Top. Quantum Electron., Volume: 26 , Issue: 1 , Jan.-Feb. 2020
[158] Y. LeCun, L. Bottou, Y. Bengio, and P. Haffner, “Gradient-based learning applied to document recognition”, Proceedings of the IEEE, 1998
[159] C. Zhang, Zhenman Fang, Peipei Zhou, Peichen Pan and Jason Cong, "Caffeine: Towards uniformed representation and acceleration for deep convolutional neural networks," ICCAD, 2016
[160] F. Sunny, A. Mirza, M. Nikdast, and S. Pasricha, “CrossLight: A Cross-Layer Optimized Silicon Photonic Neural Network Accelerator,” IEEE/ACM Design Automation Conference (DAC), 2021
[161] W. Liu, W. Liu, Y. Ye, Q. Lou, Y. Xie, L. Jiang, “HolyLight: A Nanophotonic Accelerator for Deep Learning in Data Centers”, Design, Automation Test in Europe Conference Exhibition (DATE), 2019
[162] K. Shiflett, D. Wright, A. Karanth, and A. Louri, “PIXEL: Photonic Neural Network Accelerator”, IEEE HPCA 2020

[163] I. Chakraborty , G. Saha, A. Sengupta and K. Roy, “Toward Fast Neural Computing using All-Photonic Phase Change Spiking Neurons”, Nature, August 2018
[164] M. Wuttig, N. Yamada, “Phase-change materials for rewriteable data storage”, Nat. Mater. 6, 824–832, 2007
[165] S. R. Ovshinsky, Reversible electrical switching phenomena in disordered structures”, Phys. Rev. Lett. 21, 1450–1453, 1968
[166] W. H. P. Pernice, H. Bhaskaran, “Photonic non-volatile memories using phase change materials”, Appl. Phys. Lett. 101, 171101, 2012
[167] C. Rios, P. Hosseini, C. D. Wright, H. Bhaskaran, W. H. P. Pernice, “On-chip photonic memory elements employing phase-change materials”, Adv. Mater. 26, 1372–1377, 2014
[168] T. Van Vaerenbergh, M. Fiers, P. Bienstman and J. Dambre, "Towards integrated optical spiking neural networks: Delaying spikes on chip," 2013 Sixth "Rio De La Plata" Workshop on Laser Dynamics and Nonlinear Photonics, Montevideo, 2013
[169] C. Mesaritakis, V. Papataxiarhis, and D. Syvridis, “Micro ring resonators as building blocks for an all-optical high-speed reservoir-computing bit-pattern-recognition system,” J. Opt. Soc. Amer. B, vol. 30, no. 11, pp. 3048–3055, Nov. 2013
[170] F. D. Coarer et al., “All-Optical Reservoir Computing on a Photonic Chip Using Silicon-Based Ring Resonators,” IEEE J. Sel. Top. Quantum Electron., vol. 24, no. 6, Nov/Dec 2018
[171] K. Vandoorne et al., “Experimental demonstration of reservoir computing on a silicon photonics chip”, Nature March 2014
[172] C. Mesaritakis, M. Skontranis, G. Sarantoglou and A. Bogris, “Micro-Ring-Resonator Based Passive Photonic Spike-TimeDependent-Plasticity Scheme for Unsupervised Learning in Optical Neural Networks”, OFC 2020
[173] M. Reck, A. Zeilinger, H. J. Bernstein, and P. Bertani, “Experimental realization of any discrete unitary operator”, Phys. Rev. Lett., vol. 73, no. 1, pp. 58–61, Jul. 2002
[174] F. Shokraneh, S. Geoffroy-Gagnon, M. S. Nezami and O. Liboiron-Ladouceur, "A Single Layer Neural Network Implemented by a 4×4 MZI-Based Optical Processor", IEEE Photonics Journal, vol. 11, no. 6, pp. 1-12, Dec. 2019
[175] D. A. B. Miller, “Self-configuring universal linear optical component [Invited],” Photon. Res., vol. 1, no. 1, p. 1, Jun. 2013
[176] W. R. Clements, P. C. Humphreys, B. J. Metcalf, W. S. Kolthammer, and I. A. Walsmley, “Optimal design for universal multiport interferometers”, Optica, vol. 3, no. 12, pp. 1460–1465, Dec. 2016
[177] J. Gu et al., “Towards Area-Efficient Optical Neural Networks: An FFT-based Architecture”, ASPDAC 2020
[178] Z. Li, S. Wang, C. Ding et al., “Efficient recurrent neural networks using structured matrices in fpgas,” ICLR Workshop, 2018
[179] J. Friedman, T. Hastie, and R. Tibshirani, “A note on the group lasso and a sparse group lasso,” arXiv preprint arXiv:1001.0736, 2010
[180] B. Shi, D. Bunandar, D. Englund and R. Stabile, “WDM Weighted Sum in an 8x8 SOA-Based InP Cross-Connect for Photonic Deep Neural Networks”, Photonics in Switching and Computing (PSC), 2018
[181] J. W. Cooley and J. W. Tukey, “An algorithm for the machine calculation of complex Fourier series,” Math. Comput., vol. 19, no. 90, 1965
[182] D. Brunner and I. Fischer, ``Reconfigurable semiconductor laser networks based on diffractive coupling,'' Opt. Lett., vol. 40, no. 16, pp. 3854_3857, 2015
[183] J. Bueno, S. Maktoobi, L. Froehly, I. Fischer, M. Jacquot, L. Larger, and D. Brunner, “Reinforcement learning in a large-scale photonic recurrent neural network,” Optica, vol. 5, pp. 756–760, 2018
[184] M. C. Mackey, L. Glass, “Oscillation and chaos in physiological control systems,” Science, vol. 197, pp. 287-289, 1977
[185] J. Dong, M. Rafayelyan, F. Krzakala, and S. Gigan, “Optical reservoir computing using multiple light scattering for chaotic systems prediction,” IEEE J. Sel. Top. Quantum Electron. Vol. 26, pp. 1–12, 2020
[186] G. Bi and M. Poo, “Synaptic modifications in cultured hippocampal neurons: Dependence on spike timing, synaptic strength, and postsynaptic cell type,” J. Neurosci., vol. 18, no. 24, pp. 10464–10472, Dec. 1998
[187] L. F. Abbott and S. B. Nelson, “Synaptic plasticity: Taming the beast,” Nature Neurosci., vol. 3, pp. 1178–1183, Nov. 2000
[188] G. Q. Bi and M. M. Poo, “Synaptic modification by correlated activity: Hebb’s postulate revisited,” Annu. Rev. Neurosci., vol. 24, pp. 139–166, Mar. 2001
[189] B. Shi, N. Calabretta and R. Stabile, “Deep Neural Network Through an InP SOA-Based Photonic Integrated Cross-Connect”, IEEE J. Sel. Top. Quantum Electron, VOL. 26, NO. 1, Jan/Feb 2020
[190] M. P. Fok, Y. Tian, D. Rosenbluth, and P. R. Prucnal, “Pulse lead/lag timing detection for adaptive feedback and control based on optical spike timing-dependent plasticity,” Opt. Lett., vol. 38, no. 4, pp. 419–421, Feb. 2013
[191] Q. Ren, Y. Zhang, R. Wang, and J. Zhao, “Optical spike-timing dependent plasticity with weight-dependent learning window and reward modulation,” Opt. Express, vol. 23, no. 19, pp. 25247–25258, Sep. 2015
[192] N. Caporale and Y. Dan, “Spike timing-dependent plasticity: a Hebbian learning rule,” Annual Review of Neuroscience, vol. 31, 2008
[193] S. Xiang et al., “STDP-Based Unsupervised Spike Pattern Learning in a Photonic Spiking Neural Network With VCSELs and VCSOAs”, IEEE J. Sel. Top. Quantum Electron, vol. 25, no. 6, Nov-Dec 2019
[194] A. Hurtado, I. D. Henning and M. J. Adams, “Effects of parallel and orthogonal polarization on nonlinear optical characteristics of a 1550 nm VCSOA,” Opt. Express, vol. 15, no. 14, pp. 9084–9089, Jul. 2007
[195] M. D. Sánchez, P. Wen, M. Gross, and S. C. Esener, “Rate equations for modeling dispersive nonlinearity in Fabry-Perot semiconductor optical amplifiers,” Opt. Express, vol. 11, no. 21, pp. 2689–2696, Oct 2003
[196] A. Hurtado and M. J. Adams, “Two-wavelength switching with 1550 nm semiconductor laser amplifiers,” J. Opt. Netw., vol. 6, no. 5, pp. 434–441, May 2007
[197] Z. Song et al., “Spike Sequence Learning in a Photonic Spiking Neural Network Consisting of VCSELs-SA With Supervised Training”, IEEE J. Sel. Top. Quantum Electron, VOL. 26, NO. 5, Sept/Oct 2020
[198] K. Vandoorne, J. Dambre, D. Verstraeten, B. Schrauwen and P. Bienstman, “Parallel reservoir computing using optical amplifiers”, *IEEE Trans. Neural Netw.*, vol. 22, no. 9, 2011
[199] X. Xing Guo et al., “High-Speed Neuromorphic Reservoir Computing Based on a Semiconductor Nanolaser With Optical Feedback Under Electrical Modulation”, IEEE J. Sel. Top. Quantum Electron, VOL. 26, NO. 5, Sept/Oct 2020
[200] A. S. Weigend and N. A. Gershenfeld, “Time series prediction: Forecasting the future and understanding the past,” 1993, [Online]. Available: http: //www-psych.stanford.edu/andreas/Time-Series/SantaFe.html
[201] Y. Zhu et al., “Countering Variations and Thermal Eects for Accurate Optical Neural Networks”, IEEE ICCAD 2020
[202] M.Y.-S. Fang et al., “Design of optical neural networks with component imprecisions”, Opt. Express vol. 27, pp. 14009-14029, 2019
[203] F. Zokae et al., “LightBulb: A Photonic-Nonvolatile-Memory-based Accelerator for Binarized Convolutional Neural Networks”, Design, Automation Test in Europe Conference Exhibition (DATE), 2020
[204] M. Nikdast, G. Nicolescu, J. Trajkovic, and O. Liboiron-Ladouceur, “Modeling fabrication non-uniformity in chip-scale silicon photonic interconnects,” Design, Automation & Test in Europe Conference & Exhibition (DATE), 2016.
[205] S. V. R. Chittamuru, I. G. Thakkar, and S. Pasricha. "Analyzing voltage bias and temperature induced aging effects in photonic interconnects for manycore computing." SLIP. 2017.

[206] J. S. Orcutt, A. Khilo, C. W. Holzwarth, M. A. Popovic, H. Li, J. Sun, T. Bonifield, R. Hollingsworth, F. X. Kartner, H. I. Smith, V. Stojanovic, and R. J. Ram “Nanophotonic integration in state-of-the-art CMOS foundries,” Optics Express, vol. 19, pp. 2335-2346, 2011
[207] A. Mahendra, C. Xiong, X. Zhang, B. J. Eggleton, and P. H. W. Leong, “Multiwavelength stabilization control of a thermo-optic system with adaptive reconfiguration,” Applied Optics, vol. 56, no. 4, pp. 1113-1118, 2017
[208] W. R. Clements, William R., P. C. Humphreys, B. J. Metcalf, W. S. Kolthammer, and I. A. Walmsley, “Optimal design for universal multiport interferometers,” Optica, vol. 3, no. 12, pp. 1460–1465, 2016
[209] D. Dang, S. V. R. Chittamuru, S. Pasricha, R. Mahapatra, D. Sahoo, “BPLight-CNN: A Photonics-based Backpropagation Accelerator for Deep Learning”, to appear, ACM Journal on Emerging Technologies in Computing Systems (JETC), 2021.